\documentclass[12pt,preprint]{aastex}
\usepackage{amssymb}

\newcommand{\threej}[6]
{ \left( \begin{array}{ccc}
#1&#2&#3\\
#4&#5&#6
\end{array} \right) }

\newcommand{\sixj}[6]
{ \left\{ \begin{array}{ccc}
#1&#2&#3\\
#4&#5&#6
\end{array} \right\} }

\begin{document}
\title{A near-IR line of \ion{Mn}{1} as a diagnostic tool of the \\
average magnetic energy in the solar photosphere}
\email{aasensio@iac.es}

\author{A. Asensio Ramos\altaffilmark{1}, M. J. Mart\'{\i}nez Gonz\'alez\altaffilmark{1}, 
A. L\'opez Ariste\altaffilmark{2}, J. Trujillo Bueno\altaffilmark{1,3} \& M. Collados\altaffilmark{1}}
\altaffiltext{1}{Instituto de Astrof\'{\i}sica de Canarias, 38205, La Laguna, Tenerife, Spain}
\altaffiltext{2}{THEMIS, CNRS-UPS 853, C/ V\'{i}a L\'actea s/n, 38200, La Laguna, Tenerife, Spain}
\altaffiltext{3}{Consejo Superior de Investigaciones Cient\'{\i}ficas, Spain}


\begin{abstract}
We report on spectropolarimetric observations of a near-IR line of \ion{Mn}{1} 
located at 15262.702 \AA\ whose intensity and polarization profiles are very
sensitive to the presence of hyperfine structure. A theoretical investigation of the magnetic sensitivity
of this line to the magnetic field uncovers several interesting properties. The most important one is that
the presence of strong Paschen-Back perturbations due to the hyperfine structure produces an intensity
line profile whose shape changes according to the absolute value of the magnetic field strength. A line
ratio technique is developed from the intrinsic variations of the line profile. This line ratio technique
is applied to spectropolarimetric observations of the quiet solar photosphere in order to explore the 
probability distribution function of the magnetic field strength. Particular attention is given to 
the quietest area of the observed
field of view, which was encircled by an enhanced network region. A detailed theoretical investigation shows that the inferred
distribution yields information on the average magnetic field strength and the spatial
scale at which the magnetic field is organized. A first estimation gives 
$\sim$250 G for the mean field strength 
and a tentative value of $\sim$0.45" for
the spatial scale at which the observed magnetic field is horizontally organized.
\end{abstract}

\section{Introduction}
Most part of the solar surface is covered by quiet regions that appear non-magnetic in typical
solar magnetograms. However, the investigation of such apparently non-magnetic regions is of 
great scientific interest, and it is important to determine how much photospheric magnetic flux remains 
hidden from view. The reason is that it might have a significant 
impact on the magnetic coupling to the outer atmosphere and on chromospheric and coronal heating 
\citep[e.g., the recent reviews by][]{sanchez_solarb04,schrijver05,trujillo_esa05,priest06,trujillo_asensio_shchukina_spw4_06}.

There are at least three possibilities for obtaining empirical information on the magnetism of the 
``quiet" Sun (e.g., Stenflo 1994): the polarization induced by the Zeeman effect, the Hanle effect 
and the Zeeman broadening of the intensity profiles. These techniques have, of course, their advantages 
and disadvantages, but the important point is that they can be suitably complemented in order to obtain 
far richer information on solar surface magnetism than that provided by a high-resolution magnetogram.

The detection of polarization produced by the Zeeman effect implies the presence of a magnetic 
field. The main problem with the polarization of the Zeeman effect as a diagnostic tool is that 
it is insensitive to magnetic fields that are tangled on scales too small to be resolved.
The solar photosphere is certainly expected to have highly tangled field lines with resulting mixed 
magnetic polarities on very 
small spatial scales, well below the current spatial resolution limit 
\citep[e.g.,][]{stenflo94,cattaneo99,sanchez_emonet03,stein_nordlund03,vogler_thesis03}. Consistently, 
the analysis of high-resolution observations of the polarization induced by the Zeeman effect has led to the conclusion
that the portion of the resolution element filled with a non-tangled magnetic field that
produces the observed net flux is of the order of 2\% \citep[e.g.,][]{stenflo94,lin95,dominguez03,khomenko03,martinez_gonzalez06}.
Although there is a general consensus with this conclusion, lack of agreement exists in the exact 
probability distribution function of the magnetic field strength associated with such fields that
produces a measurable non-zero net flux. The inferred distribution function is different in different
works
\citep{lin95,lin_rimmele99,socasnavarro02,dominguez03,khomenko03,socas_pillet_lites04,lites_socas04,martinez_gonzalez_spw4_06,martinez_gonzalez06,dominguez06}.
This point remains one of the key problems to clarify in the near future because such ${\sim}2\%$ of the magnetism of 
the quiet Sun might still carry a significant fraction of the total magnetic energy 
if the probability distribution function (PDF) turns out to present an important contribution at kG fields
\citep[as concluded by][]{dominguez03,dominguez06b}. 

The contradictory results obtained over the last few years suggest that it is fundamental to use a large
number of spectral lines when analyzing the magnetic field in unresolved structures \citep[e.g.,][]{semel81}.
For example, it has been shown recently \citep{martinez_gonzalez06} that the information encoded in,
for instance, the pair of \ion{Fe}{1} lines at 630 nm may not be enough to
constrain the thermodynamical and magnetic properties of the plasma. 
This helps to explain why the probability distribution function obtained by several authors
give different results. \cite{martinez_gonzalez06} have pointed out that the exact 
value of the magnetic field strength inferred from the observations critically depends on the
rest of poorly known thermodynamical parameters. As a consequence, it is mandatory to
include additional constraints (ideally obtained from observables) to obtain reliable magnetic
and thermodynamical information from the observations.

Following this strategy, \cite{arturo02,arturo06} have recently considered the polarization
of some \ion{Mn}{1} lines to investigate the magnetism of the quiet Sun.
Apart from thorium, all the elements of the periodic table present at least
one isotope with a non-zero nuclear spin. Selected species like \ion{Mn}{1}, 
\ion{V}{1} and \ion{Co}{1} present
strong observable perturbations in the Stokes profiles produced by the presence of hyperfine structure. Since the
energy separation between hyperfine levels is much smaller than that between fine structure
levels, strong perturbations take place in the line profiles when a magnetic field is present. 
These perturbations are a consequence of the transition to the intermediate Paschen-Back regime.
\cite{arturo02} investigated in detail the possibility of detecting hyperfine features in 
several lines of \ion{Mn}{1}, locating a spectral line at 5537 \AA\ that is of diagnostic 
interest. The advantage of these lines with hyperfine structure
is that a purely morphological inspection of the Stokes profiles can 
be used to infer which is the dominant magnetic field strength in the resolution element.  
In the case of the 5537 \AA\ line, these features allow to distinguish field strengths above and
below $\sim$900 G. Although again limited by cancellation effects (i.e., magnetic fields tangled
at scales below the resolution element cannot be detected), this diagnostic tool has the advantage that it 
can be used to get information on the magnetic field strength instead of simply the 
magnetic flux through the resolution element.

Obviously, the investigation of the magnetism of the quiet Sun cannot be done by using 
only the spectral line polarization induced by the Zeeman effect. We need to complement 
it with other diagnostic tools based on physical mechanisms whose observable signatures 
are sensitive to the presence of tangled magnetic fields at sub-resolution scales. These 
are the Hanle effect and the Zeeman broadening of the intensity profiles.

The Hanle effect is the modification of the atomic-level polarization
(and of the ensuing observable effects on the emergent Stokes profiles) caused by the action of a magnetic field
{\em inclined} with respect to the symmetry axis of the pumping radiation field
\citep[e.g., the recent reviews by][]{trujillo01,trujillo_esa05}. The basic formula to estimate
the magnetic field intensity, $B_{\rm H}$ (measured in G),
sufficient to produce a sizable change in the atomic level
polarization results from equating the Zeeman splitting
with the natural width (or inverse lifetime) of the energy level under consideration, thus
$B_H = 1.137\times 10^{-7} / (t_{\rm life} g_J)$.
In this expression, $g_{J}$ and $t_{\rm life}$ stand, respectively, for the Land\'e factor
and the level's lifetime (in seconds), which can be either the upper or the lower level of the chosen spectral line.
This formula provides a reliable estimation only when radiative transitions dominate
the atomic excitation, but its application to the upper and lower levels
of typical spectral lines is more than sufficient to understand
that the Hanle effect may allow us to diagnose solar and stellar magnetic fields
having intensities between milligauss and a few hundred gauss,
i.e., in a parameter domain that is very hard to study via the Zeeman effect alone.
For instance, the sensitivity 
range of the Sr {\sc i} 4607 \AA\ line lies between 2 and 200 G, approximately. This 
means that a volume-filling microturbulent field of 200 G would produce an amount of 
depolarization similar to that caused by a microturbulent magnetic field of 1000 G. 
It is the application of the Hanle effect in atomic and molecular lines which led 
\cite{trujillo_nature04} to conclude that there must be a 
vast amount of hidden magnetic energy and unsigned magnetic flux localized in the 
(intergranular) downflowing regions of the quiet solar photosphere, carried mainly 
by tangled fields at subresolution scales with strengths between the equipartition 
field values and ${\sim}1$ kG \cite[see also][for understanding the contraints they used for
establishing this upper limit of $\sim$1 kG]{trujillo_asensio_shchukina_spw4_06}.

The Zeeman broadening of the intensity profiles is proportional to the squared modulus of 
the magnetic field strength, $B^2$, so that it can also give us information 
on the presence of tangled magnetic fields at subresolution scales. The problem is that 
it is extremely difficult to disentangle the Zeeman line broadening from that due to the 
thermal and convective motions. Nevertheless, \cite{stenflo77} applied it to 
hundreds of visible iron lines and could establish via a statistical regresion analysis 
an upper limit of about 100 G for the case of a volume-filling and single value
microturbulent field. An attractive possibility to enhance the sensitivity of this diagnostic
tool is to use lines with larger wavelengths. For this reason, Asensio Ramos \& Trujillo 
Bueno (2006; in preparation) have theoretically investigated the Zeeman line broadening
technique using the \ion{Fe}{1} lines at 1.56 $\mu$m, 
pointing out that plausible distributions of relatively strong tangled magnetic fields in 
the intergranular regions of the quiet solar photosphere would produce a measurable 
Zeeman broadening signature in the red wing of the ${\lambda}15648.52$ line.

The previous summary of the advantages and disadvantages of the various techniques
that have been proposed suggests that it would be of great diagnostic interest to identify spectral
lines whose intensity profiles show still more sensitivity to the Zeeman line broadening
effect and whose polarization profiles could be used to distinguish more easily between
weak and strong fields. This is precisely the motivation which led us to search for near-IR
lines of manganese located at wavelengths that can be observed with the Tenerife Infrared Polarimeter.

The aim of this paper is to report on the interesting properties presented by the
15262.702 \AA\ \ion{Mn}{1} line in the near-IR. The Paschen-Back perturbations are so large 
that it is possible to detect perturbations not only in the circular polarization
profiles but also in the intensity profile. Since the perturbation of the intensity profile is sensitive
to the magnetic field strength, it does not suffer from cancellation effects. Consequently,
this spectral line can be used to diagnose the net flux and the average magnetic field strength in
the resolution element. The comparison of both measurements in a magnetically enhanced region of the quiet Sun 
sheds some light on the complex magnetism of the quiet Sun.

\section{Hyperfine structure}
Almost all the elements of the periodic table present an isotope with non-zero nuclear angular 
momentum $I$. Such nuclear angular momentum couples with the 
sum of the orbital and spin angular momentum $J$. Therefore, the fine structure levels characterized by 
their value of $J$ present
a splitting due to the precesion of $J$ around $I$. The hyperfine splitting
is usually much smaller than the fine structure splitting. When a magnetic field is present,
its interaction with the atom generates a splitting of the magnetic sublevels $M_F$ belonging to 
each hyperfine level 
$F$. Since the hyperfine splitting is small, the presence of a magnetic field of weak strength 
is sufficient to produce Zeeman splittings that are of the order of the energy level separation 
between consecutive $F$ levels. As a consequence, non-linear interactions among the 
magnetic sublevels arise produced by non-diagonal coupling terms in the total Hamiltonian. This 
regime of intermediate Paschen-Back 
effect\footnote{The decoupling of the nuclear angular momentum $I$ and the total electronic angular momentum
$J$ is known as the Back-Goudsmit effect. However, in the physics and astrophysics literature this effect 
is almost always termed ``the Paschen-Back effect for the hyperfine structure''. For this reason, we will follow
the standard terminology in the rest of this paper.} leads to strong perturbations on the Zeeman 
patterns, which have an important impact on the emergent Stokes profiles.

\subsection{Quantum mechanical treatment}
Several lines of species presenting hyperfine structure are located in the near-IR. We have
focused on a \ion{Mn}{1} line that is specially interesting. This line is situated 
at a wavelength of 15262.702 \AA\ and is produced by the transition between the fine structure
levels $e^8S_{7/2} - y^8P_{5/2}$. Manganese has only one stable isotope, $^{55}$Mn, which has
a nuclear angular momentum $I=5/2$. Consequently, all the manganese lines present signatures
produced by the hyperfine structure \citep{kurucz_hfs93,arturo02}. For zero magnetic field both $J$-levels present 
six $F$ levels that arise due to the coupling between $J$ and $I$. This follows from
the standard rule for angular momentum addition, that gives $F=|J-I| \ldots J+I$. 
The upper fine structure level presents hyperfine levels from $F=0$ to $F=5$, while the lower 
fine structure level presents hyperfine levels from $F=1$ to $F=6$. The energy splitting of these 
$F$ levels with respect to the original $J$ level (the fine structure level without hyperfine structure)
is given with very good approximation by \citep{casimir63}:
\begin{eqnarray}
\Delta_\mathrm{HFS}(J,F,I) &=& \frac{1}{2} AK \nonumber \\
&+& \frac{1}{2} B \frac{(3/4)K(K+1)-I(I+1)J(J+1)}{I(2I-1)J(2J-1)},
\end{eqnarray}
where
\begin{equation}
K=F(F+1)-I(I+1)-J(J+1).
\end{equation}
The energy splitting is represented in cm$^{-1}$ when the constants $A$ and $B$ are given in
cm$^{-1}$. These constants are the magnetic-dipole and electric-quadrupole hyperfine 
structure constants, and are characteristic of a given fine structure level. The line under 
investigation in this work has both levels with strong hyperfine interactions with coupling constants
$A_\mathrm{low}=25.2 \times 10^{-3}$ cm$^{-1}$ and $A_\mathrm{up}=27.5 \times 10^{-3}$ 
cm$^{-1}$ \citep{lefebvre03}. The electric-quadrupole constants are not known with sufficient precision
and we have decided to set them to zero. In any case, their influence is always much smaller than that 
of the magnetic-dipole interaction. In the presence of a magnetic field, these levels suffer from a
rapid transition to the intermediate Paschen-Back regime. Therefore,
the numerical diagonalization of the total hamiltonian (hyperfine+magnetic) turns out to be fundamental.
The energy level separation between consecutive fine structure levels is very large in comparison to 
the typical Zeeman splitting produced by the magnetic fields we are interested in. Then, it is enough
to focus only on the coupling between the hyperfine and magnetic interactions. The total Hamiltonian 
is block-diagonal and each block can be written as \citep[e.g.,][]{landi_landolfi04}:
\begin{eqnarray}
\langle (LS)&J I F M_F | H | (LS) J I F' M_F' \rangle
= \delta_{FF'} \delta_{M_F M_F'} \Delta_\mathrm{HFS}(J,F,I) \nonumber \\
&+ \delta_{M_F M_F'} \mu_0 B g_J (-1)^{J+I-M_F} \nonumber \\
& \times  \sqrt{J(J+1)(2J+1)(2F+1)(2F'+1)} \nonumber \\
&\times \sixj{F'}{F}{1}{J}{J}{I} \threej{F}{F'}{1}{-M_F}{M_F}{0},
\end{eqnarray}
where $\mu_0$ is the Bohr magneton, $B$ is the magnetic field strength and $g_J$ is the 
Land\'e factor of the level in L-S coupling. According to the 
NIST\footnote{\url{http://physics.nist.gov/PhysRefData/ASD/index.html}} 
database, the lower level has $g_J(e^8S_{7/2})=2$ and
it is known to be in L-S coupling. There is no information about the Land\'e factor for the
upper level, but according to \cite{kurucz_atomic93}, the Land\'e factor is $g_J(y^8P_{5/2})=2.284$,
similar to the value given in L-S coupling. If we assume that both levels are well described by
L-S coupling, the effective Land\'e factor of the line is $\bar g \approx 1.64$. 
When hyperfine structure is neglected, the line presents a weak magnetic sensitivity
due to the small value of the effective Land\'e factor.
However, we will see below that the presence of hyperfine structure produces
anomalous Stokes profiles of great diagnostic potential.

The total Hamiltonian is diagonal in $M_F$, so 
that it remains a good quantum number even in the presence of a magnetic field. This is not the
case with $F$, that looses its meaning as a good quantum number because the total Hamiltonian
mixes levels with different values of $F$. After a numerical diagonalization of the Hamiltonian,
the eigenvalues are associated with the energies of the $M_F$ magnetic sublevels. The 
transition between the upper and lower fine structure levels produce many allowed transitions
following the selection rules $\Delta M_F=0,\pm 1$. The strength of each component can be
obtained by evaluating the squared matrix element of the electric dipole operator \citep{landi_landolfi04}:
\begin{equation}
S_q^{i M_F, i' M_F'} \propto | \langle (LS) J I i M_F | r_q | (LS) J I i' M_F' \rangle |^2,
\end{equation}
where $q=M_F-M_F'=0,\pm 1$ and $| (LS) J I i M_F \rangle$ are the eigenvectors of the Hamiltonian. 
The symbol $i$ is used for labelling purposes since $F$ is not a good quantum number 
\citep[e.g.,][]{landi_landolfi04}.

Figure \ref{fig:splitting} shows the splitting of the upper and lower levels of the \ion{Mn}{1} transition
at 15262.702 \AA\ when the magnetic field is increased. Note the presence of the six $F$ levels
for both fine structure levels. When the field is smaller than 100 G, all the hyperfine 
levels are in the linear Zeeman regime, where the splitting is 
roughly proportional to the magnetic field strength. Since the hyperfine levels are so close in 
energy, interferences among the magnetic sublevels arise and all the hyperfine 
levels enter into the intermediate Paschen-Back regime. As a result, the $F$ quantum
number looses its meaning and the magnetic sub-levels start to interact. 
The coupling introduced by the magnetic hamiltonian only acts between
levels with the same value of $M_F$ belonging to different $F$ levels. Due to the special
structure of the Hamiltonian, they
anticross\footnote{Only the elements of the Hamiltonian matrix associated with every value of
$M_F$ having $|F-F'| \leq 1$ can be different from zero. It can be demonstrated
that the elements of the upper and lower diagonals are never zero, so that the matrix has non-degenerate
eigenvalues \citep[e.g.,][]{numerical_recipes86}. This shows that the energy levels with equal value of 
$M_F$ can never cross, but anticross.}. The levels with different values of $M_F$ are those that cross at the values of the
magnetic field shown in Fig. \ref{fig:splitting}. When the field strength is
sufficiently strong (stronger than the ones shown in the figure), the sublevels enter into the complete 
Paschen-Back regime, where a linear relation between the splitting of the levels and the 
magnetic field strength is again encountered.

The Zeeman patterns for two different magnetic field strength are shown in Fig. 
\ref{fig:zeeman_patterns}. These figures indicate the splitting and strength of each of the
transitions permitted by the selection rules between the two fine structure levels. Since the
$M_F$ levels of each $F$ level have the same energy for $B=0$ G, the Zeeman pattern is very simple.
When a magnetic field is present, this degeneracy is broken and a large number of components
following the $\Delta M_F=0,\pm 1$ selection rules appear. These figures clearly indicate the
complexity of the Zeeman pattern.

\subsection{Milne-Eddington synthesis}
The large number of level crossings that we find for magnetic field strengths typical of the solar atmosphere
translate into strong perturbations in the observed polarization spectrum.
In order to clarify this behavior, we have calculated the synthetic profiles emerging from 
a Milne-Eddington atmosphere in which a constant vertical magnetic field pointing away 
from the observer is present. This field topology is chosen for an easier comparison between
the synthetic profiles and the observations we present below. The results are shown in 
Fig. \ref{fig:milne} for four values of the magnetic field strength that cover approximately the expected
range in the quiet Sun: 100, 600, 900 and 1300 G. The thermodynamical parameters of 
the Milne-Eddington atmosphere are 
chosen so that the Stokes $I$ profile observed in weak magnetized regions is approximately fitted.
The figure clearly shows that the presence of hyperfine structure and level crossings produce 
strong perturbations in the emergent Stokes profiles. The intensity profile for 
zero magnetic field contains two lobes. The ratio between the relative absorptions in the two
lobes is close to 2. When the field increases 
the relative absorption in the two lobes tends to be similar. If the field increases even more, 
the wavelength distance between both lobes starts to increase. This great sensitivity of the 
Stokes $I$ profile to the magnetic field strength is produced by two effects. Firstly,
the line is in the incomplete Paschen-Back regime in a range of field strengths typical 
of the solar atmosphere. This produces perturbations in the Zeeman pattern that translates
into perturbations in the emergent Stokes parameters. Secondly, the Zeeman splitting 
$\Delta \lambda_B$ is proportional to $\lambda^2$, while the thermal broadening of the line is 
proportional to $\lambda$. Therefore, since the Zeeman broadening increases faster than the 
Doppler broadening, it is possible to detect the separation of the Zeeman components in these near-IR lines
for relatively weak fields.

Concerning the Stokes $V$ profile, we detect the presence of two peaks in the blue lobe of the line 
when the magnetic field is very weak. Additionally, the blue lobe is systematically broader and
shallower than the 
red lobe. This behavior is contrary to the one found from the correlation between
stronger fields (those that produce broader profiles) and redshifted material (in the intergranular regions).
When the field increases, the two peaks in the blue lobe change their relative strength.
For fields of the order of $\sim$700 G, the two peaks have the same amplitude and it is impossible
to detect them for fields above $\sim$1200 G because of the broadening of the profile. For stronger 
fields, the Stokes $V$ profile has the
typical antisymmetric shape. The amplitude of the Stokes $V$ lobes saturates at around 800 G, indicating
that the line is surely in the weak field regime for fields well below 800 G. For a spectral line without hyperfine 
structure and in the weak field regime, the amplitude of the Stokes $V$ signal is proportional to the 
magnetic flux density $\alpha B_\parallel$, where $\alpha$ is a magnetic filling factor and $B_\parallel$ is
the longitudinal component of the magnetic field. In the case of a line with hyperfine structure, the 
intrinsic morphological changes produced by the presence of hyperfine structure can help us 
distinguish the value of the magnetic field strength itself.

\subsection{Diagnostic capabilities}
It is interesting to analyze the variation of several intrinsic properties of the line for different
values of the magnetic field strength. These simple diagnostic tools illuminate the possible diagnostic capabilities of this 
\ion{Mn}{1} line. We have focused on two different properties: peak ratios and peak separations.
Peak ratios are the most conspicuous properties of the line because they present a large variation with the 
magnetic field strength (see Fig. \ref{fig:milne}). It is possible to define two different ratios, one for 
Stokes $V$ and another one for Stokes $I$. Their behavior are shown in Fig. \ref{fig:peak_ratio}. The ratio
defined for the blue lobe of Stokes $V$ is:
\begin{equation}
r_\mathrm{V} = \frac{\mathrm{V}(\lambda_\mathrm{rp})}{\mathrm{V}(\lambda_\mathrm{bp})},
\end{equation}
where $\lambda_\mathrm{rp}$ is the wavelength of the hyperfine feature in the blue lobe of Stokes $V$ profile 
that has a larger wavelength, while $\lambda_\mathrm{bp}$ is the wavelength of the hyperfine feature that presents
a shorter wavelength. This ratio is close to 2 when the field is weak and decreases monotonically when 
the field increases. As already noted above, the ratio becomes 1 for fields of the order of $\sim$700-800 G.
The left panel of figure \ref{fig:peak_ratio} presents the ratio for fields below 1000 G because the two peaks 
cannot be identified above this value and the ratio is no more reliable. Concerning Stokes $I$, we have defined the 
following ratio:
\begin{equation}
r_\mathrm{I} = \frac{\mathrm{I}(\lambda_\mathrm{rp})}{\mathrm{I}(\lambda_\mathrm{bp})}.
\end{equation}
This ratio, shown in the right panel of Fig. \ref{fig:peak_ratio} rapidly decreases when the field increases 
until arriving at a saturation above $\sim$700 G. Stronger fields present a ratio that goes again above 1. 
The curve exhibits a minimum, so that the ratio $r_\mathrm{I}$ is only reliable for fields below 600 
G. Assuming a linear functional form for the calibration curve, we can estimate the magnetic field
strength using $B \approx 600(2-r_\mathrm{I})$. 

It is of interest to consider what happens when the magnetic field within the resolution 
element is chaotic and presents random inclinations\footnote{The azimuth of the field does not
affect Stokes $I$ and can be considered random or deterministic.}. In this case, the peak ratio $r_\mathrm{I}$
remains quite similar to the one calculated for a vertical magnetic field and shown in the right
panel of Fig. \ref{fig:peak_ratio}. In order to obtain this result, the propagation matrix of the
Stokes-vector transfer equation has to be
averaged over all directions \cite[see \S9.25 of][]{landi_landolfi04}. The non-diagonal terms of the propagation
matrix vanish because of the isotropy of the magnetic field and because the absorption coefficient
for the intensity is equivalent to that obtained with a magnetic field with an inclination equal to the
Van Vleck angle ($\cos \theta_\mathrm{VV}=1/\sqrt{3}$). Therefore, we use the curve
shown in Fig. \ref{fig:peak_ratio} to estimate the magnetic field from the observations.

Finally, we have investigated the peak separation in the Stokes $I$
and Stokes $V$ profiles. The separation in \AA\ between the two peaks in the Stokes $I$ profile is shown 
in the left panel of 
Fig. \ref{fig:peak_separation_stokesi}. This separation remains constant for fields below 
$\sim$600 G and then increases almost linearly with the magnetic field. This behavior can
be understood because the centers of gravity of the $\sigma$ and $\pi$ components
are linear functions of the magnetic field strength even in the case that hyperfine structure is
present \citep[see \S3.5 of][]{landi_landolfi04}. This strong sensitivity of the peak separation to the magnetic
field strength can be used to discard between the two possible solutions that are consistent
with similar values of the ratio between the peaks of the Stokes $I$ profile. 
A similar behavior is found for Stokes $V$ and it is plotted in the right panel of Fig. 
\ref{fig:peak_separation_stokesi}. This is the so-called strong field regime of the Zeeman
effect in which the peak separation of the two lobes of the Stokes $V$ profile increases.
The plot is limited to fields above $\sim$900 G
because the peak separation cannot be defined for weaker fields due to the
appearance of the hyperfine features. The Stokes $V$ peak separation also
increases linearly. 

\section{Observations}
We have performed exploratory spectropolarimetric observations of the spectral region around 
15260 \AA\ with the Tenerife Infrared Polarimeter \citep[TIP; ][]{martinez_pillet99} 
mounted on the Vacuum Tower Telescope (VTT) at Iza\~na (Tenerife). 
The observations were taken during June 8$^\mathrm{th}$ 2002. We performed a 
scanning of a magnetically enhanced region of the quiet solar photosphere containing an
enhanced network region of circular shape with an internetwork region 
inside. The seeing conditions were not especially good and the spatial resolution 
was of the order of $\sim$1.4".
The region was situated very close to the disk center ($\mu \approx 1$) and three small pores were also 
in the field of view. The left panel of Figure \ref{fig:totalV} shows an image taken in the local continuum
of the 15262.702 \AA\ line. The
right panel of the figure shows the map of the integrated absolute value of the circular 
polarization signal. The annular enhanced network region is clearly seen in 
white. The polarimetric signals of the annular region are quite strong in the
\ion{Mn}{1} line. Except for very few points, the linear polarization signals could not be 
detected. However, we detected Stokes $V$ profiles produced by the longitudinal component of the magnetic field vector. 
Outside the annulus, the circular polarization signals were very 
weak.

The value of the signal-to-noise in the observed internetwork region is poor. In order to 
increase it, we have carried out a de-noising procedure based on the PCA
decomposition of the Stokes $V$ data. After having obtained the eigenvectors along the 
directions of maximum covariance, the Stokes $V$ data has been projected on a 
manifold of reduced dimension using the first 8 eigenvectors (those that carry the
statistically most relevant information). We have verified that this approach highly reduces the 
noise without introducing strong perturbations in the Stokes profiles with respect to the original ones.

\section{Results and Discussion}
\subsection{Stokes $V$}
Even after applying the above-mentioned noise reduction technique, the observations do not allow us to 
investigate in detail the Stokes $V$ profiles 
pixel by pixel. Therefore, we have focused on a statistical approach by means
of a Stokes $V$ profile classification. In order to perform this classification, we have 
used a self-organizing map \citep[SOM;][]{kohonen_SOM01}, also known as Kohonen's network. 
The SOM is a special kind of neural network that is usually applied for visualization 
and classification purposes \citep[e.g.,][]{maehoenen_SOM95,brett_SOM04}. It can 
be considered as a nonlinear mapping between the high-dimensional
input space of Stokes parameters (in fact the space spanned by the first 8 eigenvectors of the PCA
decomposition) and a regular two-dimensional grid. It consists 
of a two-dimensional grid of neurons where each neuron is associated with a feature in the 
original high-dimensional input space (the Stokes profiles observed for all
the pixels in our field of view), with nearby neurons containing features that present 
similarities. In a way similar to a PCA classification, the SOM is an unsupervised
classification method and it is capable of clustering data in the input space parameter.
The PCA classification can only search for linear features in the high-dimensional
input space. In this sense, the SOM classification produces much better results because
it can detect nonlinearities in the input space. Since the classification is performed in an iterative
scheme with initially random profiles in all the neurons, the final result typically
depends on the number of neurons and on the initial configuration. This is produced by
local minima in the learning process. Therefore, it is fundamental to
carry out the learning process a sufficiently large number of times and verify that the
same solution is obtained in a large fraction of such runs. A more detailed description
is presented in Appendix \ref{sec:app_som}.

The Stokes $V$ profiles are first normalized so that each profile has unit length if 
considered as a vector. We have carried out two different classifications, one with
a 4$\times$4 map and another one with a 6$\times$6 map, obtaining quantitatively
similar results. For clarity purposes, we only
show the results of the 4$\times$4 map.
The resulting profiles assigned to each neuron of the map 
are shown in Fig. \ref{fig:classes_4x4}. It presents profiles with the
typical antisymmetric shape in one corner, while the profiles become 
more and more distorted when moving towards the opposite corner. 
The profile associated with each neuron that the SOM classification produces may not
have physical sense. In order to overcome this problem, we have calculated the Stokes $V$ profile
corresponding to each neuron by averaging all the pixels that have been associated with each class.
The results are shown in Fig. \ref{fig:average_profile_4x4}.

It is immediate to verify that the SOM has generated a classification in which profiles are ordered 
according to the field strength (see Fig. \ref{fig:milne}). In each row, profiles associated with stronger
fields (profiles in which the two peaks in the blue lobe are absent or hardly present) are located more to the right. The magnetic field decreases for the
profiles located to the left because the two peaks in the blue lobe present a value
of $r_\mathrm{I}$ that increases monotonically. One can argue that this may be 
produced by the presence of noise. As an argument in favor of the physical significance of the SOM classification, we can
say that it has been carried out without any knowledge of the physical 
ingredients that produce these 
hyperfine signatures. As a result, since the final classification shows profiles that
follow the trend obtained from the theory (Fig. \ref{fig:milne}), we are inclined to think
that these features are real and not produced by noise. The ratio of the two peaks 
is smaller than 1 for large fields (above $\sim$900 G) and they become larger
than 1 for weaker fields. Of course, such a one-to-one relation between peak ratio
and magnetic field strength is only possible in the case that only one field strength is
present within the resolution element. In the more general situation where a combination
of magnetic fields is present in the resolution element, this association is not possible. 
In spite of this, the presence of the hyperfine signature in the blue lobe of the Stokes $V$ 
profiles indicates the presence of a field predominantly weaker than $\sim$700 G within 
the resolution element. In this sense, the 15262.702 \AA\ line presents a diagnostic 
potential similar to that of the \ion{Mn}{1} lines investigated by \cite{arturo02,arturo06}. 
The profiles in the top left part of the map are associated with 
weak fields in which noise and the presence of mixed polarities plays an important 
role. For this reason, these profiles are highly distorted. For a fixed column in the SOM, the vertical 
variation of the profiles appears to be also related to the strength of the 
field, with the field decreasing when moving upwards in each column.

For comparison purposes, Fig. \ref{fig:classes_4x4} indicates the percentage of points in the 
observed map that are associated with each neuron of the SOM. A large portion of the 
field-of-view is covered by the enhanced network region, so that the most abundant 
profiles are associated with those of stronger fields. 
In order to better understand the classification carried out by the SOM, we indicate the location
of each class in the observed map in Fig. \ref{fig:location_4x4}. The background
images in the figures represent (in grayscale) the class of each point as obtained with the SOM.
Each panel represents one of the rows of the classification network. Each color 
is associated with the profiles belonging to the columns of each row.
The neurons of the fourth row of the map are clearly associated with the
points belonging to the magnetically enhanced region that include the three pores.
The neurons of the first row are clearly associated with the internetwork regions.
The neurons belonging to the second and third row are associated with regions
located between the network and the internetwork. The classification has detected 
the presence of ring-like regions that have been classified in the same row suggesting 
some kind of segregation of the magnetic field strength and a relation between the network
and in the internetwork. Inside each ring-like structure, it is
possible to find a distribution of magnetic fields whose distribution clearly shows 
smaller values for the inner parts than for the outer parts.

A magnetic field can be approximately associated with the profiles of the classes with the aid of 
the Milne-Eddington calculation and the calibration curve presented in 
Fig. \ref{fig:peak_ratio}. Obviously, this can only be done for those profiles that present the 
hyperfine signature in the blue lobe of the Stokes $V$ profile. The values used in this work are 
indicated in Table \ref{tab:peak_ratio_4x4}.
The profiles for which the hyperfine features cannot be easily identified because 
the profile shows a clear antisymmetric shape are assigned an arbitrary field larger than 1 kG.
We do not assign any magnetic field strength to the noisy profiles.

Assuming only one magnetic component in the resolution element, it is possible
to plot maps of the magnetic field strength. Figure \ref{fig:magnetic_field_map} shows the 
map of field strength resulting from associating each class with a magnetic field strength. The kG fields
have been saturated to 1000 G (they can be larger, but not smaller), while the noisy profiles 
are set to 0 G. It is interesting to note that the strong field strengths profiles are only placed in 
the magnetically enhanced region, specially in the pores and the surrounding regions. 
When we move towards the center of the internetwork region through the 
network-internetwork separation layer, the field strength rapidly decreases to 
the sub-kG regime.

Although the majority of the profiles coming from the network region have been classified as
kG, it is possible to clearly identify the shape and position of the peaks of the Stokes $V$ profiles.
Consequently, we have calculated the field strength associated with each profile measuring the
wavelength separation between the peaks of the blue and red lobes and using the calibration curve
presented in the right panel of Fig. \ref{fig:peak_separation_stokesi}.
The results are shown in the right panel of Fig. \ref{fig:magnetic_field_map}. 
The strongest fields are associated with the pores, with field strengths above 
1600 G and with 2000 G at some points. An estimation of the filling factor can
be carried out assuming two components with the same thermodynamics, one
magnetic and the other non-magnetic. The amplitude of the peak in the strong field regime 
arrives to 10 \% for a filling factor of 1 (see Fig. \ref{fig:milne}). The amplitude of the strongest Stokes $V$ 
signals (those of the pores) is $\sim$3 \%. This suggests that the maximum filling factor of
the kG elements in the pores might be of the order of 30 \%.

\subsection{Stokes $I$}
The presence of hyperfine structure allows us to use the intensity profile of the 
15262.702 \AA\ \ion{Mn}{1} line as a powerful diagnostic tool. According to the 
Milne-Eddington results, the line presents two
peaks whose ratio is close to 2 for zero magnetic field. This ratio can be easily measured
in the observations. The most important source of problems is that the ratio critically
depends on the exact value of the continuum. Figure \ref{fig:TIP_vs_FTS} 
shows the comparison between one of the profile in the internetwork regions and the
FTS atlas. The peak ratio given by the FTS atlas is $\sim$1.75,
equivalent to a field of the order of $\sim$250 G using the calibration curve shown in Fig.
\ref{fig:peak_ratio}. A comparison between the observed TIP spectrum and the FTS atlas 
indicates that the observed Stokes $I$ spectrum contains several spurious features. These
features make it difficult to fix a value for the continuum. Following a conservative approach, we 
have selected two different values of the continuum. These values are indicated with vertical lines in the upper panel of 
Fig. \ref{fig:ratio_peak_intensity}. The two values of the continuum are chosen so that they
represent an upper and lower limit of the continuum. The weaker continuum (vertical
dotted line) is chosen so that the largest value of the ratio is smaller than 2. The reason is that ratios 
larger than 2 are not compatible with the results presented in Fig. \ref{fig:milne}. As a consequence, the 
magnetic field strengths obtained from these ratios represent a lower boundary for the true 
magnetic field strength distribution. The larger value of the continuum (vertical dashed 
line), chosen as an upper limit of the continuum, will give smaller ratios and, consequently, larger values of
the magnetic field strength. This might be considered as an upper limit for the true
magnetic field strength distribution.

The upper panel of Figure \ref{fig:ratio_peak_intensity} shows 
the value of the ratio of the two peaks of the intensity profile for all the points in the field-of-view
for the two values of the continuum. It is clearly seen that the smaller values of
the ratios are correlated with the magnetically enhanced region, as expected from the
previous considerations. Due to the presence of two solutions for peak
ratios below $\sim$1.14, the points that fulfill this criterion are indicated in the map 
with contours. Fortunately, this criterion is only fulfilled in those regions of the map that
coincide with the pores, where we expect the strongest fields. The 
smooth variation of the ratio between these strong field regions and the internetwork 
regions suggests a smooth variation in the distribution of fields. This smooth behavior
also discards the noise as the responsible for the variations of the peak ratio. 

Additionally, we have verified that the observed peak separation in Stokes $I$ is constant over the whole
field-of-view, even for the enhanced network and the pores. The profiles of 
these points present peak ratios close to 1, which is indicative of moderately strong fields 
(of the order or above $\sim$500 G). However, since we do not detect any peak separation
from the zero-field case ($\sim 0.23$ \AA), we have to conclude that the average magnetic field that we
are measuring with Stokes $I$ cannot be much stronger than $\sim$600 G.

We apply the calibration curve to transform the observed ratios into magnetic field strength.
Since this measurement of the magnetic field strength has been carried out with Stokes $I$, it 
does not suffer from cancellation effects. Therefore, it represents an average value of unknown 
nature of the magnetic field per resolution element (see 
\S\ref{sec:flux_cancellation}). The lower panels of Fig. \ref{fig:ratio_peak_intensity} show 
the magnetic field strength maps. The maximum value of the magnetic field strength
shown in the maps is 700 G. Stronger fields are represented with the same color coding. 
This maximum value are obtained at the points where
the peak ratio is smaller than $\sim$1.14.

The previous analysis allows us to calculate some statistical properties of the magnetic
field traced with the Stokes $I$ profiles. The left panels of Fig. \ref{fig:histograms} represent the 
histogram of the ratios obtained for the whole field-of-view using the two assumed values of the
continuum. The behavior is very well represented by an exponential for ratios below
$\sim$1.6. These histograms clearly show that the maximum ratio obtained for the smallest value
of the continuum is 2, while the maximum value of the ratio turns out to be smaller when the second
value of the continuum is chosen. We point out that almost all the points with
ratios below 1.6 ($\sim$30\% of the field of view) are located in the most active 
region. The points in the internetwork region have systematically ratios above 1.6. 

In order to see this behavior in more detail, we have separated the points belonging to the 
network and those belonging to the internetwork. A pixel is assigned to the network when the
maximum of the $V/I_c$ profile is above 1.6$\times$10$^{-3}$. We have verified that this 
value provides a good isolation of the network region. The histograms of the network points
are shown in Figure \ref{fig:histograms} as dotted lines while the histograms of the internetwork points are
shown as dashed lines. The histogram of the internetwork points peaks at $\sim$1.75 and it
rapidly falls when moving towards higher ratios. The contribution of the internetwork to the 
points with ratios smaller than 1.6 is also small. The histogram of the internetwork region closely 
resembles a gaussian centered at a peak ratio of $\sim$1.75, equivalent to a gaussian centered
at a magnetic field strength of $\sim$200 G. Concerning the network, the histogram has a less gaussian
shape, with a peak at a ratio of $\sim$1.5, equivalent to a magnetic field strength of $\sim$350 G.


In an effort to understand the previous histograms, we have separated the internetwork histograms
in granules and intergranules. Our observations were carried out with quite bad
seeing conditions, so that the spatial resolution is about 1.4". Consequently, the separation of the
granular and intergranular regions is not very reliable. The granules are obtained as the set
of points whose continuum intensity is larger than 1$\sigma$ above the mean continuum 
intensity of the whole internetwork. The intergranular regions are obtained as the set of pixels 
whose continuum is smaller than 1$\sigma$ below.
The percentage of points assigned to granules is 
15.8\% while 15.3\% of the points are classified as intergranules. 
The results are similar to those
obtained by \cite{socas_pillet_lites04}, who find 18\% of granules and 17\% of intergranules. The histograms of 
the peak ratios for these points are plotted 
in Fig. \ref{fig:histograms_granule_intergranule}, indicating no apparent difference between granules
and intergranules. Two possible explanations can be given. Firstly, the poor resolution 
produces a contamination between the light coming
from the granular and the intergranular material, thus masking any possible intrinsic difference
between them. Secondly, it is possible that there is no intrinsic difference and the background 
field that we measure with the Stokes $I$ profile is not spatially 
structured. 
The only way of investigating this in more detail
is by performing observations with much better spatial resolution.

\subsection{Incompatibility of Stokes $I$ and Stokes $V$ information}
From the previous discussion, it is clear that the Stokes $V$ profiles that emerge from the network 
regions are associated with a quite strong field. We find profiles in the pores and nearby regions that 
we associate with fields above 1 kG due to the lack of hyperfine features in the blue lobe and
that we confirm to be kG fields due to the peak separation in Stokes $V$.
However, the ratio of Stokes $I$ peaks clearly indicates the presence of weak fields with an 
upper limit of $\sim$700 G. Furthermore, we do not detect enhanced Stokes $I$ peak separations in
the pore with respect to what we see in the internetwork regions. An explanation 
might be the presence of relatively weak 
fields in the network (and in the pore) that have a large filling factor 
in comparison with the kG fields that are detected by the Stokes $V$ profile.

\subsection{PDFs}
Is it possible to retrieve the underlying probability distribution function of the magnetic 
field strength from the Stokes $I$ peak ratio measurements? Is the weak-field tail that is 
detected both in the internetwork and network histogram real? Does the measurement
introduce any bias? In order to investigate these questions, a numerical experiment has been carried out.
We have generated a set of $n=30000$ realizations of Stokes profiles with different values
of a vertical magnetic field\footnote{The results are similar when a turbulent magnetic field (random
inclinations) is considered instead of a vertical magnetic field. The vertical field case has been
chosen for simplifying the calculations.}.
The value of the magnetic field for each realization is selected following a given probability distribution
function (PDF). Technically, the
field strength for each one of the 30000 realizations is obtained
by picking a random number $x$ following a uniform distribution in
the interval $[0,1]$.  By an appropriate variable transformation $y=f(x)$, the
resulting numbers $y$ turn out to be distributed according to the desired
distribution.  
Motivated by the results found in different works \citep{khomenko03,martinez_gonzalez_spw4_06},
we have chosen to use exponential and maxwellian PDFs in this numerical experiment.
A ME synthetic profile is obtained for each value of the field and the ratio between Stokes $I$ peaks 
is calculated. This procedure is also repeated by averaging the Stokes profiles emerging from
$f$ consecutive realizations. Consequently, we end up with $30000/f$ average profiles for which
the ratio is obtained. This mimics the loss of spatial resolution in the observation.

We synthesized 30000 Stokes profiles with a vertical field.  The
> > strength of the magnetic field strength is chosen to follow
> > several probability distribution functions.  In our case, we
> > focus on an exponential and maxwellian PDFs.  Technically, the
> > field strength for each one of the 30000 realizations is obtained
> > by picking a random number x following a uniform distribution in
> > the interval [0,1].  By an appropriate transformation y=f(x), the
> > resulting numbers y are distributed according to the desired
> > distribution.  For instance, for an exponential PDF, the number
> > of synthetic Stokes profiles with weak fields is far larger than
> > the number of Stokes profiles with strong fields.  Summarizing,
> > the 30000 realizations are in fact representative of the
> > underlying PDF that we have assumed.

We will focus on the exponential PDFs of the form $P(B) \propto e^{-B/B_0}$. In such an exponential
PDF, the probability of finding a point with a weak field is far larger than the probability of finding
a strong fields. In the averaging process, the weak fields have more weight and the resulting 
Stokes $I$ profile presents systematically larger values of the ratio. As a 
consequence, the PDF retrieved from the ratio will have the fewer strong fields the larger the value
of $f$. This is indeed observed in the experiment, as shown in the left panels of Fig. \ref{fig:experiment} for an
exponential PDF with $B_0=200$ G and different values of $f$. With $f=1$, we are able to reproduce the
slope of the original PDF and the exponential behavior for small fields. When $f$ increases, the 
strong field tail looses weight and moves towards weaker fields, increasing its slope. Thus, the slope 
of the recovered PDF combines information about the average field strength and the resolution.

A similar effect is also detected for the weak field tail, which moves towards higher fields when 
the spatial resolution gets worse. The explanation for this is the following. Only for extremely weak fields 
the peak ratio is close to 2. Larger fields always present smaller ratios. Therefore, after averaging
several profiles, the ratio will always be smaller than 2 unless all the averaged pixels have $B=0$. A
reduction of the number of profiles showing ratios close to 2 appears and the weak field tail moves 
towards stronger fields. This behavior dominates when $f$ increases because we average 
more points. It also dominates when $B_0$ increases because it reduces the amount of very 
weak fields in the averaging process.

The two processes together (reduction of the strong and weak field tails) tend
to transform the retrieved PDF into a gaussian when the spatial resolution is not 
perfect. This behavior is also found in the experiment carried out with the
maxwellian and seems to be a general property of the PDF when the resolution decreases.
In the limit of a perfect resolution, it is possible to correctly recover the PDF.
In the limit of an infinitely poor resolution, in which
we only have one resolution element covering the whole observed area, the retrieved PDF is a Dirac delta function
centered at the average field of the original PDF. In an intermediate regime, the retrieved PDF 
tends to a Gaussian that finally converges to a Dirac delta. Fortunately, the average field given by the
distorted PDF is still the average field of the original PDF. This is a 
particularly important conclusion.

According to the previous calculations, the histograms presented in Fig. \ref{fig:histograms}
indicate that the field distribution in the internetwork has a PDF whose average is
in the range $200-350$ G for the two values of the continuum. The gaussian shape of the
observed PDF suggests (at the light of the previous numerical experiments) that we are not 
resolving the fields\footnote{This conclusion relies on the numerical experiments carried out
with only two functional forms for the PDF. A more exhaustive analysis is needed in order to
verify whether this convergence towards a gaussian shape happens irrespectively of the 
underlying PDF.}. If we assume that the true PDF of the internetwork is an exponential, a
direct comparison with the numerical experiment tends to suggest that $f=10$ is a plausible 
value for our observations. Striking is the fact that the inferred internetwork PDF is very similar to the
one obtained in the numerical experiment for $B_0=200$ G and $f=10$. Not only the mean value of the field
but the individual details of the recovered PDF are quite similar.
If we assume that the field distribution does not change with height, the previous result implies that the horizontal 
scale at which the field is organized in the internetwork has to be of the order of $\sqrt{10}$ times smaller than
our resolution element. For our resolution elements of the order of 1.4"-1.5", this yields a value 
between 0.44"-0.48". 

Concerning the
network field distribution, the peak gives an estimation of $B_0 \sim 350$ G for the mean field strength.
In the network, the shape of the PDF has a less gaussian shape than for the internetwork. The fields
between 350 G and 550 G still present an exponential decay. This might
imply that the scales of the magnetic field in the network are better resolved than in the internetwork.
Note that the slope of the strong field tail of the network distribution is consistent with what is found
in the numerical experiment using $f=10$ for an exponential PDF with $B_0 = 300-350$ G.

The previous numerical experiments have been also carried out with a maxwellian PDF. However,
the observational results are not so favourably compared with the results obtained with such a 
maxwellian PDF. In this case, when the spatial resolution worsens, the field distribution converges 
to a gaussian centered at the mean field of the maxwellian given by $2B_0/\sqrt{\pi}$.

\subsection{Agreement with other diagnostic tools}
Are the previous results 
in agreement with those obtained by 
\cite{trujillo_nature04,trujillo_asensio_shchukina_spw4_06} via analysis of the Hanle effect in \ion{Sr}{1} and C$_2$?
With the investigation of the Hanle effect in the 4607 \AA\ line of \ion{Sr}{1}, \cite{trujillo_nature04} have 
concluded that, assuming an exponential PDF that does not distinguish between granules and 
intergranules, the best fit between the synthetic and observed linear polarization profiles 
is obtained for $\langle B \rangle \approx 130$ G. As it can be seen in Fig. 1 of Trujillo Bueno et al. (2004), 
this PDF with $\langle B \rangle \approx 130$ G gives a good fit to the inferred Hanle depolarization at 
$\mu=0.3$ (which corresponds to an atmospheric height of $\sim$300 km). However, 
in Fig. 1 of
\cite{trujillo_nature04} it can also be seen that for explaining the \ion{Sr}{1} observations at $\mu=0.6$
(which corresponds to an atmospheric height of $\sim$200 km), the
exponential PDF should have $\langle B \rangle \approx 200$ G. Our observations are close to disk center 
and, because the \ion{Mn}{1} is weak, we expect it to be formed deeper than the \ion{Sr}{1} line. 
Furthermore, the \ion{Sr}{1} observations were performed in very quiet regions close to the North solar 
limb. Contrarily, our results are representative of an internetwork area that was encircled by
an enhanced network region. Therefore, our results support those of \cite{trujillo_nature04,trujillo_asensio_shchukina_spw4_06}.

\subsection{Flux cancellation}
\label{sec:flux_cancellation}
One of the most interesting diagnostic capabilities of this \ion{Mn}{1} line is that Stokes $I$ and Stokes
$V$ are sensitive to different aspects of the magnetic field distribution inside the resolution element. Stokes 
$I$ is sensitive to the
total field distribution, and gives an idea of the total unsigned flux and the mean field strength in the 
resolution element. On the contrary, Stokes $V$ is 
sensitive to the part of the magnetic field distribution that is not cancelled out, that is, to the net
flux. As a consequence, we can estimate the ratio between the net flux and
the total flux, thus giving an idea of the amount of flux that cancels out. The net flux in the internetwork
region has been obtained with the aid of the SOM classification. A net flux is assigned to each class of the SOM 
classification through the application of the weak-field approximation. We do not take into account those profiles 
marked as ``noisy'' in the SOM classification.
Therefore, the flux cancellation $F_\mathrm{net}/F_\mathrm{tot}$ can be estimated
directly. On the contrary, since the Stokes $V$
profiles in the network region are entering into the saturation regime, the amplitude of the $V$ profile is
no longer proportional to the net flux. However, a rough estimation of the filling factor can be
obtained by taking into account that the 15262.702 \AA\
line is weak. In this case, the emergent Stokes parameters are proportional to the absorption coefficients, 
so that $(I_\mathrm{c}-I) \propto h_I$ and $V \propto f h_V$ \citep[see, e.g.,][]{landi_landolfi04}, with
$I_\mathrm{c}$ being the continuum intensity. Making the additional assumption 
that the intensity profile is approximately
equal to that obtained for the zero-field case (no enhanced peak separation is found even in the
strong magnetized regions), we find that the emergent intensity at line center $I_0$ fulfills 
$(I_\mathrm{c}-I_0) \propto H(0,a) / \sqrt{\pi}$.
When Stokes $V$ is approximately described in the strong
field regime, its value in one of the lobes can be written as
$V_0 \propto H(0,a) f \cos \theta / (2\sqrt{\pi})$. Finally, we obtain the following relation
for the filling factor \citep{khomenko03}:
\begin{equation}
f \cos \theta \approx \frac{2 V_0}{(I_\mathrm{c}-I_0)},
\label{eq:ff_strong}
\end{equation}
where $\theta$ is the field inclination and $I_0$ is assumed to be not modified by
the magnetic field.
In spite of the strong simplifications assumed for the
development of Eq. (\ref{eq:ff_strong}), we have verified via ME synthetic profiles that the previous 
approximate formula holds approximately for 
magnetic field strength values greater than $\sim$ 1 kG. 
From the previous estimation of the filling factor, it is 
possible to calculate the net magnetic flux $F_\mathrm{net}$ per resolution element as:
\begin{equation}
F_\mathrm{net} = f \cos \theta B_\mathrm{V} R^2,
\end{equation}
where $R^2$ is the area of the resolution element, in our case of the order of 1.4"$\times$1.4". The magnetic field
strength $B_\mathrm{V}$ has been obtained from the peak separation of the Stokes $V$ profile (represented
in the the right panel of Fig. \ref{fig:magnetic_field_map}). The left panel of Fig.
\ref{fig:net_total_flux} presents the flux in logarithmic scale. The obtained values are in accordance with the
typical values found in the network and in small pores \citep[e.g.,][]{zwaan87}. These results can be
compared against those obtained from the peak ratio in Stokes $I$. Assuming that the value of the 
magnetic field obtained from the peak ratio (we call it $B_\mathrm{I}$) is representative of a volume 
filling field (with filling factor unity), the ratio
\begin{equation}
\frac{F_\mathrm{net}}{F_\mathrm{tot}} =\frac{2V_0}{(I_\mathrm{c}-I_0)} \frac{B_\mathrm{V}}{B_\mathrm{I}}
\end{equation}
gives us information about the cancellation of magnetic flux in those regions where Eq. (\ref{eq:ff_strong})
can be applied (i.e., the network, pores, etc.).

The ratio between the net and total fluxes is shown in the right panel of Fig. \ref{fig:net_total_flux}. 
The network presents ratios always above $\sim$10\%. This indicates that the cancellation
of flux in the network is always below 90\%. The variation of the flux cancellation in the 
network is smooth. The pores present values of the ratios close to 100\%, giving the indication that
almost all the field in these strongly magnetized regions is in the form of a field with a well established 
direction. In these regions, the magnetic field strength diagnosed with Stokes $I$ and Stokes $V$
coincides.
The cancellation ratio gets close to zero when moving towards the internetwork
and it presents patches of equal value of the flux density (always below
10 Mx cm$^{-2}$). The contour shown in Fig.  \ref{fig:net_total_flux} indicates the separation between
the points that belong to the network and those that belong to the internetwork. It is important to
remind that, as already
mentioned, the net flux in these regions has been obtained following a different approximation.
However, it is interesting to point out that a smooth transition between the network and internetwork regions
is found. The flux cancellation in the internetwork is such that the ratio between the net flux and the total flux
is below 10\% for more than 90\% of the points.

In view of the approximations we have been forced to use due to the low Stokes $V$ signal of the
\ion{Mn}{1} line, it will be of great help to observe simultaneously this \ion{Mn}{1} line together with other
more magnetically sensitive lines so that the net flux can be estimated with more accuracy.

\section{Conclusions}
\label{sec_conclusion}
We have presented the results of a theoretical and observational investigation of a \ion{Mn}{1} line located in the near-IR, 
which shows very attractive properties for magnetic field diagnostics.
The combined effects of the larger Zeeman splitting present in the near-IR and the presence of 
strong perturbations on the Zeeman patterns due to the Paschen-Back effect, makes this line
ideal for the diagnostic of magnetic fields in unresolved structures. The hyperfine structure
produces modifications in the intensity profile of the line when a magnetic field is present, with a 
response that covers a large range of magnetic field strengths. The advantage over other Zeeman
diagnostic tools is that it does not suffer from cancellation effects.

We have shown that the analysis of the ratio of the two peaks of the Stokes $I$ profiles measured in the 
quietest area of the observed field of view leads to 
a PDF of gaussian shape that is centered at 250-350 G, being unable to distinguish between granules and 
intergranules with our spatial resolution. This 
is a value that agrees with typical equipartition values in the photosphere and supports the 
Hanle-effect conclusion of \cite{trujillo_nature04,trujillo_asensio_shchukina_spw4_06} that
there is a substantial amount of hidden magnetic energy and unsigned flux in the quiet Sun.
A theoretical investigation assuming different
PDFs shows that the center of the inferred PDF gives information about the real PDF and that
its gaussian shape is related to a degradation of the spatial resolution. 
Although higher signal-to-noise ratio observations are needed for obtaining better Stokes $V$ profiles, 
we have shown for the first time a map of how the flux cancels in unresolved magnetic elements.

In addition to the presence of pores and a
magnetically active network, we must emphasize that
the field of view of our observation contains an
internetwork region where the net flux seldom amounts to 10\% of the total flux.
While the spatially average net flux found in very quiet internetwork regions
\citep[e.g.,][]{khomenko03,martinez_gonzalez_spw4_06}
is very close to zero \citep[see, however,][]{dominguez03,dominguez06},
we have a unipolar net flux in our
internetwork region in accordance with the polarity
of the surrounding network. The numerical experiments
on turbulent dynamos and magnetoconvection carried out by Cattaneo and coworkers
\citep{emonet_cattaneo01,cattaneo03,cattaneo_emonet04}
are initialized by a seed magnetic
field that is then tangled up
producing a mixed-polarity field with a distribution of field strengths such that the
magnetic energy density is a significant fraction of the kinetic energy density.
The field topology conserves any imbalance of net flux already present in
the initial magnetic seed
\citep{emonet_cattaneo01}. In very quiet
regions one can assume a random seed field resulting in
a zero net magnetic flux. Our observations match, on the other hand, a
scenario with an initial seed field
with a non-zero net magnetic flux. Since the surrounding network of the observed field of view is
magnetically enhanced and
mostly unipolar, it is apparent that such a unipolar field may be
ubiquitous through the encircled internetwork
area and will play the role of a seed magnetic field with a non-zero net
flux. Such a seed field might then be tangled up through dynamo action and/or
magnetoconvection into the mixed-polarity fields that we find
to constitute more than 90\% of the total
flux found in the observed internetwork region.
The high level of mixing (90\% at least) can be
now qualitatively compared with simulations of magnetoconvection.
Particularly instructive is Fig. 2 of \cite{emonet_cattaneo01}
where their
case 0 (whose seed field is a completely random field) results in a 100\%
cancellation, while their case 4 (with a
seed field with a net flux equivalent to 200 G) results in a level of mixing
smaller than 75\%. The measured level of
90\% would therefore translate into a seed field of a few G in apparent
contradiction with the enhanced field
found all around the internetwork area in our data. 
Such an apparent contradiction can be explained in two different ways. The first one is that
the actual seed field through the internetwork is actually of just a few G and, therefore, results in a cancellation
of 90\% as predicted by simulations.
The second possibility, and the one we prefer, is that
solar magnetoconvection is more successful in mixing fields than present
simulations, perhaps just because
today's numerical experiments are made with insuficient spatial resolution corresponding to relatively low
Reynolds number as compared
to the real values of the solar photosphere, at least in what concerns the magnetic description.

Another interesting conclusion is that the uncancelled flux reaches very high 
percentages (approaching 100\%) in the pores and other magnetic concentrations over the network. In these 
regions Stokes V pointed toward kG fields, but 
Stokes I unveiled a greater contribution of weak fields under the 700 G threshold. Since both components 
appear to share the same polarity and there is not much cancellation, our analysis suggests
that weak fields coexist with kG fields (carrying most part of the flux) even in these magnetically-enhanced 
regions.

The potential of this line has to be improved by simultaneously observing it with other lines, like the 
\ion{Fe}{1} lines at 525 nm, 630 nm and 1.56 $\mu$m and the \ion{Mn}{1} line at 5537 \AA. 
Furthermore, other lines with hyperfine structure can be found in the red and near-IR that can be
of great diagnostic potential if they can be observed simultaneously.

Finally, this IR \ion{Mn}{1} line may be of great interest for diagnosing magnetic fields in other stars apart from the
Sun. The line ratio we have developed can be obtained provided one observes with sufficiently high spectral resolution 
observation and for sufficiently slow rotators. With our line ratio technique, it should be possible to estimate 
the average magnetic field strength of the star.

\acknowledgements We thank R. Manso Sainz for helpful discussions. This article is based on observations taken
with the VTT telescope operated on the island of Tenerife by the Kiepenheuer-Institut f\"ur Sonnenphysik in the 
Spanish Observatorio del Teide of the Instituto de Astrof\'{\i}sica de Canarias. This research has been partly 
funded by the Ministerio de Educaci\'on y Ciencia through project AYA2004-05792.

\appendix
\section{Self-organizing maps}
\label{sec:app_som}
Self-organizing maps (SOM) are one of the multiple variants of artificial neural networks. They are trained using
unsupervised learning techniques and they produce a low-dimensional representation of the training sample. One of 
the most important characteristics is that the low-dimensional representation tries to preserve the topological
properties of the input space (training samples), so that nearby samples are mapped into nearby neurons 
\citep[][]{kohonen_SOM01}. They have been mostly applied to the visualization of high-dimensional data.

The self-organizing map consists on a single layer feedforward neural network. The output layer is arranged in a 
low-dimensional (usually 2D or 3D) grid of $N_\mathrm{out} \times N_\mathrm{out}$ neurons. The input layer is 
formed by a set of $N_\mathrm{in}$ neurons, where $N_\mathrm{in}$ is the dimensionality of the input training space.
The network is fully connected, in the sense that every neuron of the output layer is connected to every neuron 
of the input layer. Each neuron of the output layer is associated with a weight vector 
$\mathbf{W} \in \mathbb{R}^{N_\mathrm{in}}$, i.e., with the dimensionality of the input space.

Motivated by how sensory information is handled in different parts of the brain, the goal of the unsupervised
learning algorithm is to associate different parts of the map to different parts of the input space. Therefore, 
similar parts of the map will produce larger response when several input data that share a given property is
shown to the neural network.

The iterative scheme is started by initializing the weights with small random numbers (it is also possible to initialize 
by uniformly sampling the subspace generated by the first two largest principal components
of the input space). The iteration number is accounted for by the discrete index $t$. Each iteration 
needs to repeat the following update rule for each input data \citep{kohonen_SOM01}:
\begin{equation}
\mathbf{W}_i(t+1) = \mathbf{W}_i(t) + h_i(t) \Big[ \mathbf{X}(t) -\mathbf{W}_i(t) \Big],
\end{equation}
where $\mathbf{W}_i(t+1)$ is the weight associated with each neuron $i$ at iteration $t+1$,
$\mathbf{X}(t)$ represents the input vector, while $\mathbf{h}_i(t)$ is the so-called neighborhood
function. The neighborhood function is built in the following way. First of all, the euclidean distances
between all the neuron's weights and the input vector are calculated. The neuron with the smallest distance
is known as the \emph{best-matching node}, whose distance with the input vector is $r_c$. The neighborhood 
function is used for propagating the information of the input vector to the surrounding neurons, and it
is usually written as:
\begin{equation}
h_i(t) = \alpha(t) \exp \Bigg( -\frac{\parallel r_c - r_i \parallel^2}{2 \sigma^2(t)} \Bigg),
\end{equation}
where $r_i$ is the euclidean distance between the weight of neuron $i$ an the input vector. The functions
$\alpha(t)$ (learning rate) and $\sigma(t)$ (width of the kernel) are some monotonically decreasing function 
of the index $t$ that control the spread of the neighborhood kernel. For the initial iterations, the large
width of the kernel leads to a convergence in the global scale. When the width of the kernel is reduced, the
convergence tends to be more local.

After a large number of iterations, convergence of the map is obtained. The weights of the neurons tend
to be associated with \emph{patterns} in the input data, with similar patterns being located in nearby
neurons. Finally, it is possible to use the map to classify any input vector, whether it was in the learning
set or not. The euclidean distance is calculated between the input vector and all the neuron's weights. The
neuron whose weight lies closest to the input vector will give the location in the map of the group in which
the input vector has been classified.



\begin{thebibliography}{41}
\expandafter\ifx\csname natexlab\endcsname\relax\def\natexlab#1{#1}\fi

\bibitem[{{Brett} {et~al.}(2004){Brett}, {West}, \& {Wheatley}}]{brett_SOM04}
{Brett}, D.~R., {West}, R.~G., \& {Wheatley}, P.~J. 2004, \mnras, 353, 369

\bibitem[{{Casimir}(1963)}]{casimir63}
{Casimir}, H. B.~G. 1963, {On the Interaction Between Atomic Nuclei and
  Electrons} (San Francisco: Freeman)

\bibitem[{{Cattaneo}(1999)}]{cattaneo99}
{Cattaneo}, F. 1999, \apj, 515, L39

\bibitem[{{Cattaneo} \& {Emonet}(2004)}]{cattaneo_emonet04}
{Cattaneo}, F., \& {Emonet}, T. 2004, in 35th COSPAR Scientific Assembly, 4443

\bibitem[{{Cattaneo} {et~al.}(2003){Cattaneo}, {Emonet}, \&
  {Weiss}}]{cattaneo03}
{Cattaneo}, F., {Emonet}, T., \& {Weiss}, N. 2003, \apj, 588, 1183

\bibitem[{{Dom\' inguez Cerde\~na} {et~al.}(2003){Dom\' inguez Cerde\~na},
  {S\'anchez Almeida}, \& {Kneer}}]{dominguez03}
{Dom\' inguez Cerde\~na}, I., {S\'anchez Almeida}, J., \& {Kneer}, F. 2003,
  \aap, 407, 741

\bibitem[{{Dom\' inguez Cerde\~na} {et~al.}(2006{\natexlab{a}}){Dom\' inguez
  Cerde\~na}, {S\'anchez Almeida}, \& {Kneer}}]{dominguez06}
---. 2006{\natexlab{a}}, \apj, 646, 1421

\bibitem[{{Dom\' inguez Cerde\~na} {et~al.}(2006{\natexlab{b}}){Dom\' inguez
  Cerde\~na}, {S\'anchez Almeida}, \& {Kneer}}]{dominguez06b}
---. 2006{\natexlab{b}}, \apj, 636, 496

\bibitem[{{Emonet} \& {Cattaneo}(2001)}]{emonet_cattaneo01}
{Emonet}, T., \& {Cattaneo}, F. 2001, \apjl, 560, L197

\bibitem[{{Khomenko} {et~al.}(2003){Khomenko}, {Collados}, {Solanki}, {Lagg},
  \& {Trujillo Bueno}}]{khomenko03}
{Khomenko}, E.~V., {Collados}, M., {Solanki}, S.~K., {Lagg}, A., \& {Trujillo
  Bueno}, J. 2003, \aap, 408, 1115

\bibitem[{{Kohonen}(2001)}]{kohonen_SOM01}
{Kohonen}, T. 2001, {Self-organizing maps} (Berlin: Springer)

\bibitem[{{Kurucz}(1993{\natexlab{a}})}]{kurucz_hfs93}
{Kurucz}, R. 1993{\natexlab{a}}, Phys. Scr., 47, 110

\bibitem[{{Kurucz}(1993{\natexlab{b}})}]{kurucz_atomic93}
---. 1993{\natexlab{b}}, Atomic data for opacity calculations (Kurucz CD-ROM
  No.1)

\bibitem[{{Landi Degl'Innocenti} \& {Landolfi}(2004)}]{landi_landolfi04}
{Landi Degl'Innocenti}, E., \& {Landolfi}, M. 2004, Polarization in Spectral
  Lines (Kluwer Academic Publishers)

\bibitem[{{Lef\`ebvre} {et~al.}(2003){Lef\`ebvre}, {Garnir}, \&
  {Bi\'emont}}]{lefebvre03}
{Lef\`ebvre}, P.-H., {Garnir}, H.-P., \& {Bi\'emont}, E. 2003, \aap, 404, 1153

\bibitem[{{Lin}(1995)}]{lin95}
{Lin}, H. 1995, \apj, 446, 421

\bibitem[{{Lin} \& {Rimmele}(1999)}]{lin_rimmele99}
{Lin}, H., \& {Rimmele}, T. 1999, \apj, 514, 448

\bibitem[{{Lites} \& {Socas-Navarro}(2004)}]{lites_socas04}
{Lites}, B.~W., \& {Socas-Navarro}, H. 2004, \apj, 613, L600

\bibitem[{{L\'opez Ariste} {et~al.}(2002){L\'opez Ariste}, {Tomczyk}, \&
  {Casini}}]{arturo02}
{L\'opez Ariste}, A., {Tomczyk}, S., \& {Casini}, R. 2002, \apj, 580, 519

\bibitem[{{L\'opez Ariste} {et~al.}(2006){L\'opez Ariste}, {Tomczyk}, \&
  {Casini}}]{arturo06}
---. 2006, A\&A, 454, 663

\bibitem[{{Maehoenen} \& {Hakala}(1995)}]{maehoenen_SOM95}
{Maehoenen}, P.~H., \& {Hakala}, P.~J. 1995, \apj, 452, L77

\bibitem[{{Mart\' inez Gonz\'alez} {et~al.}(2006{\natexlab{a}}){Mart\' inez
  Gonz\'alez}, {Collados}, \& {Ruiz Cobo}}]{martinez_gonzalez06}
{Mart\' inez Gonz\'alez}, M.~J., {Collados}, M., \& {Ruiz Cobo}, B.
  2006{\natexlab{a}}, \aap, 456, 1159

\bibitem[{{Mart\' inez Gonz\'alez} {et~al.}(2006{\natexlab{b}}){Mart\' inez
  Gonz\'alez}, {Collados}, \& {Ruiz Cobo}}]{martinez_gonzalez_spw4_06}
{Mart\' inez Gonz\'alez}, M.~J., {Collados}, M., \& {Ruiz Cobo}, B.
  2006{\natexlab{b}}, in Solar Polarization 4, ed. R.~{Casini} \& B.~W.
  {Lites}, ASP Conf. Ser., in press

\bibitem[{{Mart{\'{\i}}nez Pillet} {et~al.}(1999){Mart{\'{\i}}nez Pillet},
  {Collados}, {Bellot Rubio}, {Rodr{\'{\i}}guez Hidalgo}, {Ruiz Cobo}, \&
  {Soltau}}]{martinez_pillet99}
{Mart{\'{\i}}nez Pillet}, V., {Collados}, M., {Bellot Rubio}, L.~R.,
  {Rodr{\'{\i}}guez Hidalgo}, I., {Ruiz Cobo}, B., \& {Soltau}, D. 1999, in
  Astronomische Gesselschaft Meeting Abstracts, vol. 15

\bibitem[{{Press} {et~al.}(1986){Press}, {Teukolsky}, {Vetterling}, \&
  {Flannery}}]{numerical_recipes86}
{Press}, W.~H., {Teukolsky}, S.~A., {Vetterling}, W.~T., \& {Flannery}, B.~P.
  1986, Numerical Recipes (Cambridge: Cambridge University Press)

\bibitem[{{Priest}(2006)}]{priest06}
{Priest}, E. 2006, in The Many Scales of the Universe, ed. J.~C. {del Toro
  Iniesta}

\bibitem[{{S{\'a}nchez Almeida}(2004)}]{sanchez_solarb04}
{S{\'a}nchez Almeida}, J. 2004, in ASP Conf. Ser. 325: The Solar-B Mission and
  the Forefront of Solar Physics, ed. T.~{Sakurai} \& T.~{Sekii}, 115

\bibitem[{{S{\'a}nchez Almeida} {et~al.}(2003){S{\'a}nchez Almeida}, {Emonet},
  \& {Cattaneo}}]{sanchez_emonet03}
{S{\'a}nchez Almeida}, J., {Emonet}, T., \& {Cattaneo}, F. 2003, \apj, 585, 536

\bibitem[{{Schrijver}(2005)}]{schrijver05}
{Schrijver}, C.~J. 2005, in ESA SP-596: Chromospheric and Coronal Magnetic
  Fields, ed. D.~E. {Innes}, A.~{Lagg}, \& S.~A. {Solanki}

\bibitem[{{Semel}(1981)}]{semel81}
{Semel}, M. 1981, \aap, 97, 75

\bibitem[{{Socas-Navarro} {et~al.}(2004){Socas-Navarro}, {Mart{\'{\i}}nez
  Pillet}, \& {Lites}}]{socas_pillet_lites04}
{Socas-Navarro}, H., {Mart{\'{\i}}nez Pillet}, V., \& {Lites}, B.~W. 2004,
  \apj, 611, 1139

\bibitem[{{Socas-Navarro} \& {S\'anchez Almeida}(2002)}]{socasnavarro02}
{Socas-Navarro}, H., \& {S\'anchez Almeida}, J. 2002, \apj, 565, 1323

\bibitem[{{Stein} \& {Nordlund}(2003)}]{stein_nordlund03}
{Stein}, R.~F., \& {Nordlund}, A. 2003, in Stellar Atmosphere Modeling, ed.
  I.~{Hubeny}, D.~{Mihalas}, \& K.~{Werner}, ASP Conf. Ser. 288 (San Francisco:
  ASP), 519

\bibitem[{{Stenflo}(1994)}]{stenflo94}
{Stenflo}, J.~O. 1994, Solar Magnetic Fields. Polarized Radiation Diagnostics
  (Dordrecht: Kluwer Academic Publishers)

\bibitem[{{Stenflo} \& {Lindegren}(1977)}]{stenflo77}
{Stenflo}, J.~O., \& {Lindegren}, L. 1977, \aap, 59, 367

\bibitem[{{Trujillo Bueno}(2001)}]{trujillo01}
{Trujillo Bueno}, J. 2001, in ASP Conf. Ser. 236: Advanced Solar Polarimetry --
  Theory, Observation, and Instrumentation, ed. M.~{Sigwarth}, 161

\bibitem[{{Trujillo Bueno}(2005)}]{trujillo_esa05}
{Trujillo Bueno}, J. 2005, in ESA SP-600: The Dynamic Sun: Challenges for
  Theory and Observations, ed. D.~{Danesy}, S.~{Poedts}, A.~{De Groof}, \&
  J.~{Andries}, 7

\bibitem[{{Trujillo Bueno} {et~al.}(2006){Trujillo Bueno}, {Asensio Ramos}, \&
  {Shchukina}}]{trujillo_asensio_shchukina_spw4_06}
{Trujillo Bueno}, J., {Asensio Ramos}, A., \& {Shchukina}, N. 2006, in Solar
  Polarization 4, ed. R.~{Casini} \& B.~W. {Lites}, ASP Conf. Ser., in press

\bibitem[{{Trujillo Bueno} {et~al.}(2004){Trujillo Bueno}, {Shchukina}, \&
  {Asensio Ramos}}]{trujillo_nature04}
{Trujillo Bueno}, J., {Shchukina}, N., \& {Asensio Ramos}, A. 2004, Nature,
  430, 326

\bibitem[{{V\"ogler}(2003)}]{vogler_thesis03}
{V\"ogler}, A. 2003, PhD thesis, G\"ottingen University

\bibitem[{{Zwaan}(1987)}]{zwaan87}
{Zwaan}, C. 1987, \araa, 25, 83

\end{thebibliography}

\clearpage

\begin{table}[!t]
\caption{Peak ratio for each profile in the 4$\times$4 SOM.}
\label{tab:peak_ratio_4x4}
\centering
\begin{tabular}{|c|c|c|c|}
\tableline Noisy & Noisy & Noisy & 0.791 \\ 
profile & profile & profile &  (457 G) \\
\tableline 0.9895  & 1.0425 & 1.129 & 1.219\\ 
(592 G) & (623 G) & (671 G) & (717 G) \\
\tableline 0.747 & 0.757  & $>$1kG & $>$1kG \\ 
(421 G) & (429 G) & & \\
\tableline 0.814  & 0.843  & $>$1 kG & $>$1 kG \\ 
(474 G) & (496 G) & & \\
\tableline 
\end{tabular}
\end{table}

\clearpage

\begin{figure*}[!t]
\caption{Splitting (in cm$^{-1}$ of the upper and lower levels of the \ion{Mn}{1} transition at 
15262.702 \AA. When the hyperfine structure is taken into account, both fine structure levels 
have six $F$ levels. Note that level crossings are present at fields as low as 200 G, thus 
leading to strong perturbations in the Zeeman patterns due to the incomplete Paschen-Back effect.\label{fig:splitting}}
\end{figure*}
\begin{figure*}[!t]
\plottwo{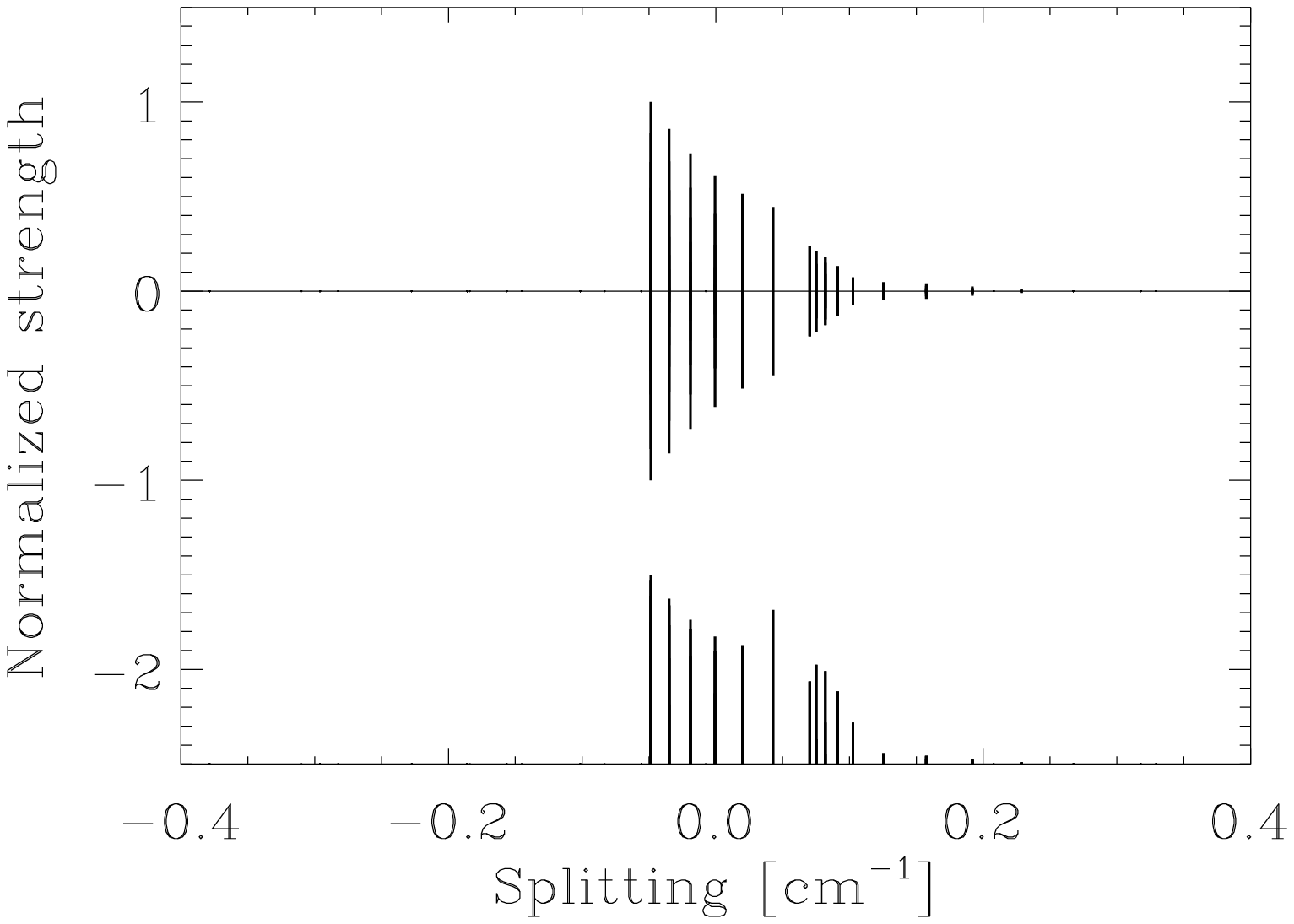}{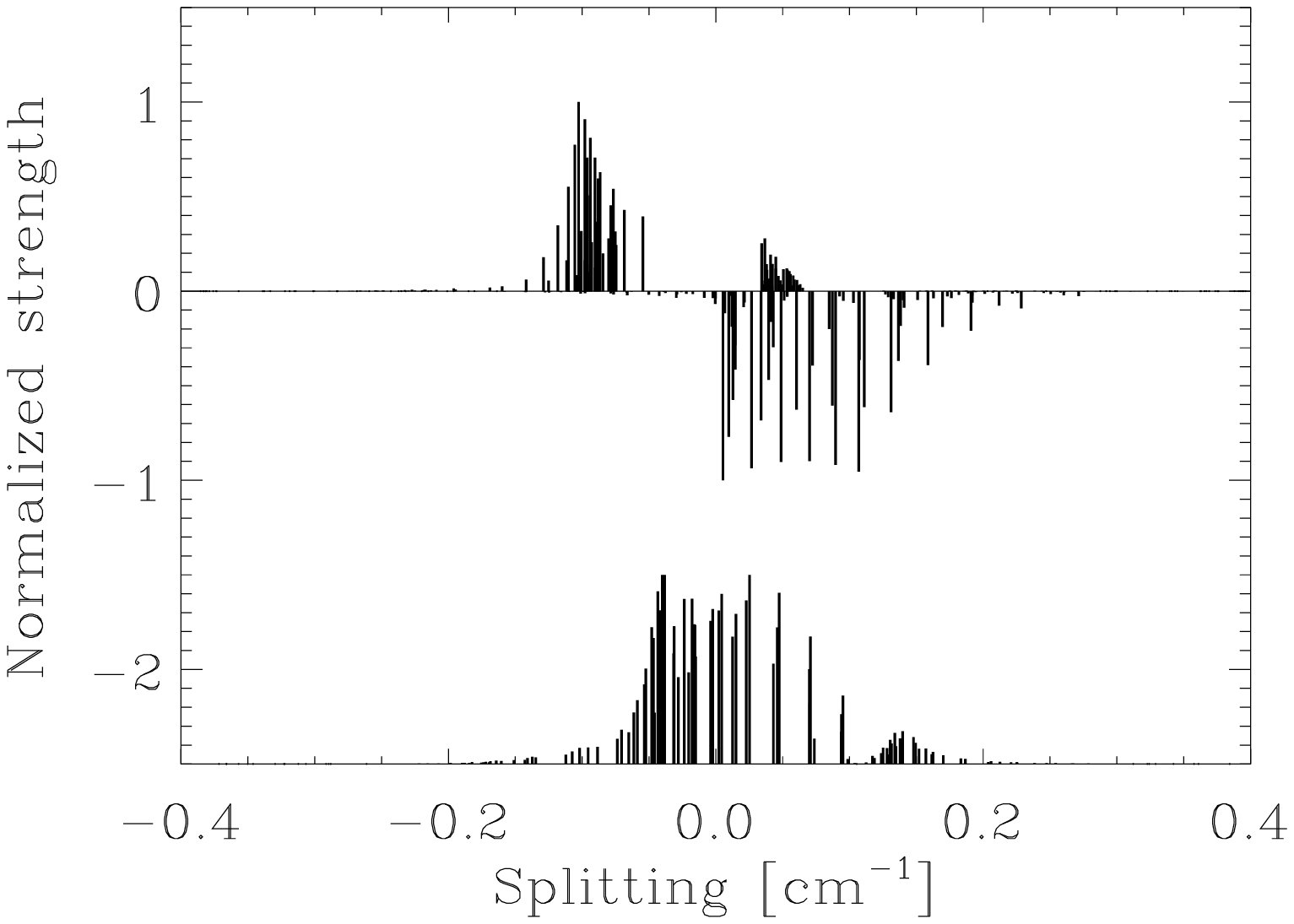}
\caption{Zeeman pattern of the 15262.702 \AA\ transition of \ion{Mn}{1} for two different values of the
magnetic field strength. The $\Delta M_F= -1$ components are in the upper panel pointing upwards,
the $\Delta M_F=+1$ components are in the upper panel pointing downwards and the $\Delta M_F=0$
components are in the lower panel.
The zero-field case shows only the allowed transitions between the $F$
levels of the upper and lower fine structure level. When the field increases, the degeneracy of the magnetic
sublevels is broken and a large amount of components appear. The number of components is 171 for the
$\Delta M_F = \pm1$ transitions and 176 for the $\Delta M_F=0$ transitions.\label{fig:zeeman_patterns}}
\end{figure*}
\clearpage
\thispagestyle{empty}
\setlength{\voffset}{-20mm}
\begin{figure}[!t]
\plottwo{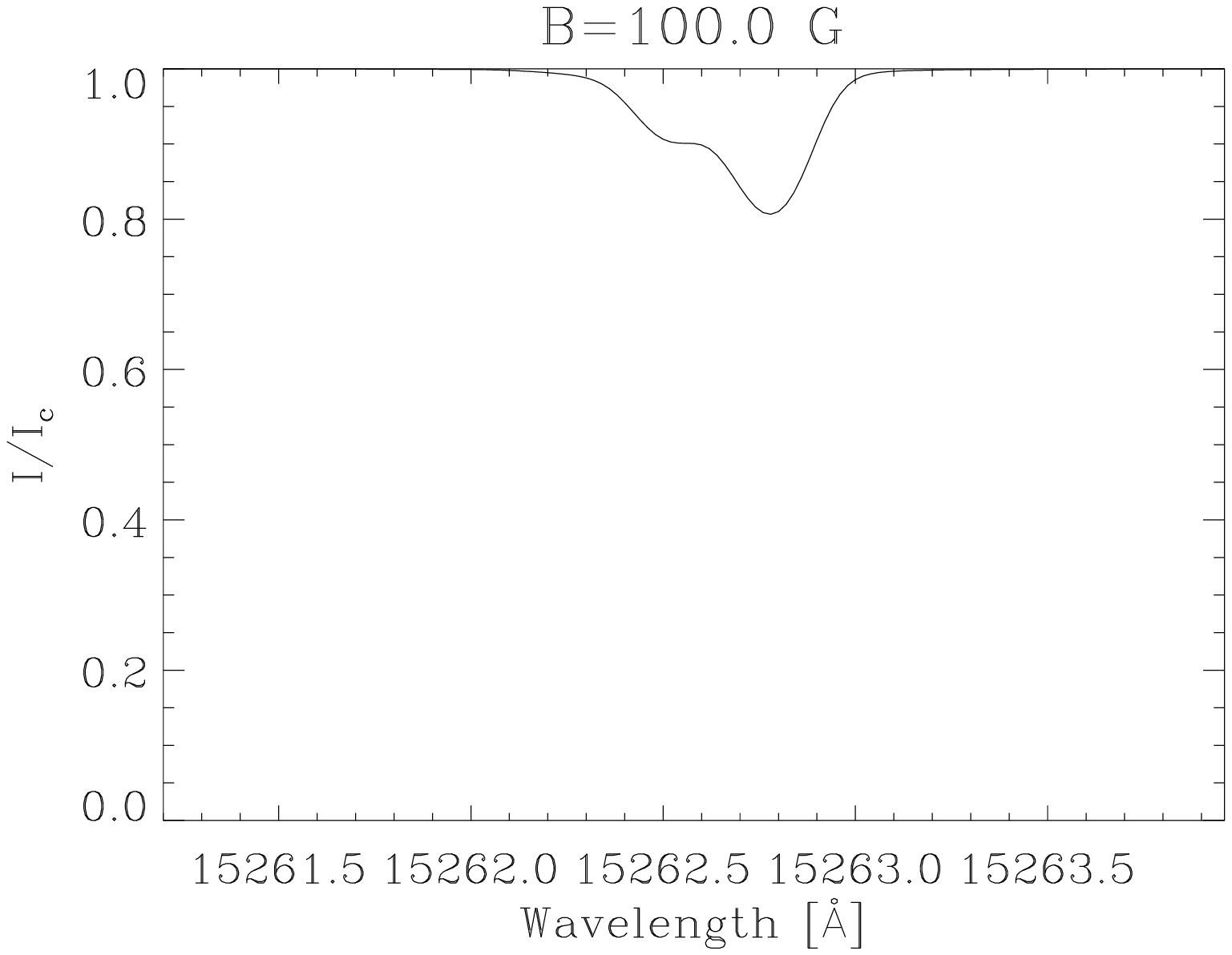}{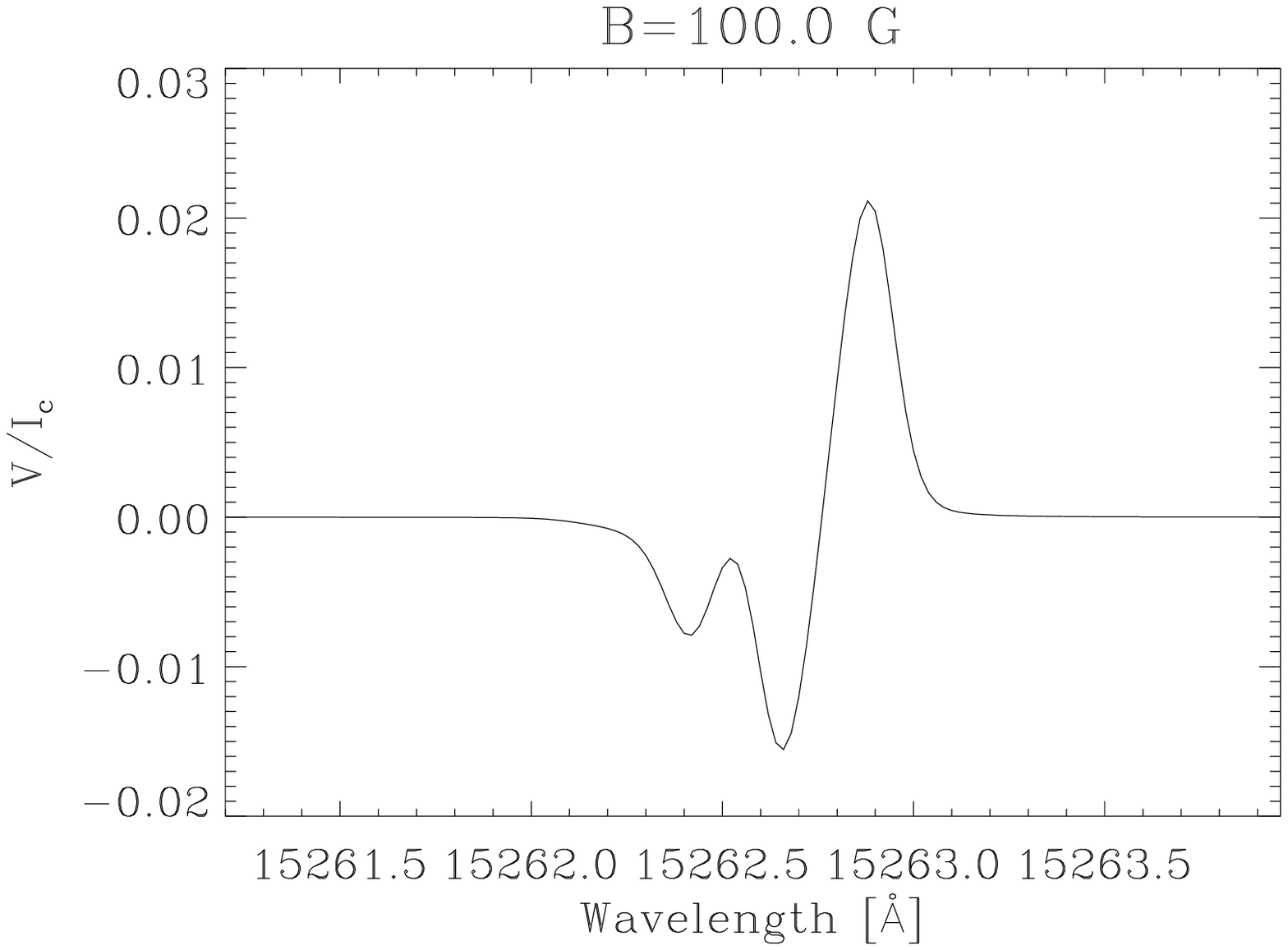}
\plottwo{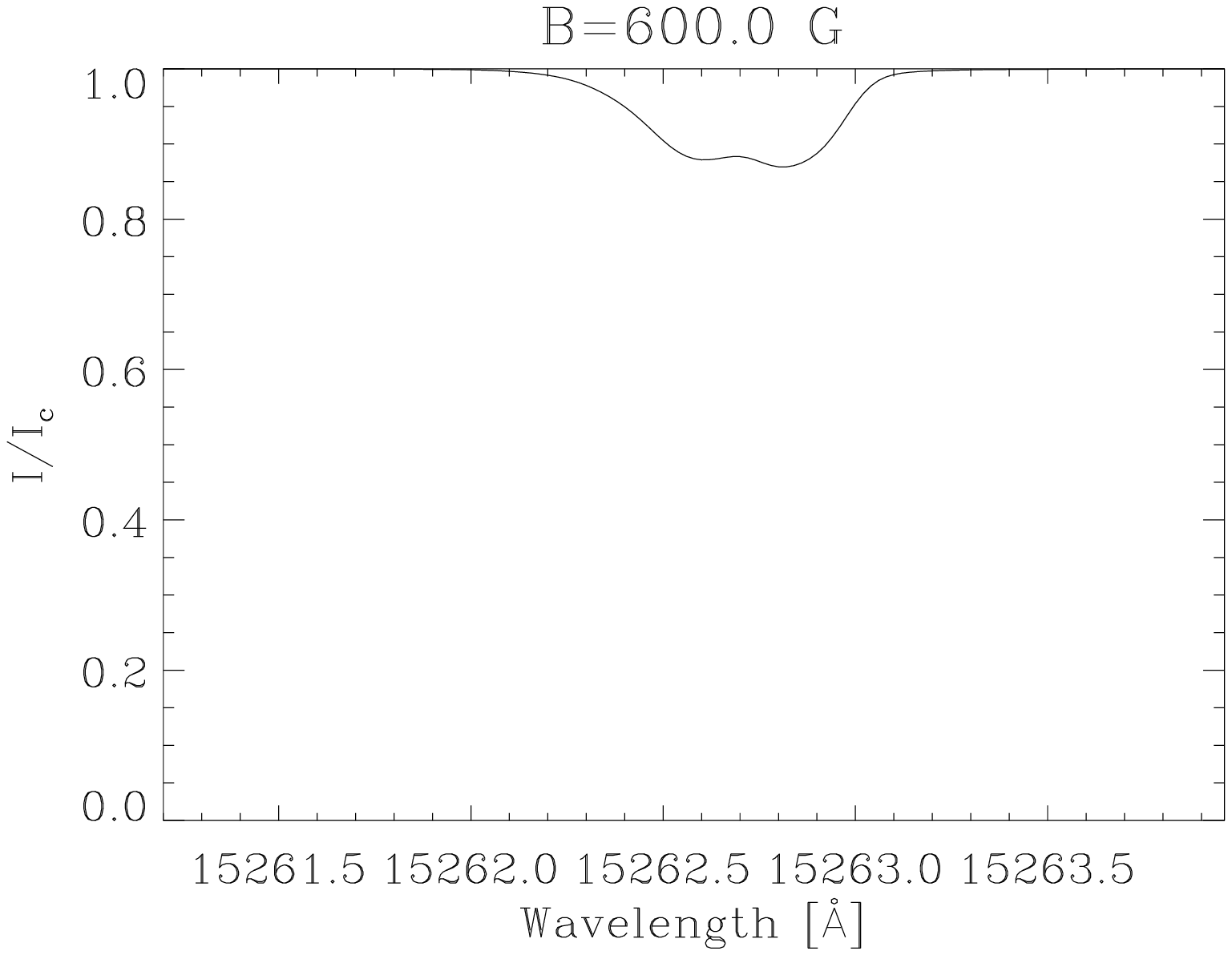}{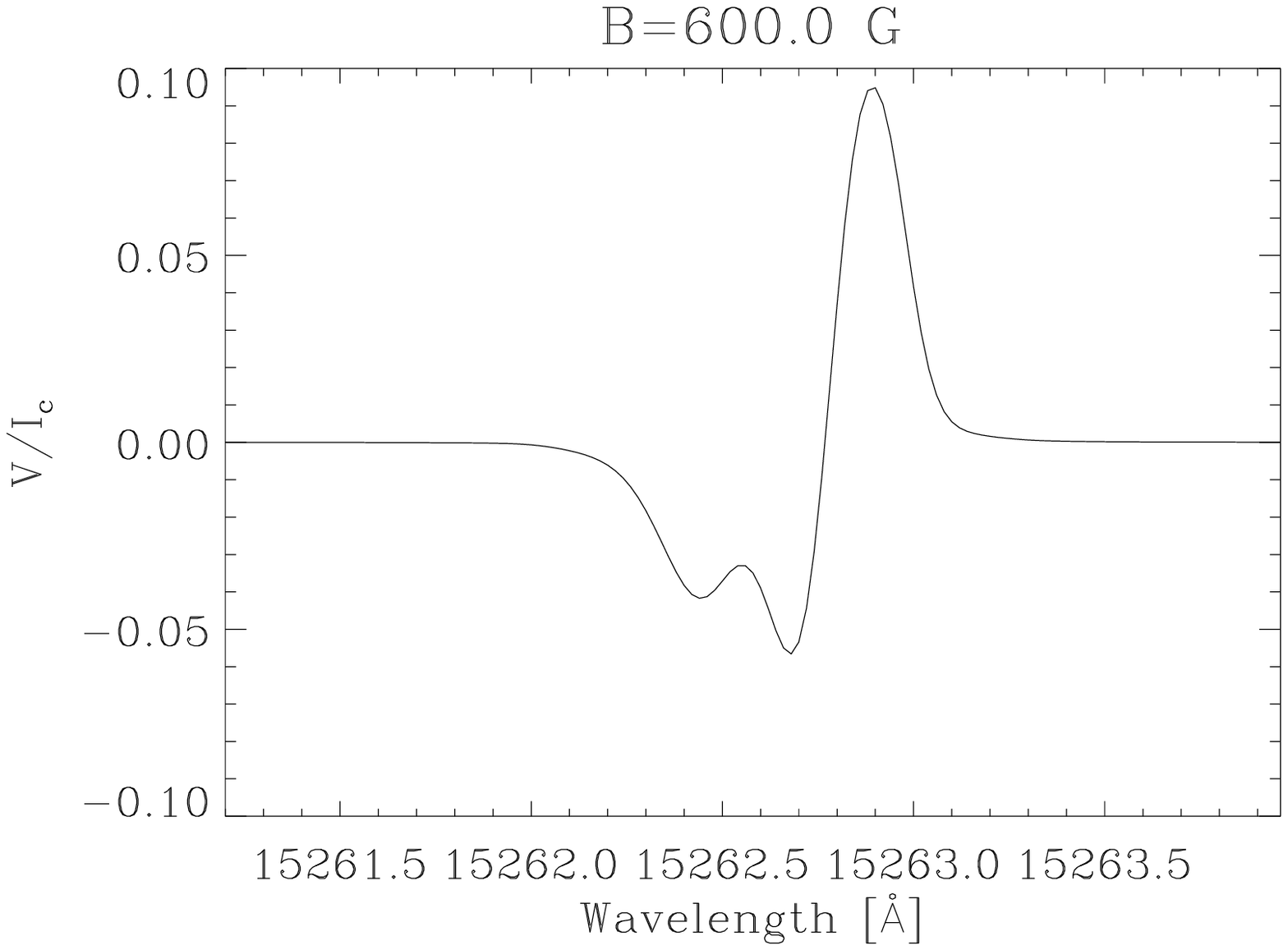}
\plottwo{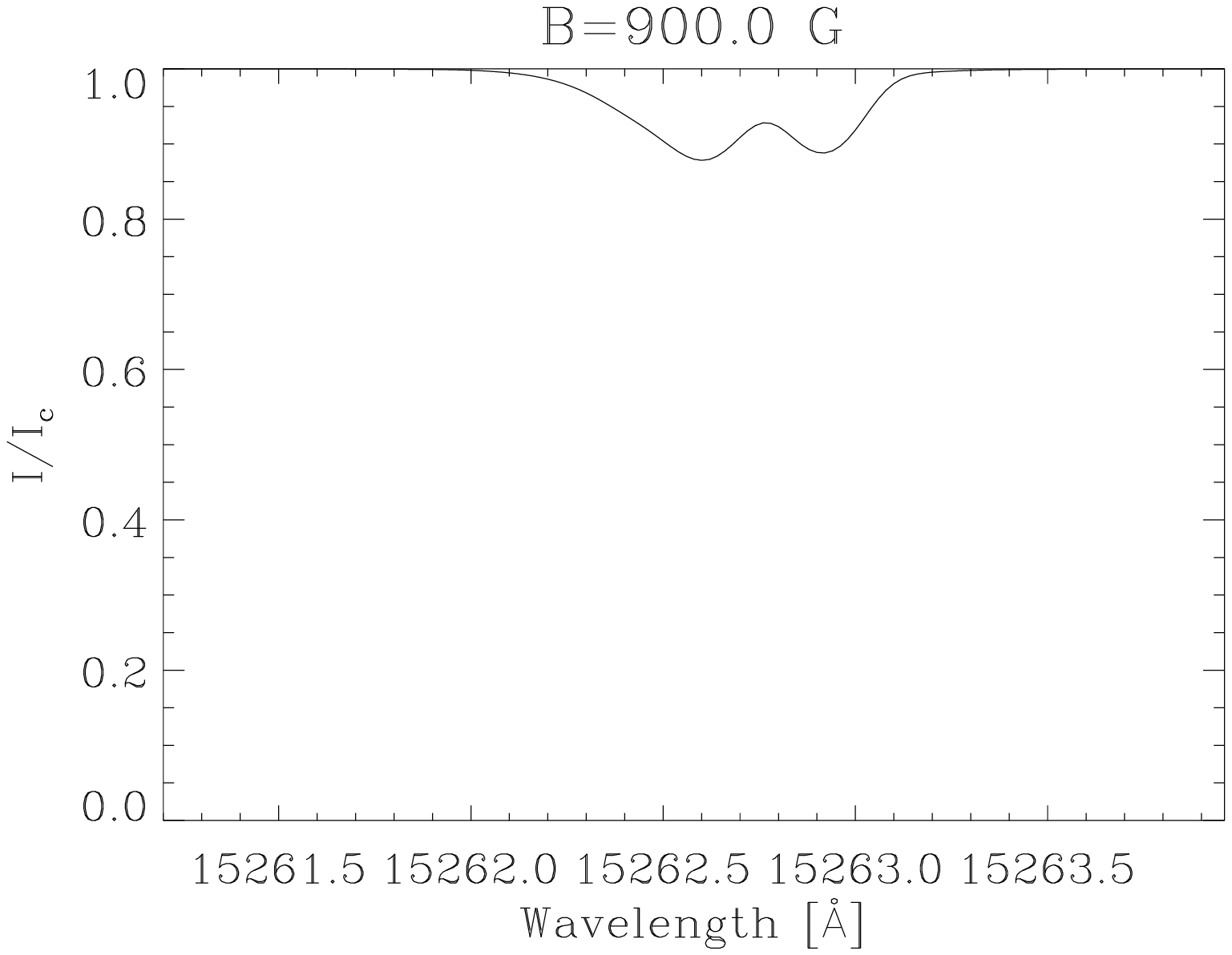}{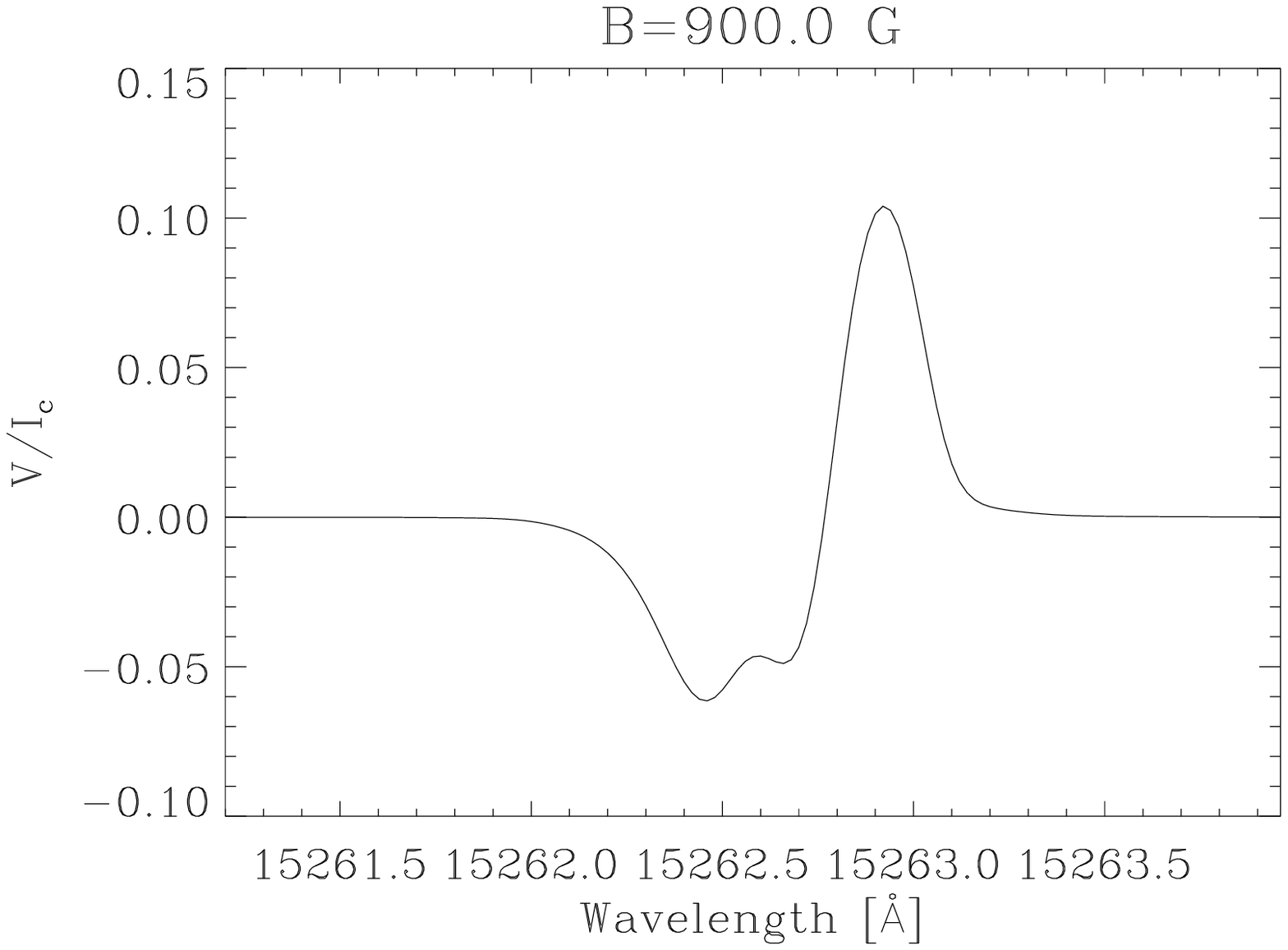}
\plottwo{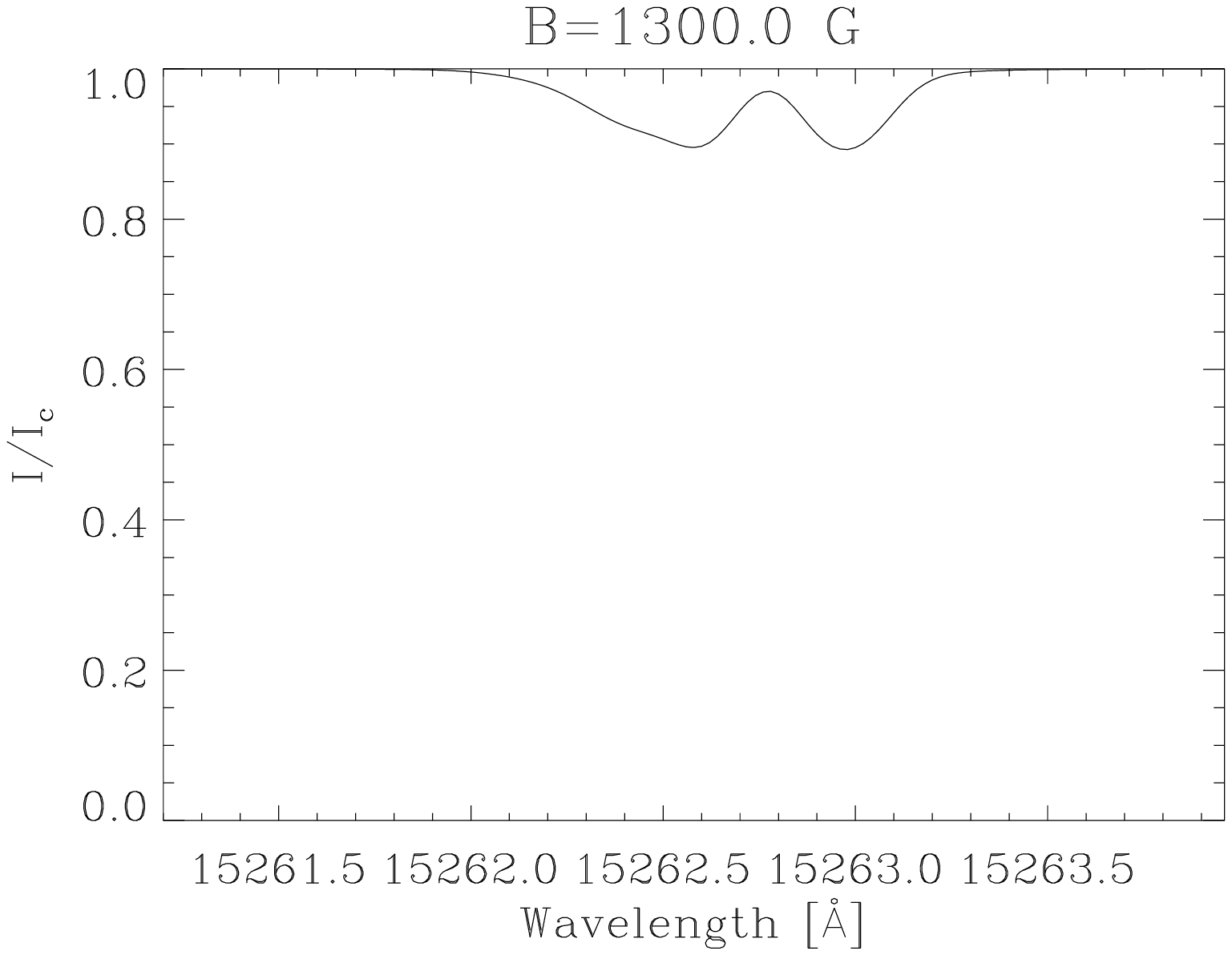}{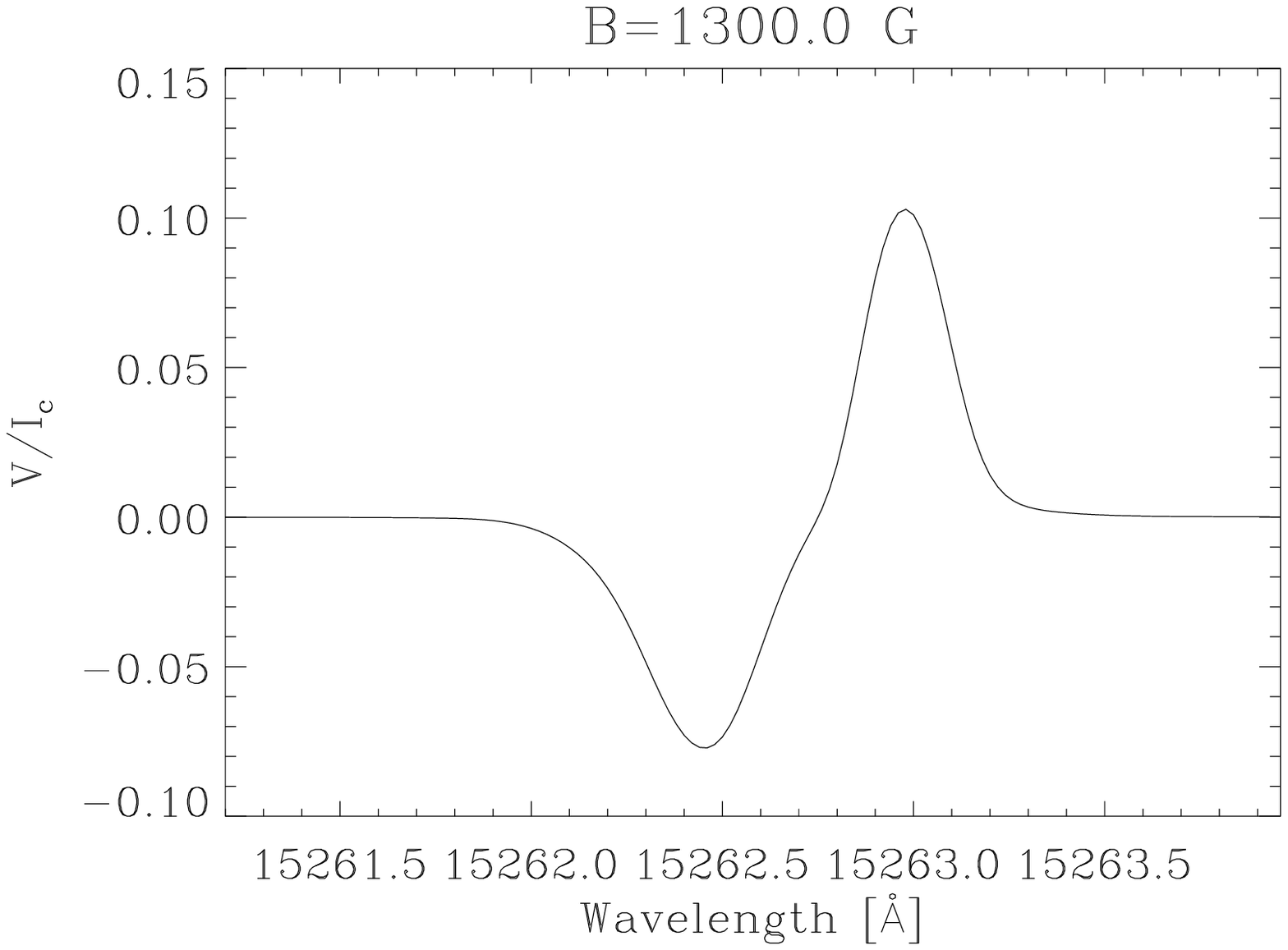}
\caption{Milne-Eddington synthetic profiles obtained for 100, 600, 900 and 1300 G. The 
thermodynamical parameters that have been employed ($v_\mathrm{th}=0.1$ \AA, $\beta=1$, $\eta_0=2$) 
approximately fit the observed Stokes $I$ profiles in the internetwork
regions. Focusing on Stokes $V$, the presence of a blue peak weaker than a red peak 
(both in the blue lobe of the Stokes $V$ profile) indicates fields below $\sim$700 G. 
Furthermore, the detection of peaks in the blue lobe indicates the presence of fields below $\sim$1000 G.\label{fig:milne}}
\end{figure}
\clearpage
\setlength{\voffset}{0mm}
\begin{figure*}[!t]
\plottwo{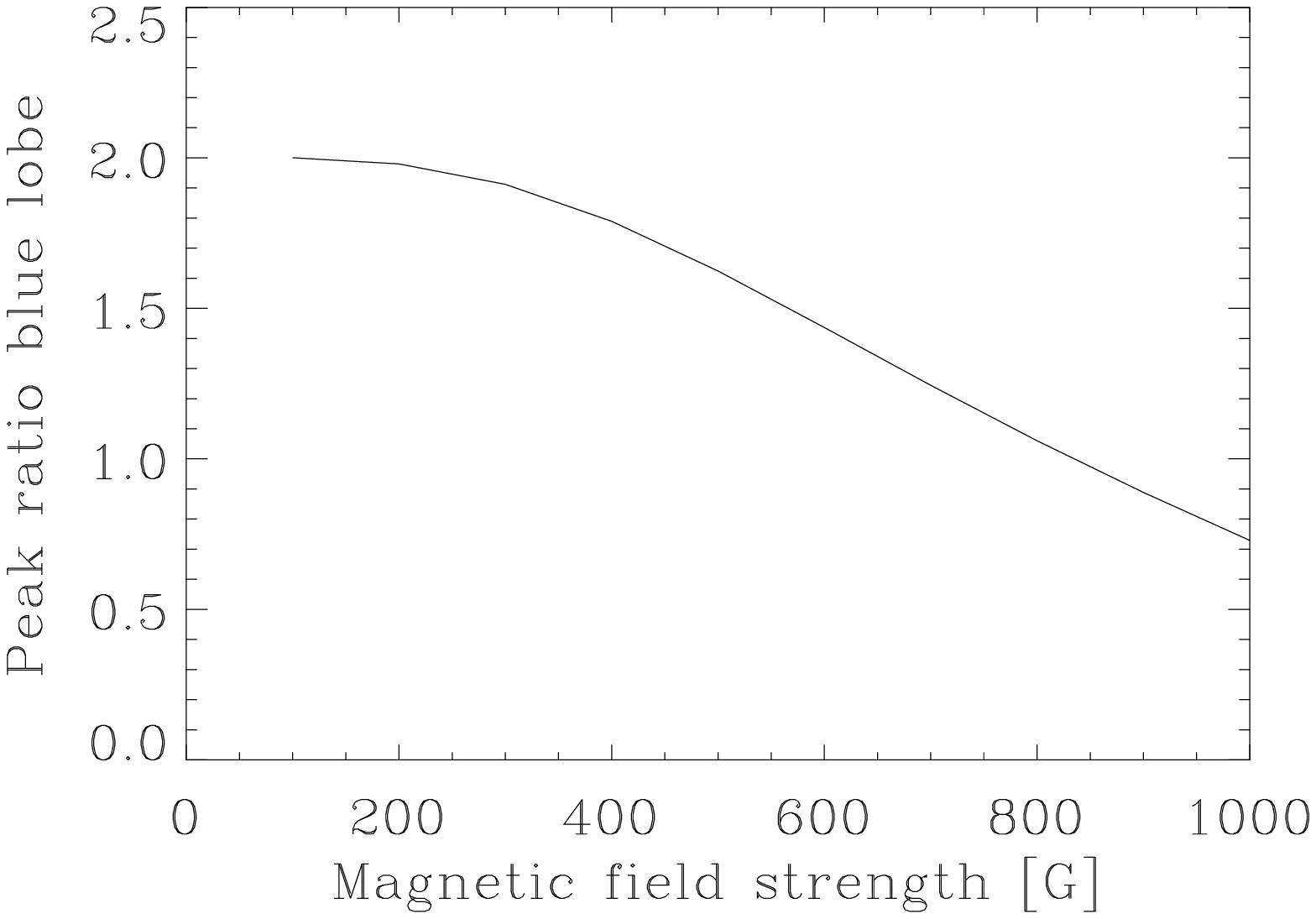}{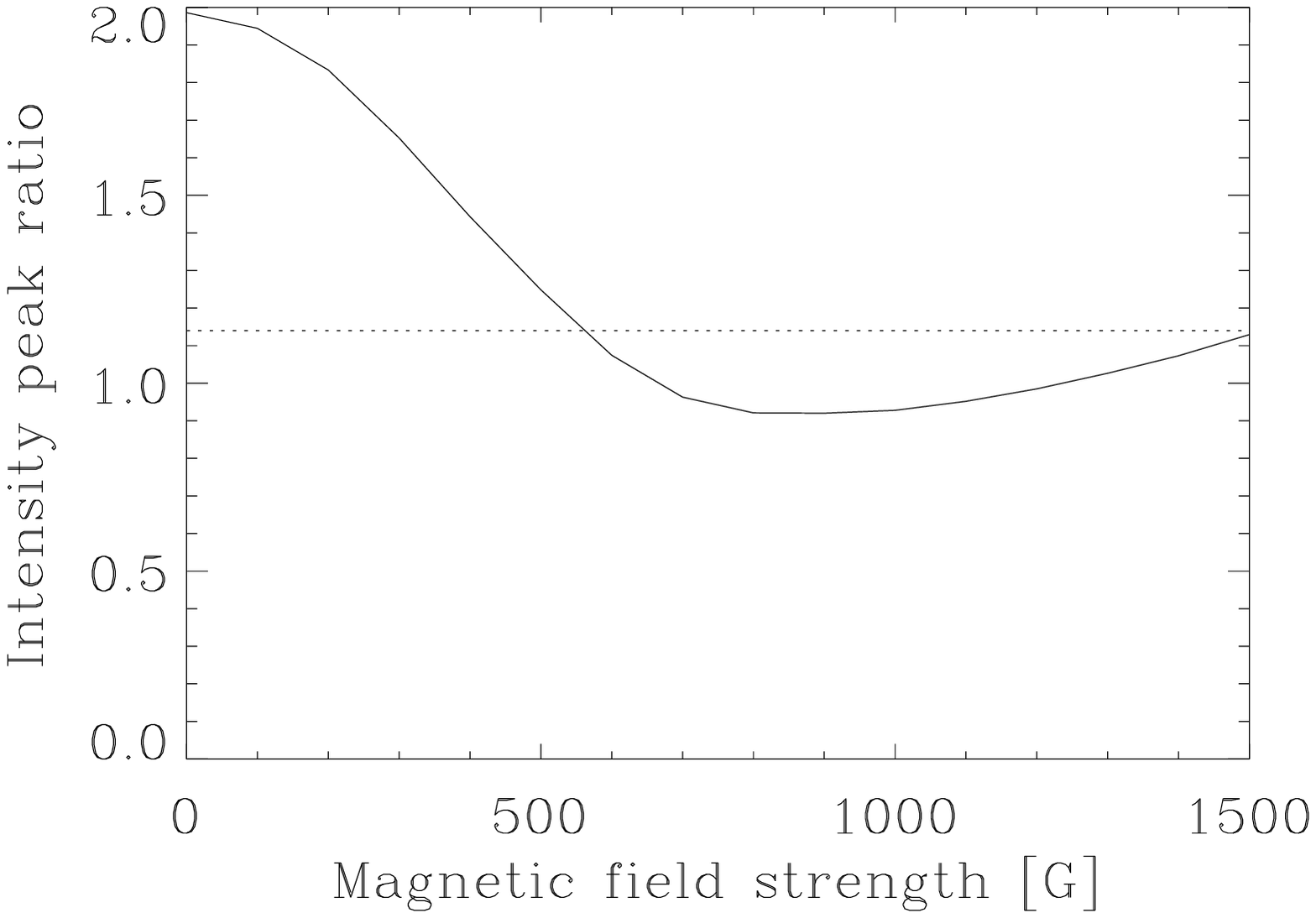}
\caption{Ratio of the peaks produced by the hyperfine structure in the Stokes $V$ profile (left panel) and
in the Stokes $I$ profile (right panel). The ratio of the peaks in Stokes $V$ cannot be correctly defined for fields 
above 1000 G because the hyperfine features disappear. Nevertheless, their presence in an observed Stokes $V$ profile is 
indicative of fields well below 1000 G. The ratio of the peaks in Stokes $I$ has a unicity problem for fields above 
$\sim$600 G, although one of the two possible solution can be chosen in terms of the peak separation
shown in Fig. \ref{fig:peak_separation_stokesi}.\label{fig:peak_ratio}}
\end{figure*}
\begin{figure*}[!t]
\plottwo{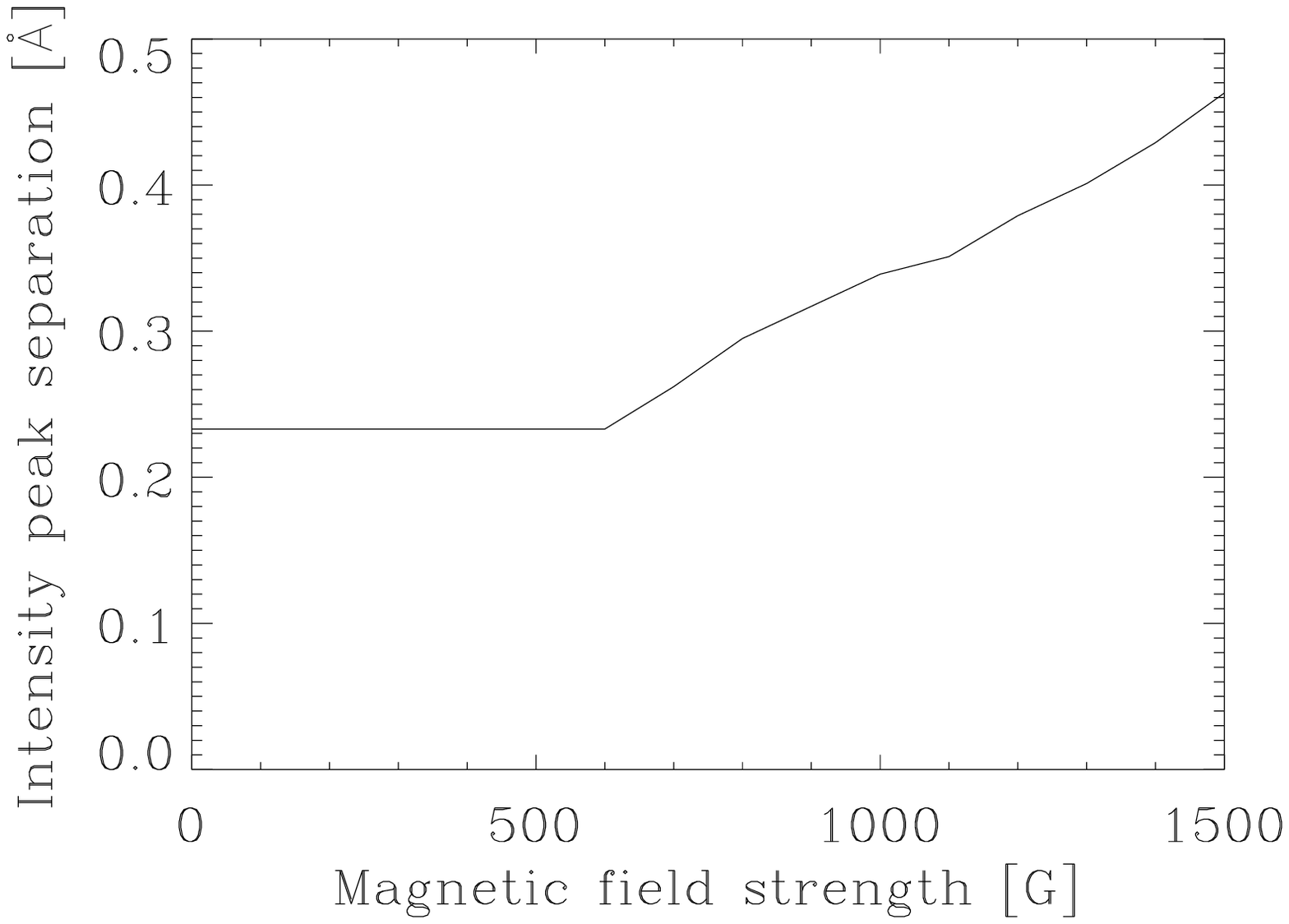}{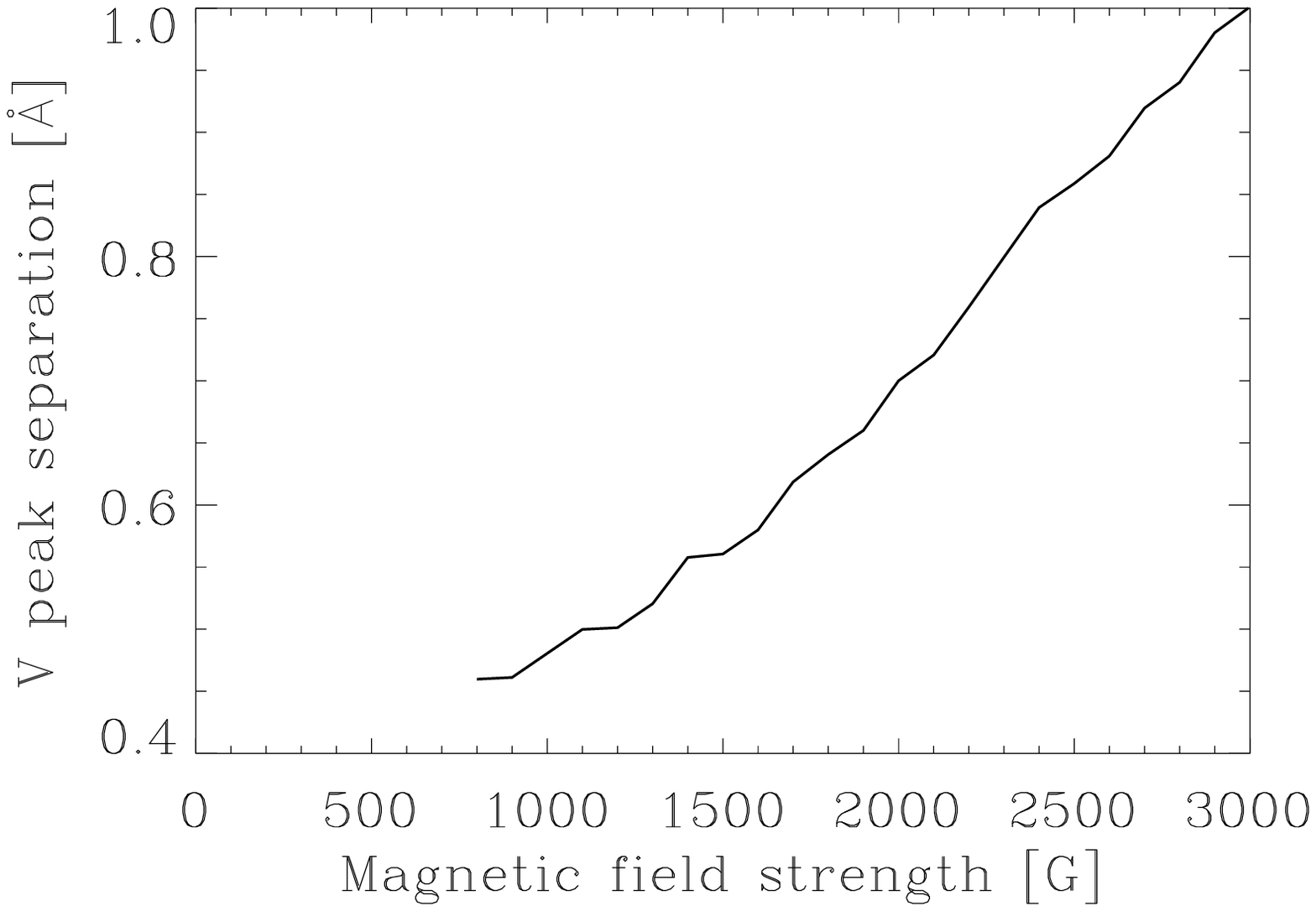}
\caption{Separation of the peaks in the Stokes $I$ profile (left panel) and in the Stokes $V$ profile (right panel). 
The peak separation for Stokes $I$ remains constant for fields below $\sim$600 G and increases linearly for 
stronger fields. For Stokes $V$, the peak separation (defined only above $\sim$900 G) increases also linearly
in the strong field regime.\label{fig:peak_separation_stokesi}}
\end{figure*}
\begin{figure*}[!t]
\plottwo{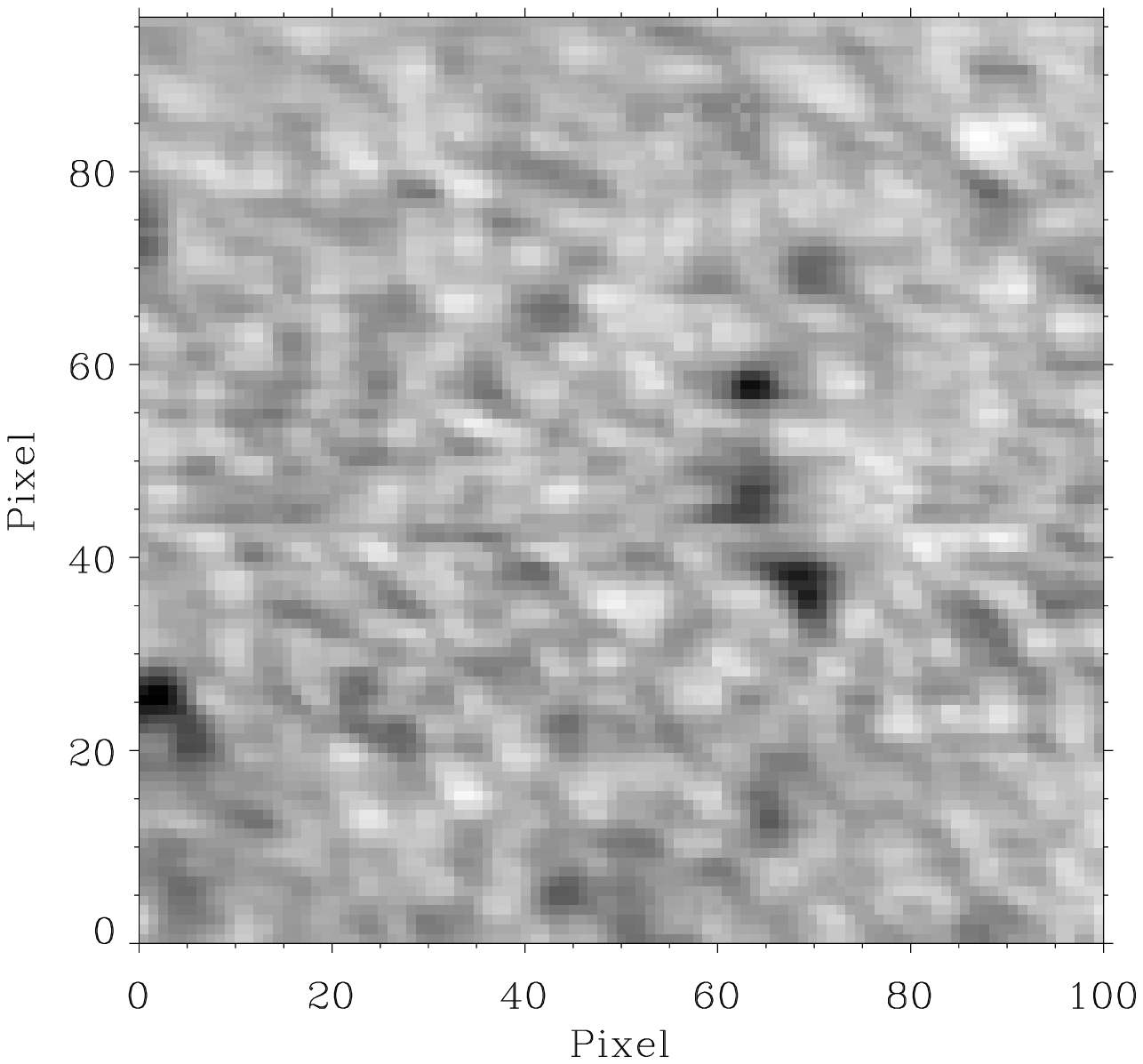}{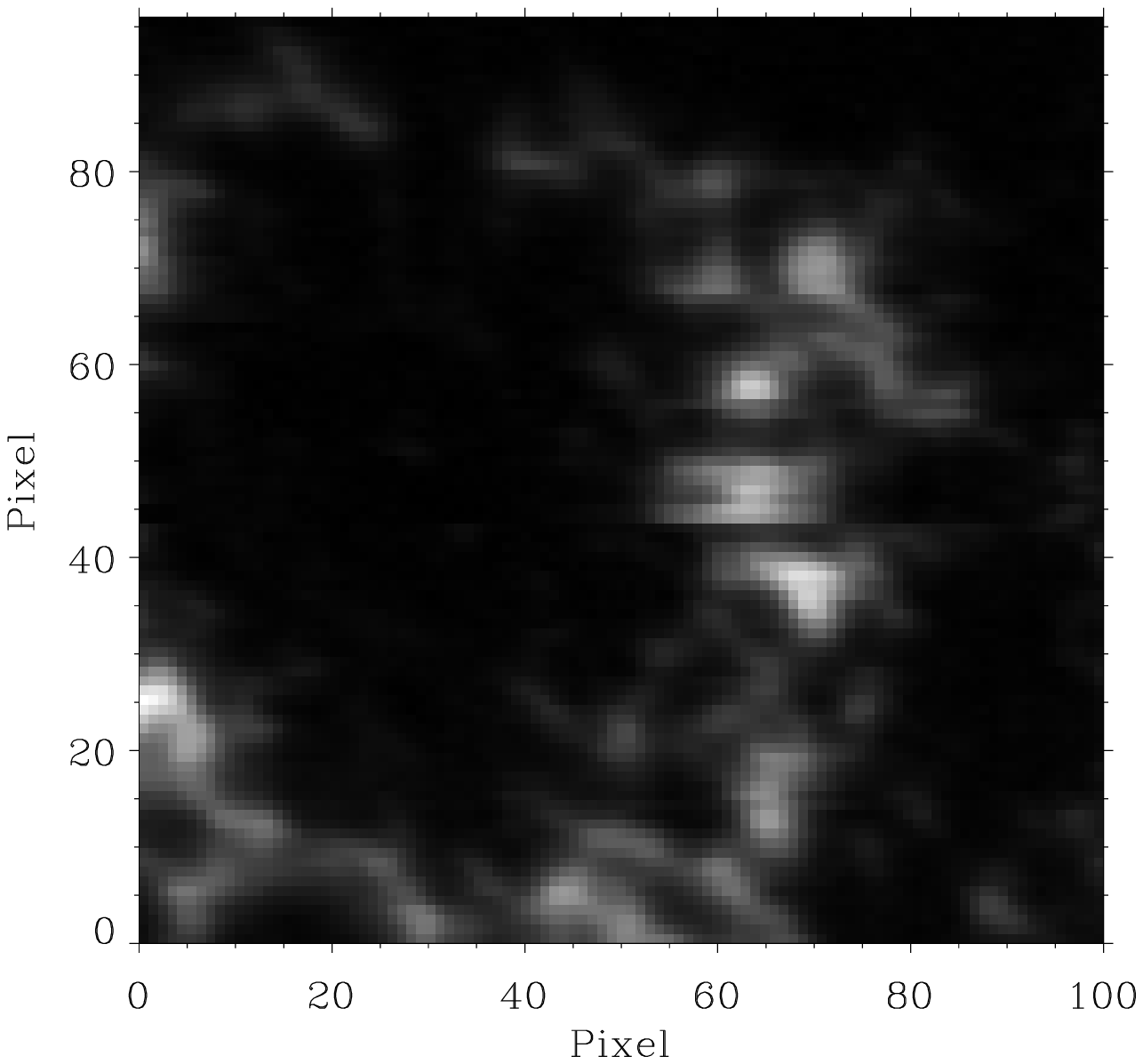}
\caption{Map of the integrated absolute value of the circular polarization signal showing 
the annular active region with the internetwork region inside it.\label{fig:totalV}}
\end{figure*}
\begin{figure*}
\centering
\plotone{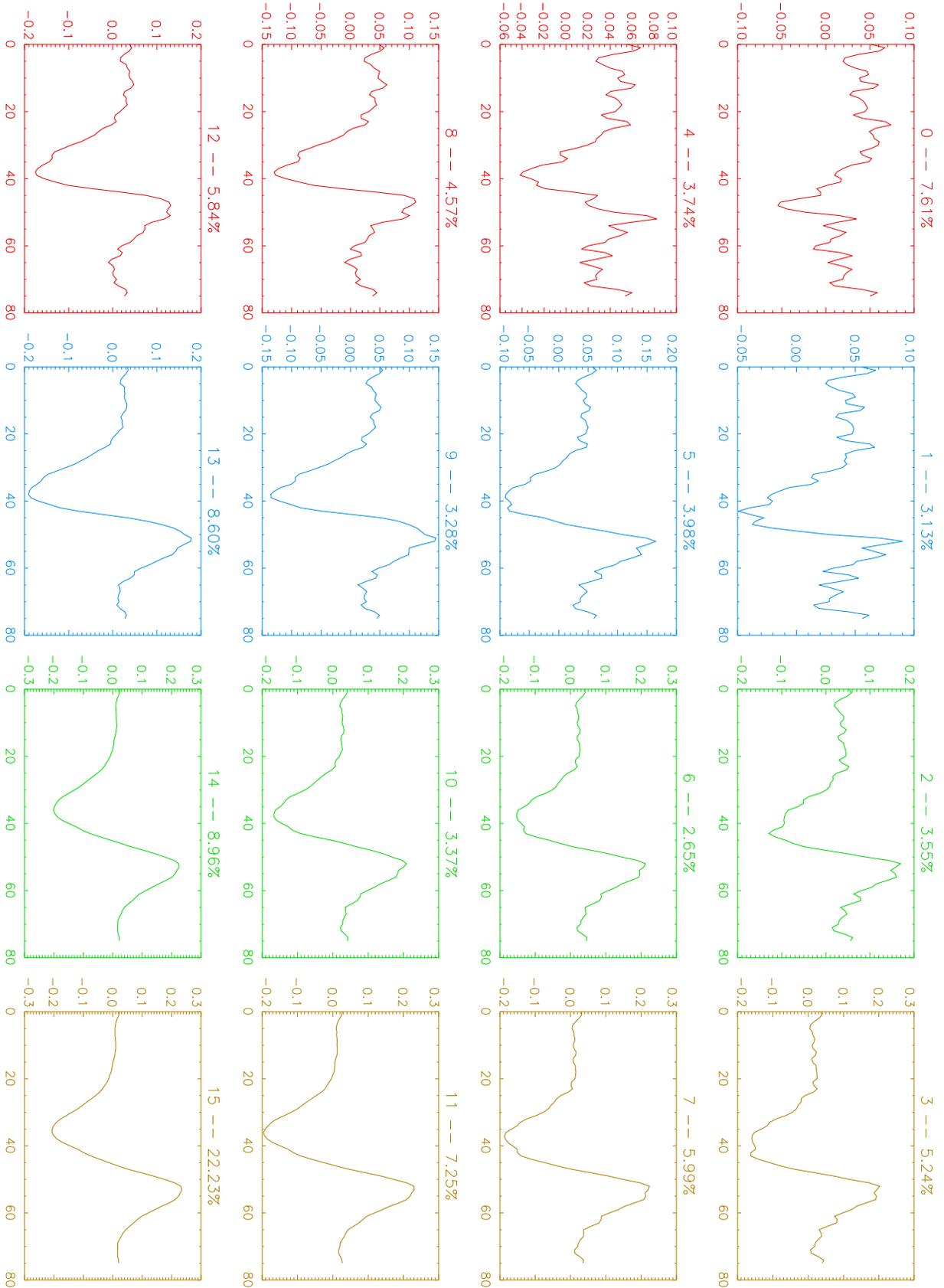}
\caption{Classification obtained using a self-organizing map with a network of 4x4 neurons.\label{fig:classes_4x4}}
\end{figure*}
\begin{figure*}
\centering
\plotone{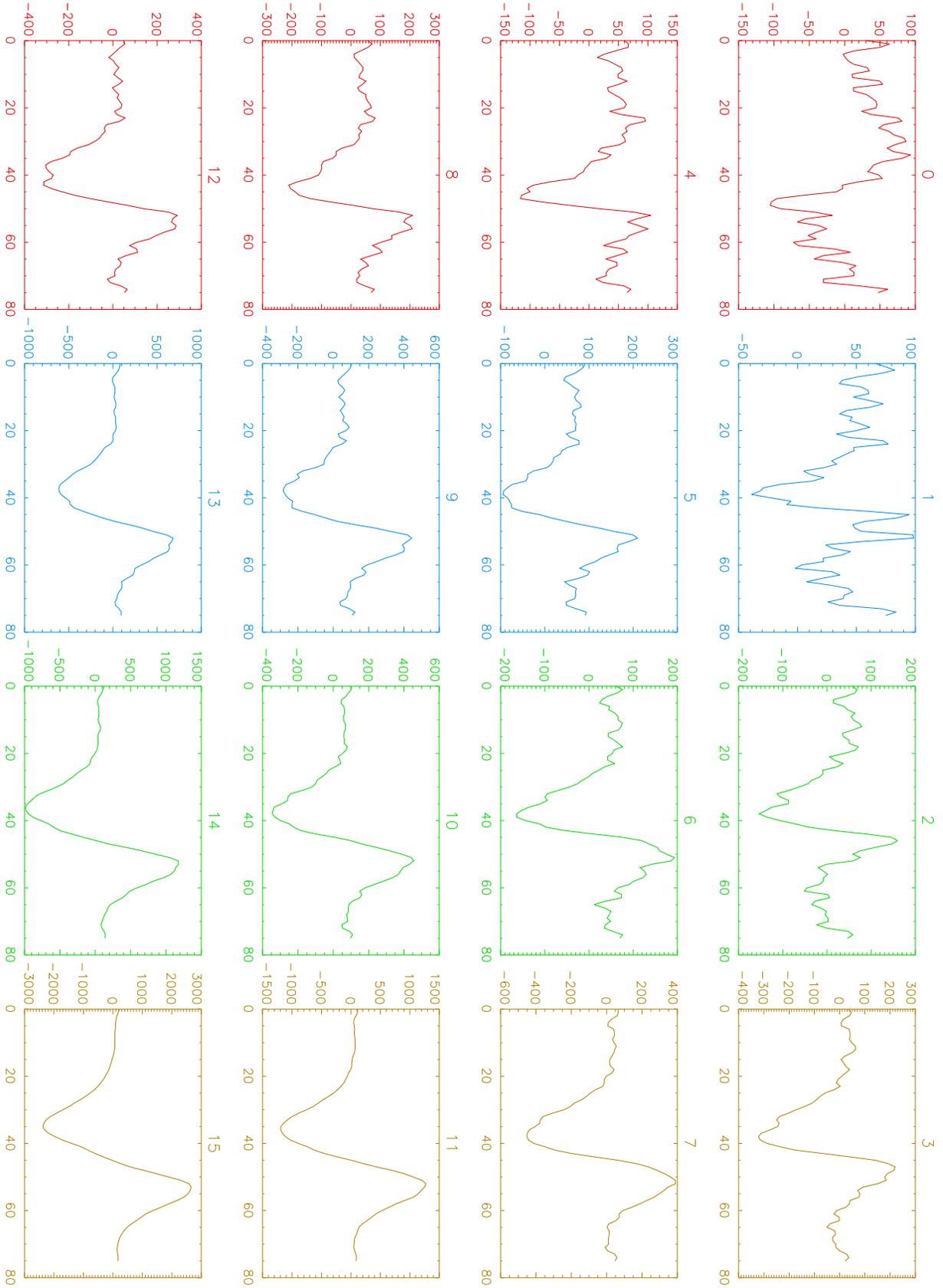}
\caption{Average profile obtained from all the profiles belonging to each class in the 4x4 case.\label{fig:average_profile_4x4}}
\end{figure*}
\begin{figure*}[!t]
\caption{Location of the profiles belonging to each class in the 4x4 classification.\label{fig:location_4x4}}
\end{figure*}
\begin{figure*}
\plottwo{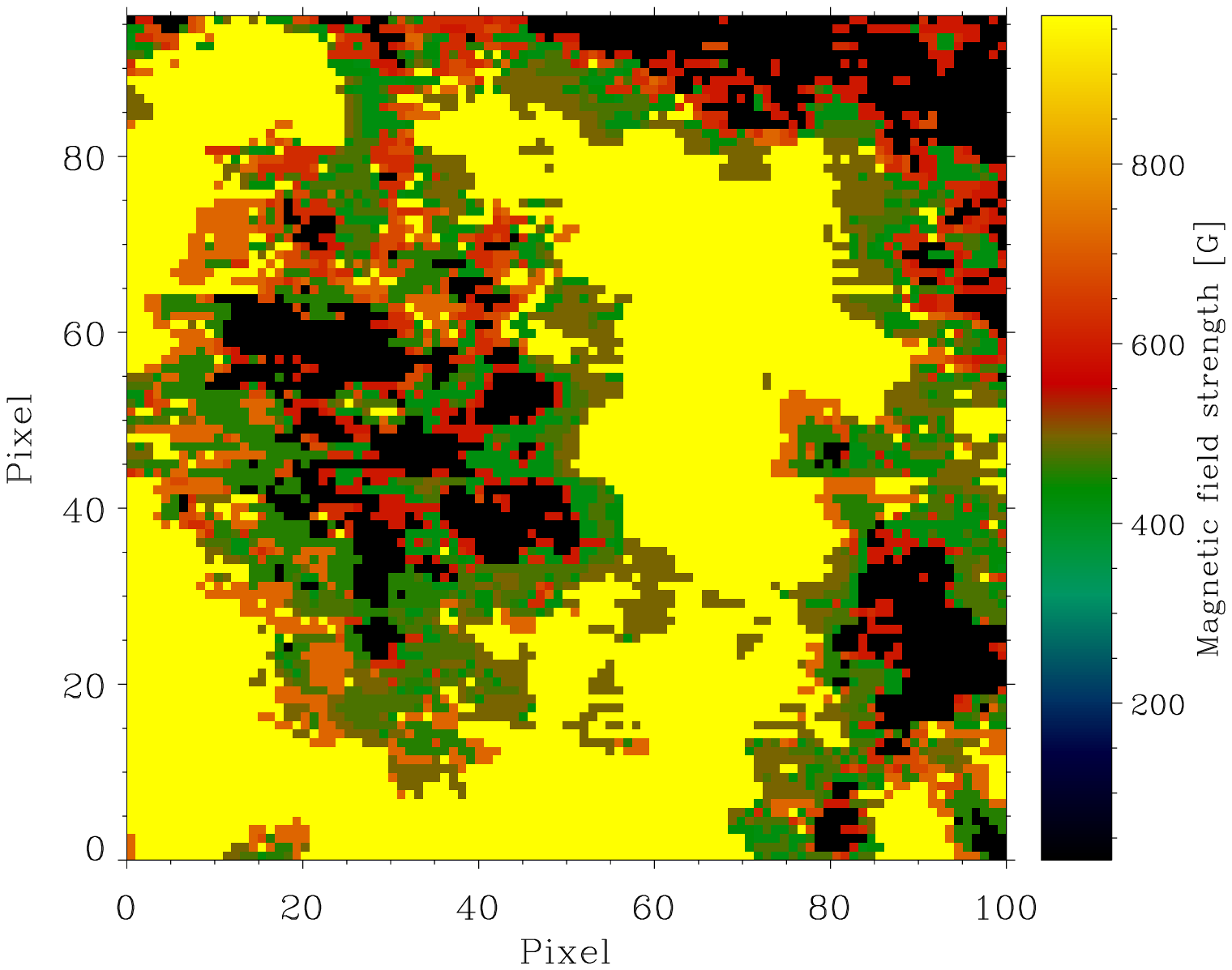}{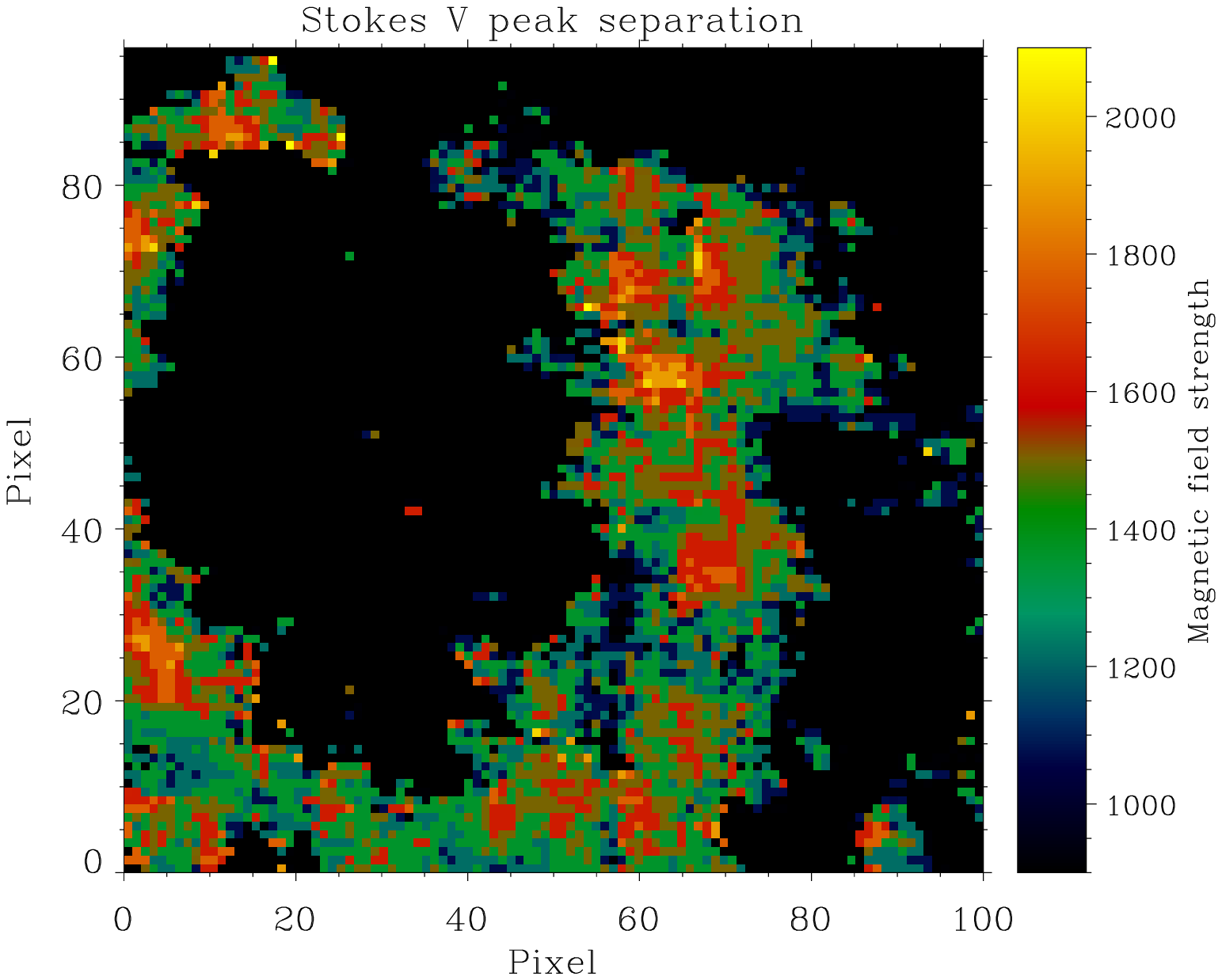}
\caption{The left panel shows the magnetic field map obtained using the transformation between Stokes $V$ peak ratio 
and magnetic field obtained from the classification with the 4$\times$4 SOM. The values for 
the transformation are summarized in Table \ref{tab:peak_ratio_4x4}. The right panel shows
the magnetic field strength obtained from the separation of the Stokes $V$ profile. We have
only taken into account the profiles that present a clear antisymmetric shape.\label{fig:magnetic_field_map}}
\end{figure*}
\begin{figure}
\plotone{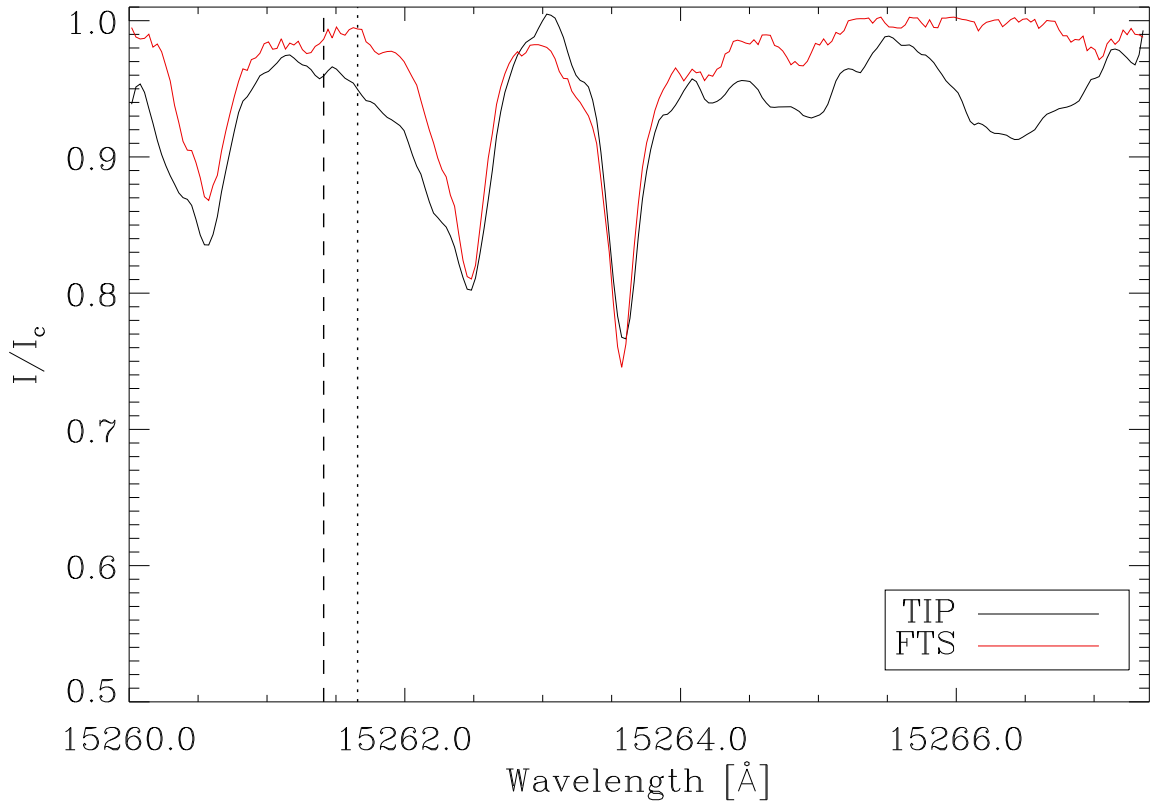}
\caption{Intensity spectrum observed with TIP (black curve) and the one from the FTS 
atlas (red curve). We indicate with a vertical dotted line the wavelength where we take 
the continuum 1. This continuum forces the largest peak ratios of the \ion{Mn}{1}
line to be close to 2, the theoretical limit. The dashed line indicates the wavelength where 
we take the continuum 2, that can be considered as an upper limit.\label{fig:TIP_vs_FTS}}
\end{figure}
\begin{figure*}[!t]
\centering
\plottwo{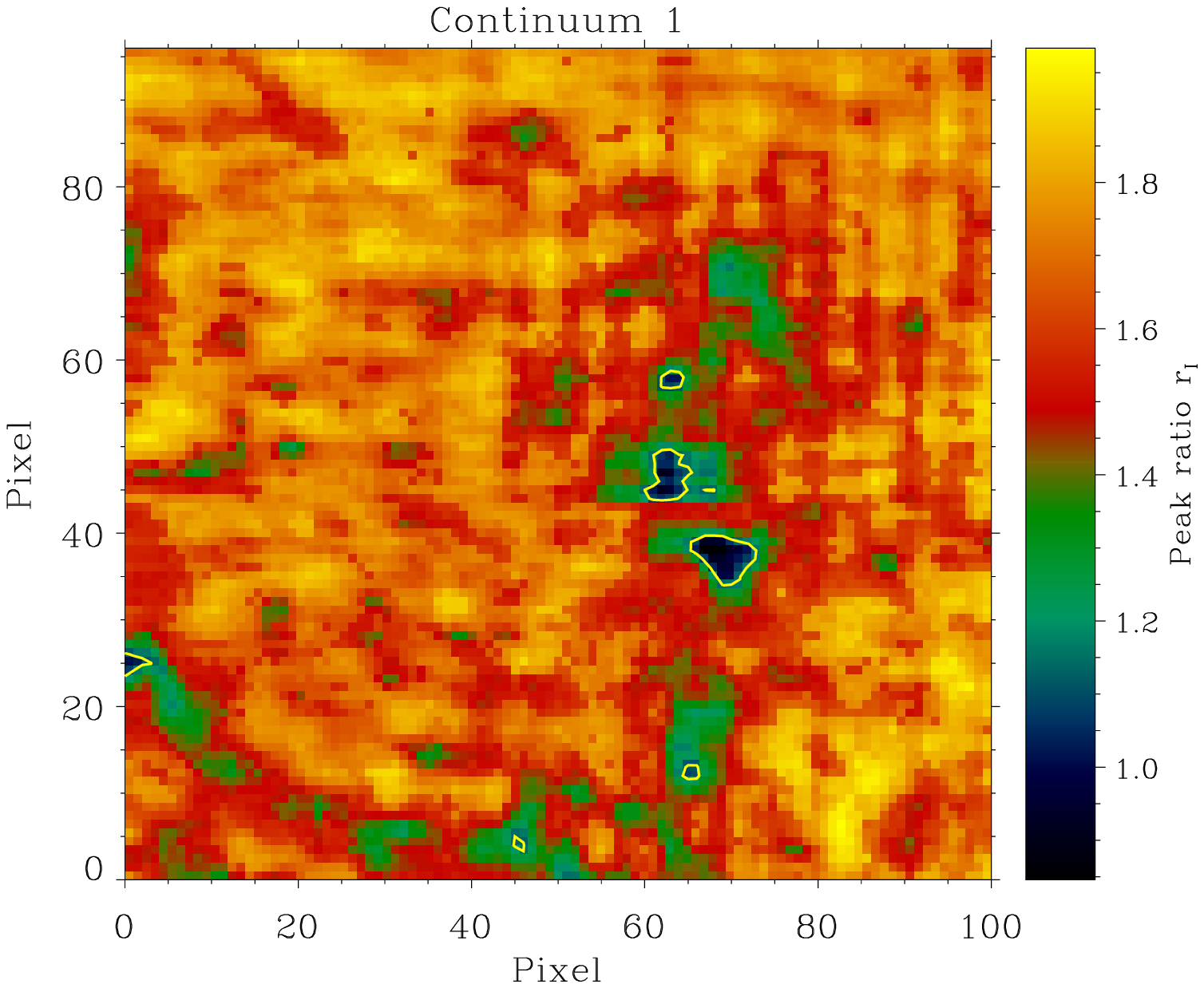}{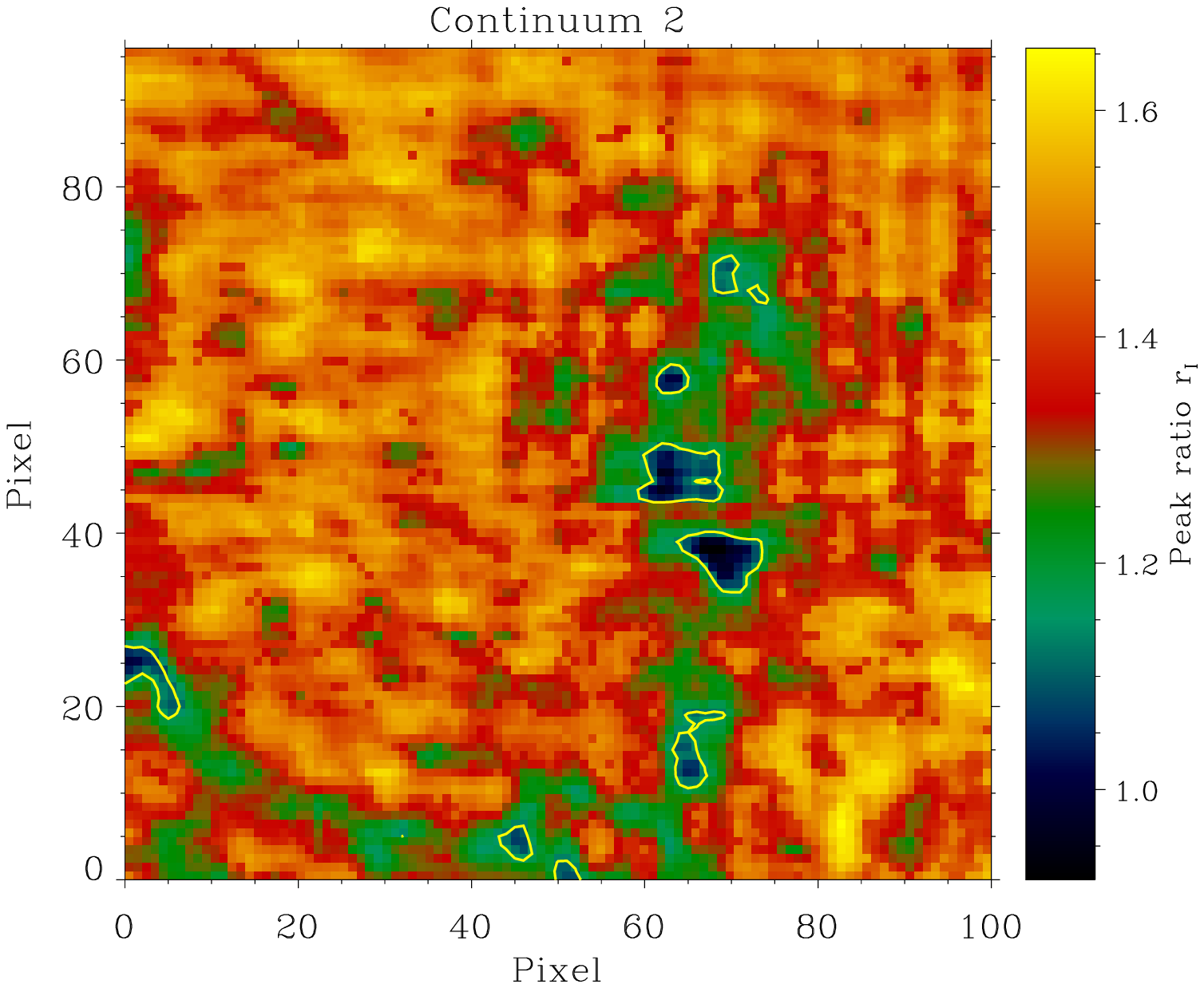}
\plottwo{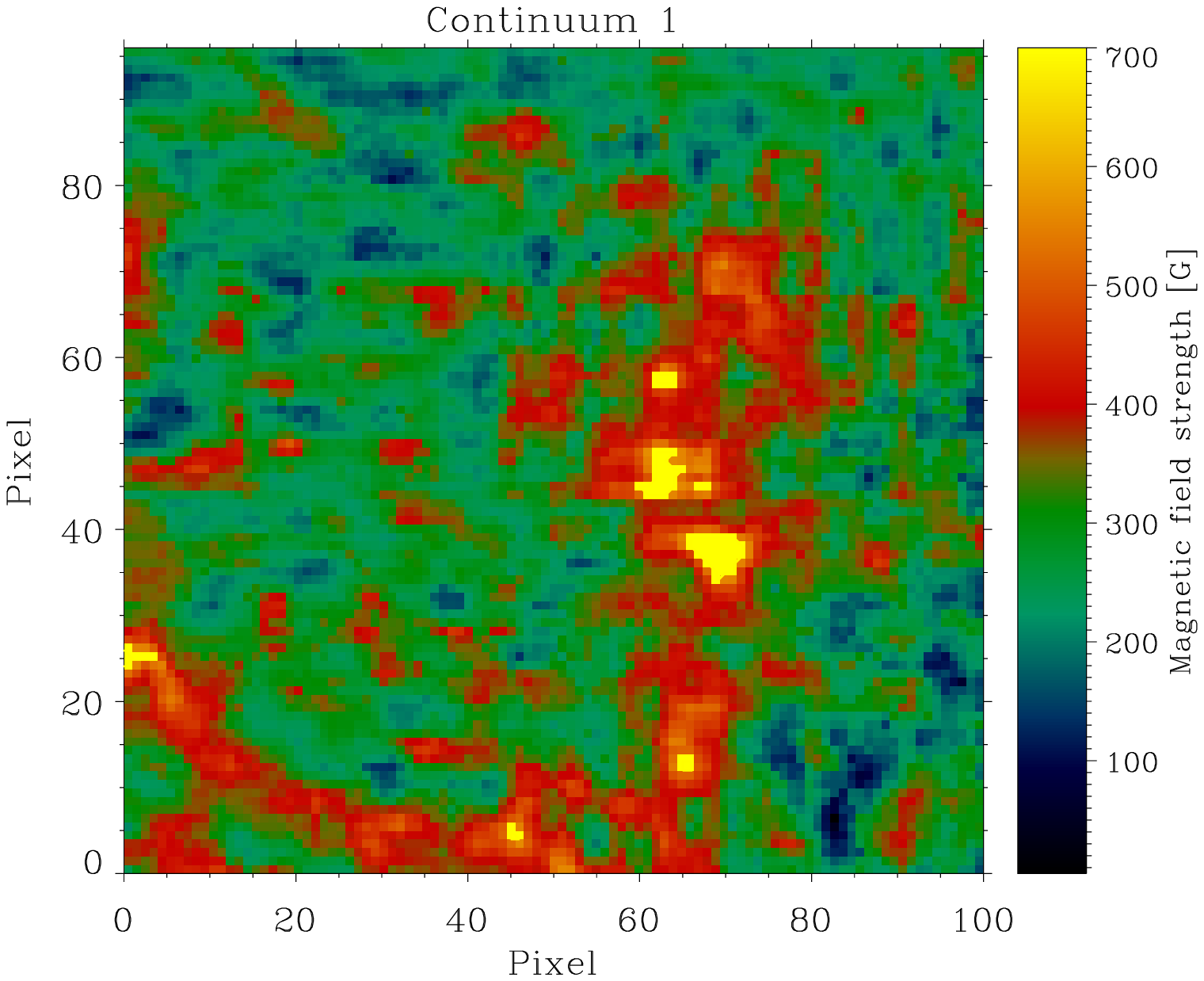}{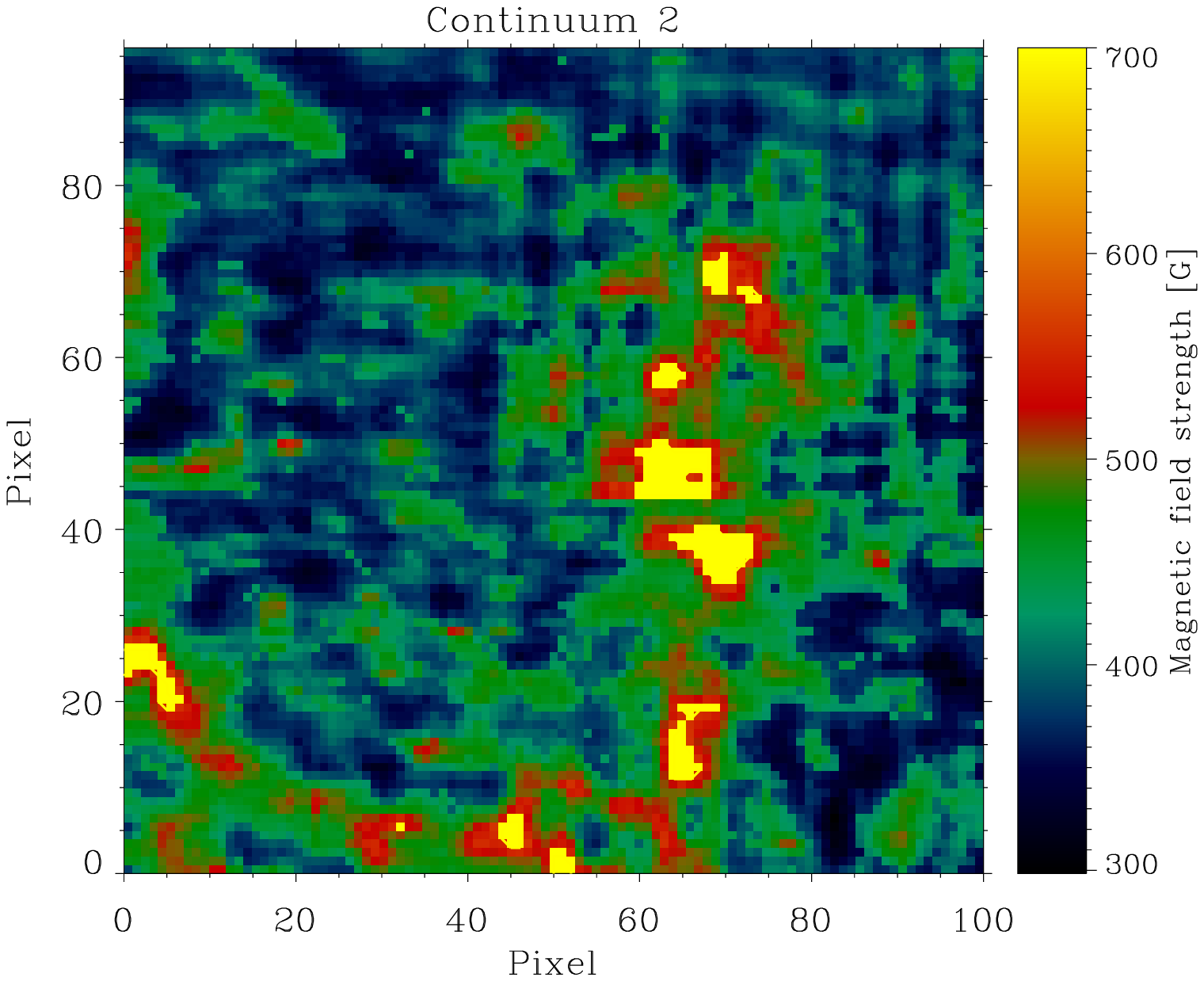}
\caption{The upper panel shows the horizontal variation of the peak ratio measured 
in the intensity profile. The left panel is the ratio obtained using the continuum indicated in
Fig. \ref{fig:TIP_vs_FTS} with a dotted line, while the right panel is the ratio obtained
using the continuum marked in Fig. \ref{fig:TIP_vs_FTS} with a dashed line. We indicate
the region where the ratio is smaller than 1.14, where two possible values of the magnetic
field are associated with the same ratio. The lower panel indicates the value of the magnetic
field obtained from the ratio using the calibration given by Fig. \ref{fig:peak_ratio}. The pixels
where the ratio is smaller than 1.14 are indicated with a constant field of 700 G.\label{fig:ratio_peak_intensity}}
\end{figure*}
\begin{figure*}
\plottwo{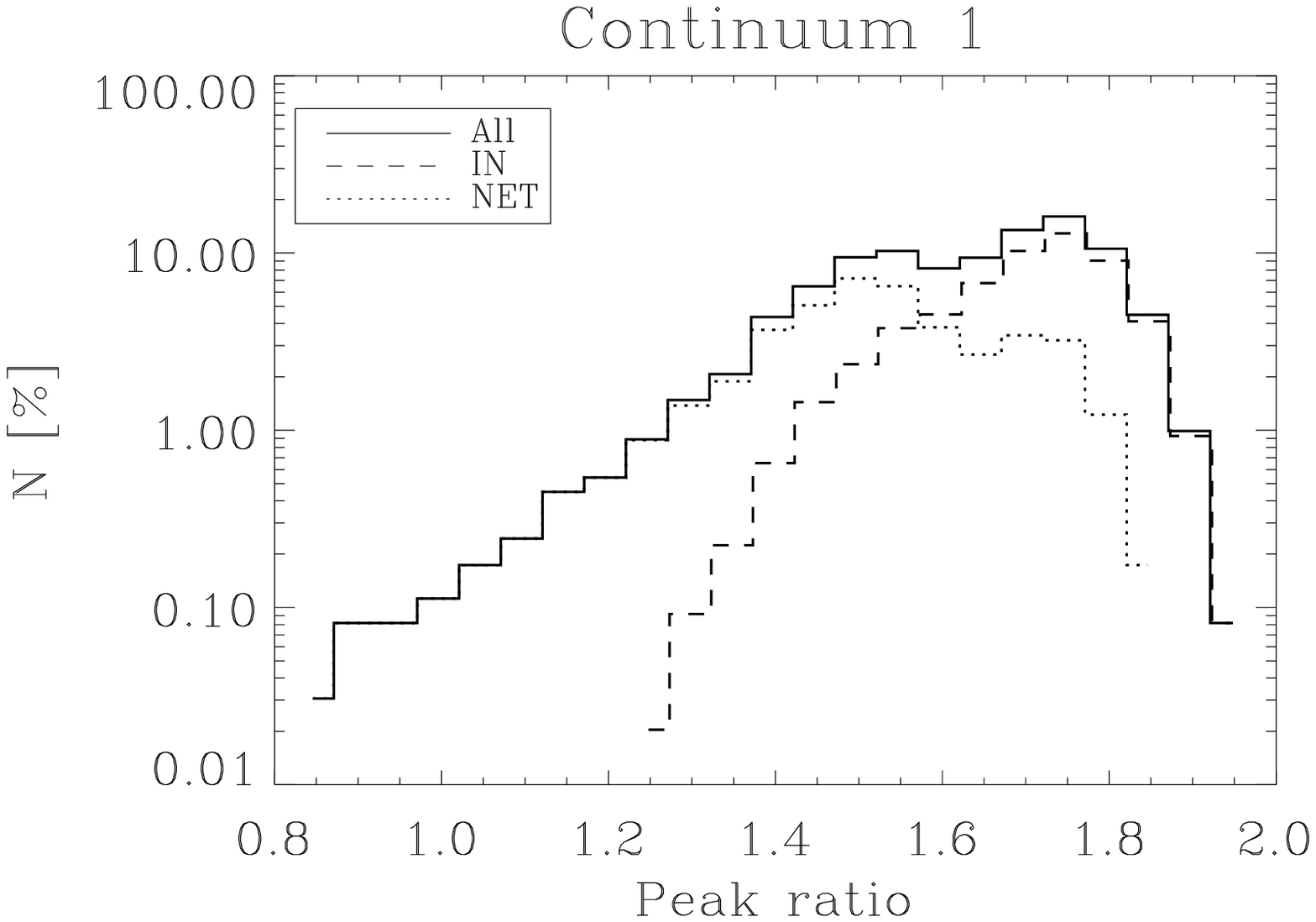}{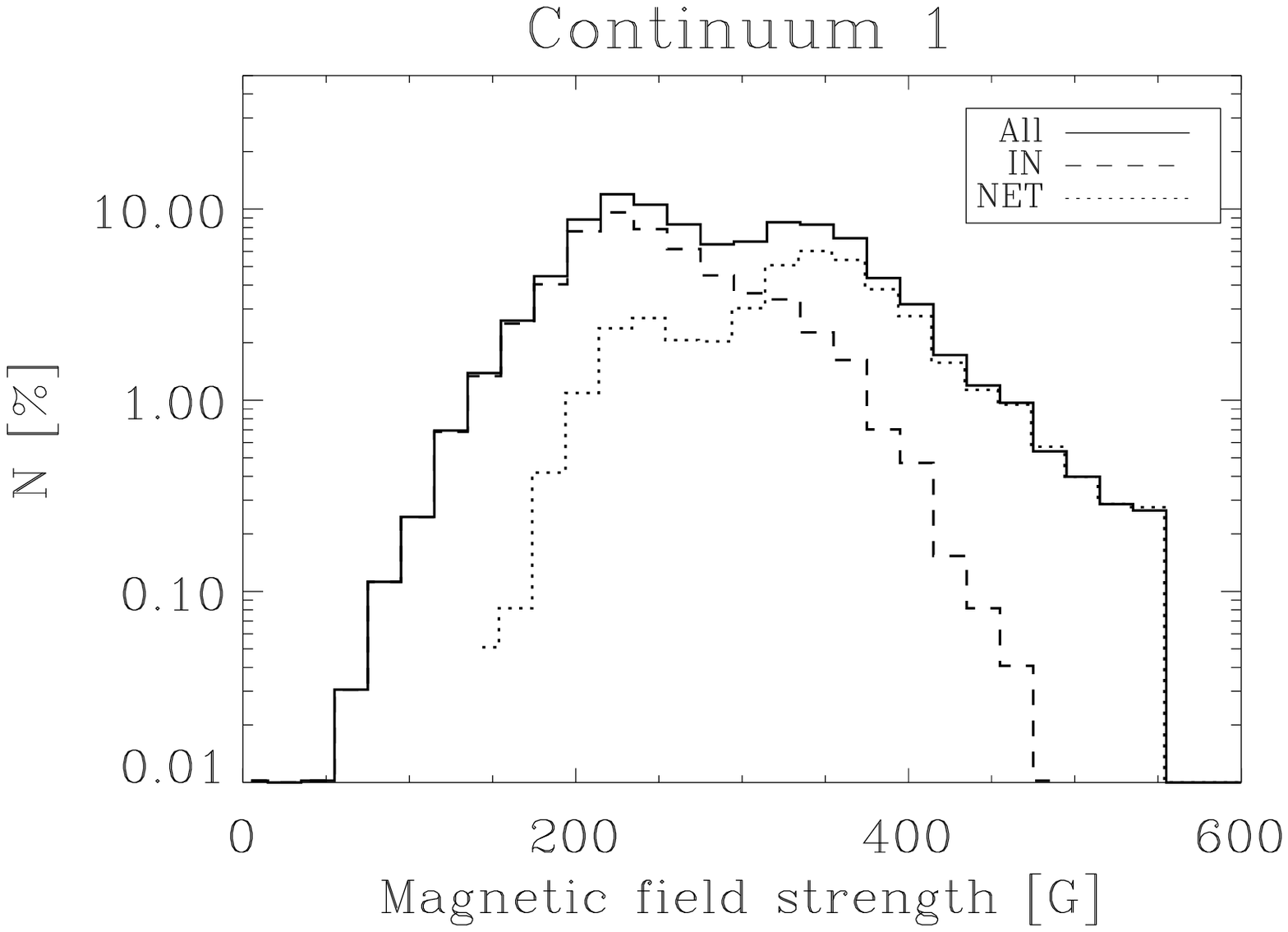}
\plottwo{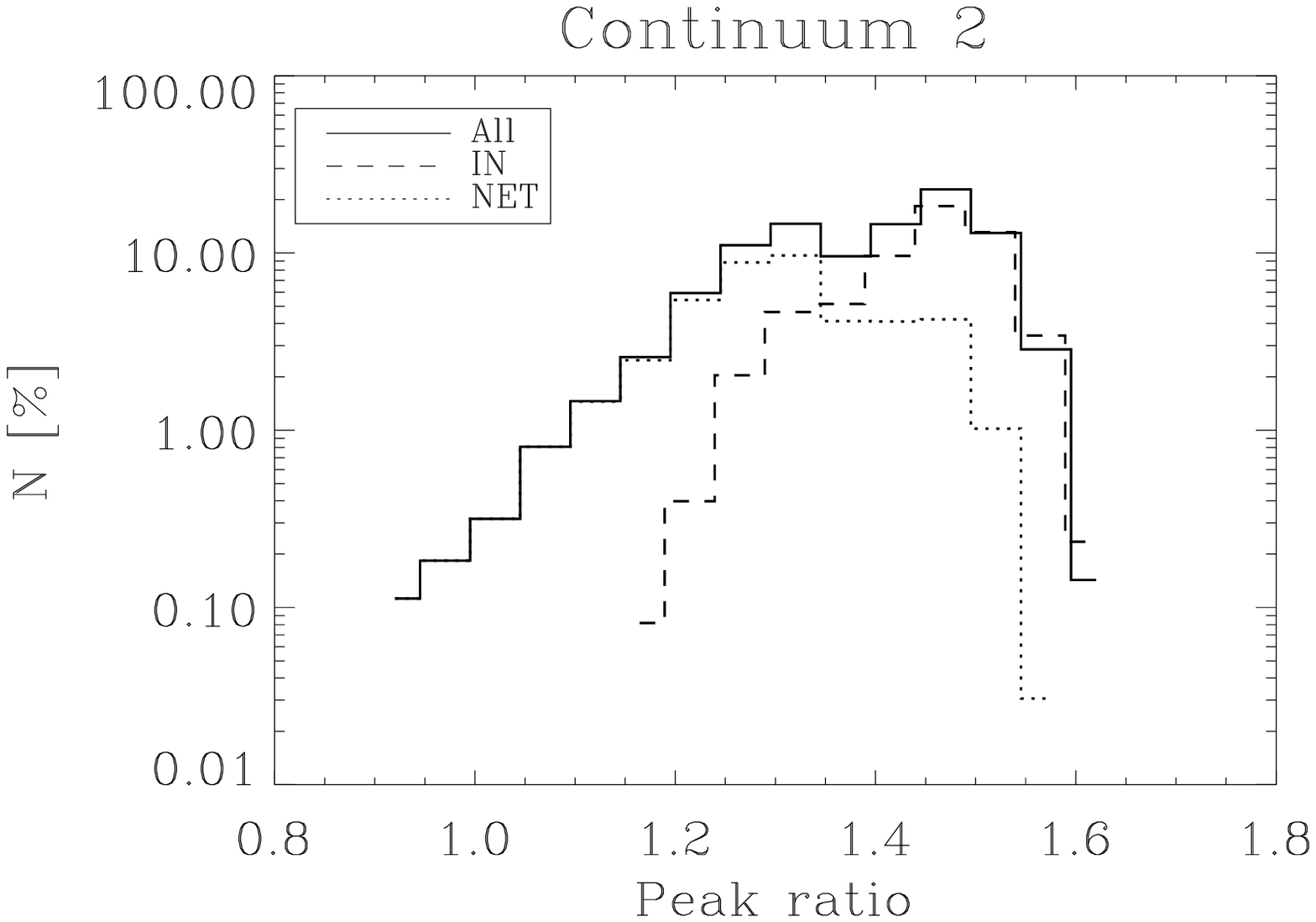}{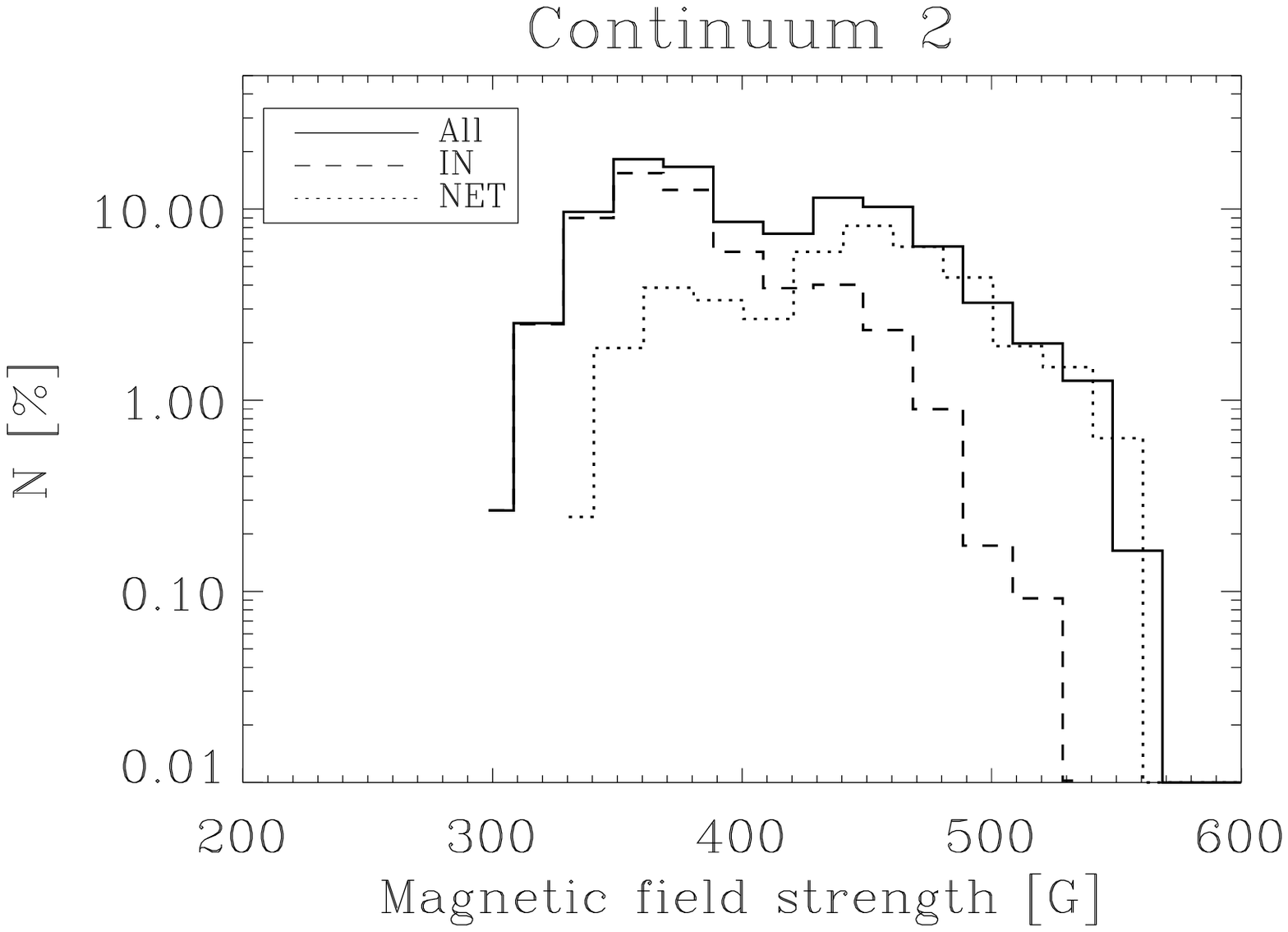}
\caption{Histograms of the peak ratio (left panels) and magnetic field strength using the calibration presented
in the paper (right panels). We present the results obtained for the two selected values of the continuum.
We also include the histogram of the points that belong to the internetwork.\label{fig:histograms}}
\end{figure*}
\begin{figure*}
\plottwo{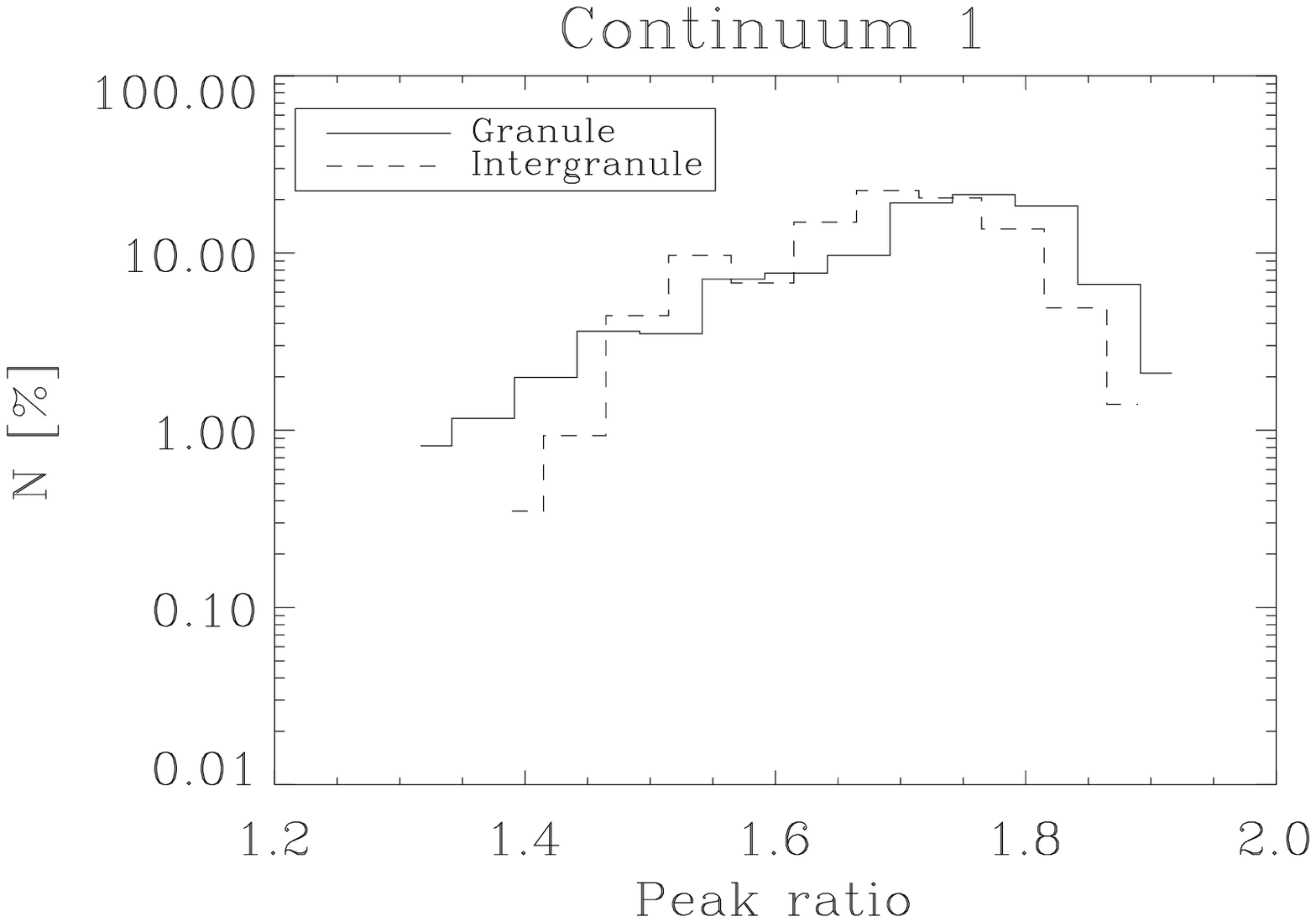}{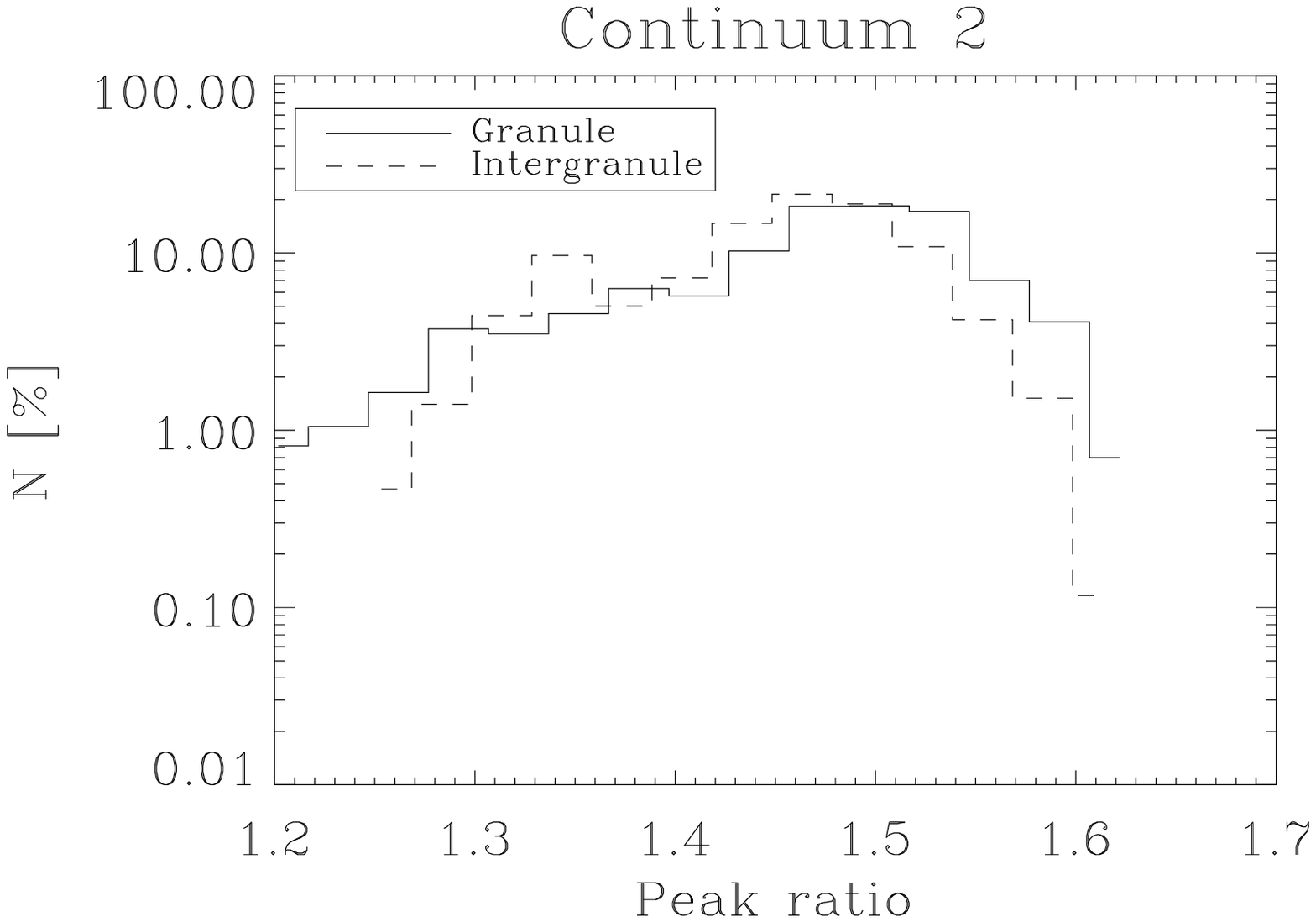}
\caption{Histograms of peak ratios for the granular and intergranular pixels separately. Both panels
have been calculated with different values of the continuum. Note that the histograms are similar,
giving an indication that we are not able to distinguish between granules and intergranules.\label{fig:histograms_granule_intergranule}}
\end{figure*}
\clearpage
\thispagestyle{empty}
\setlength{\voffset}{-25mm}
\begin{figure*}
\plottwo{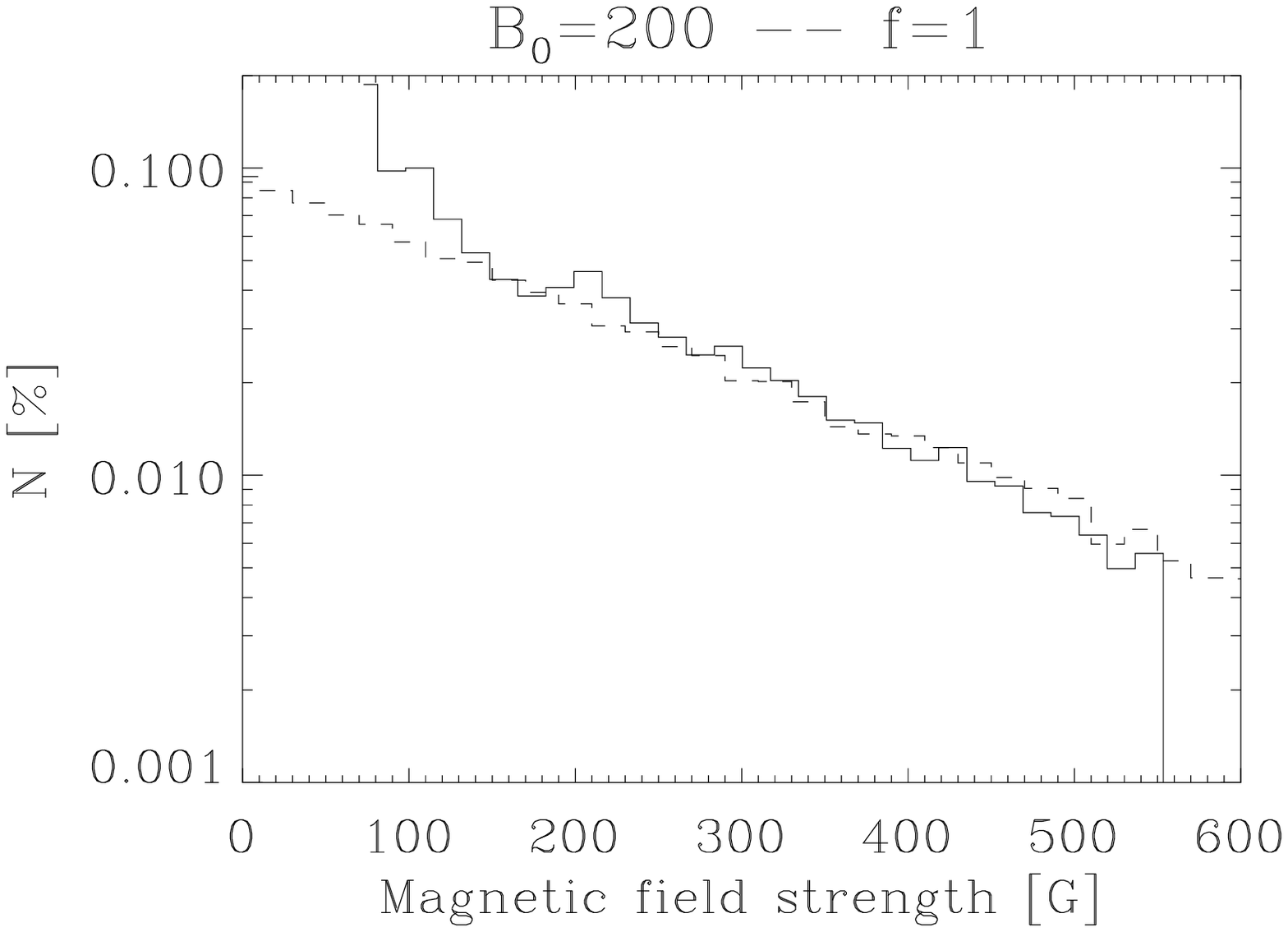}{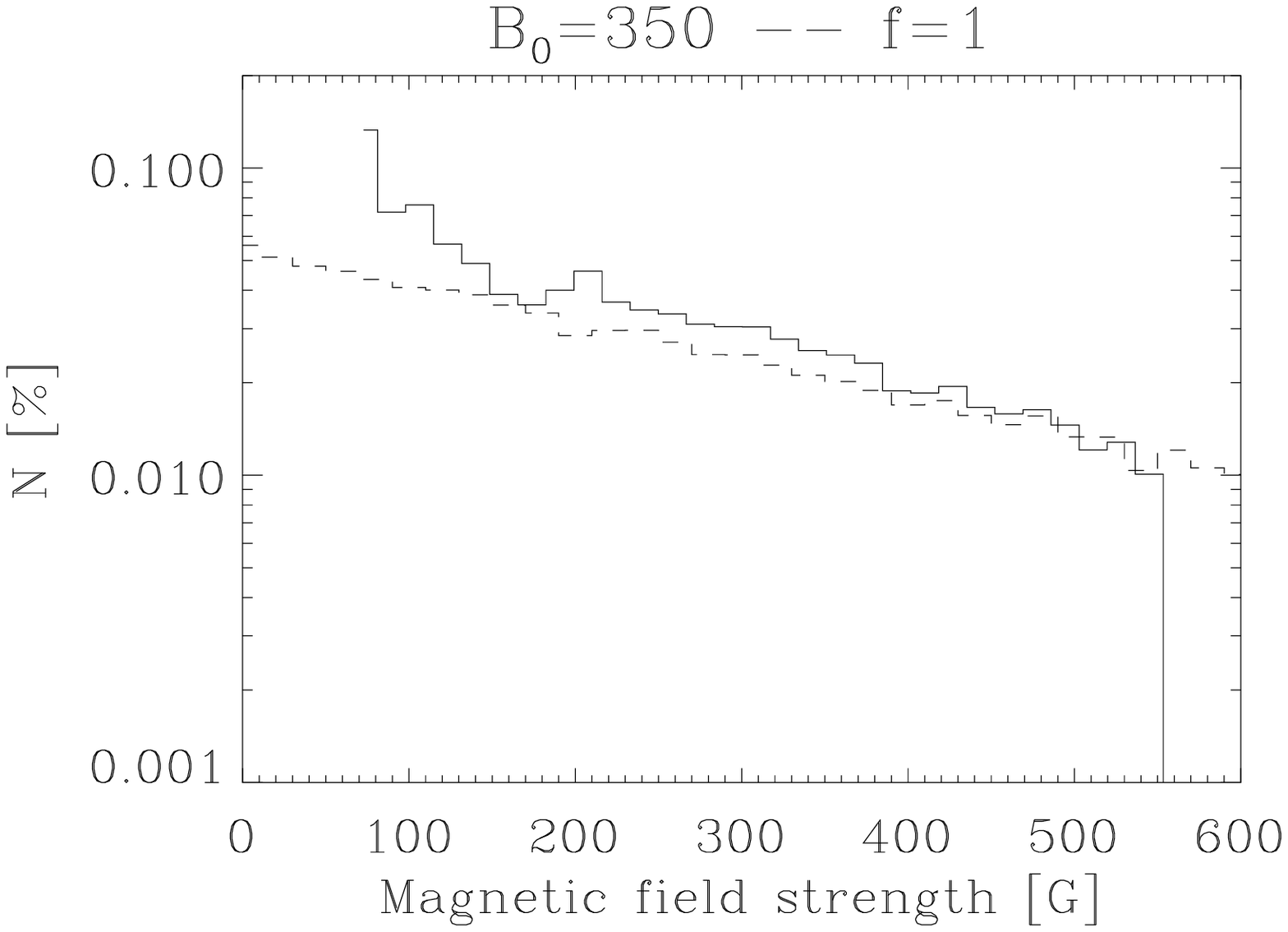}
\plottwo{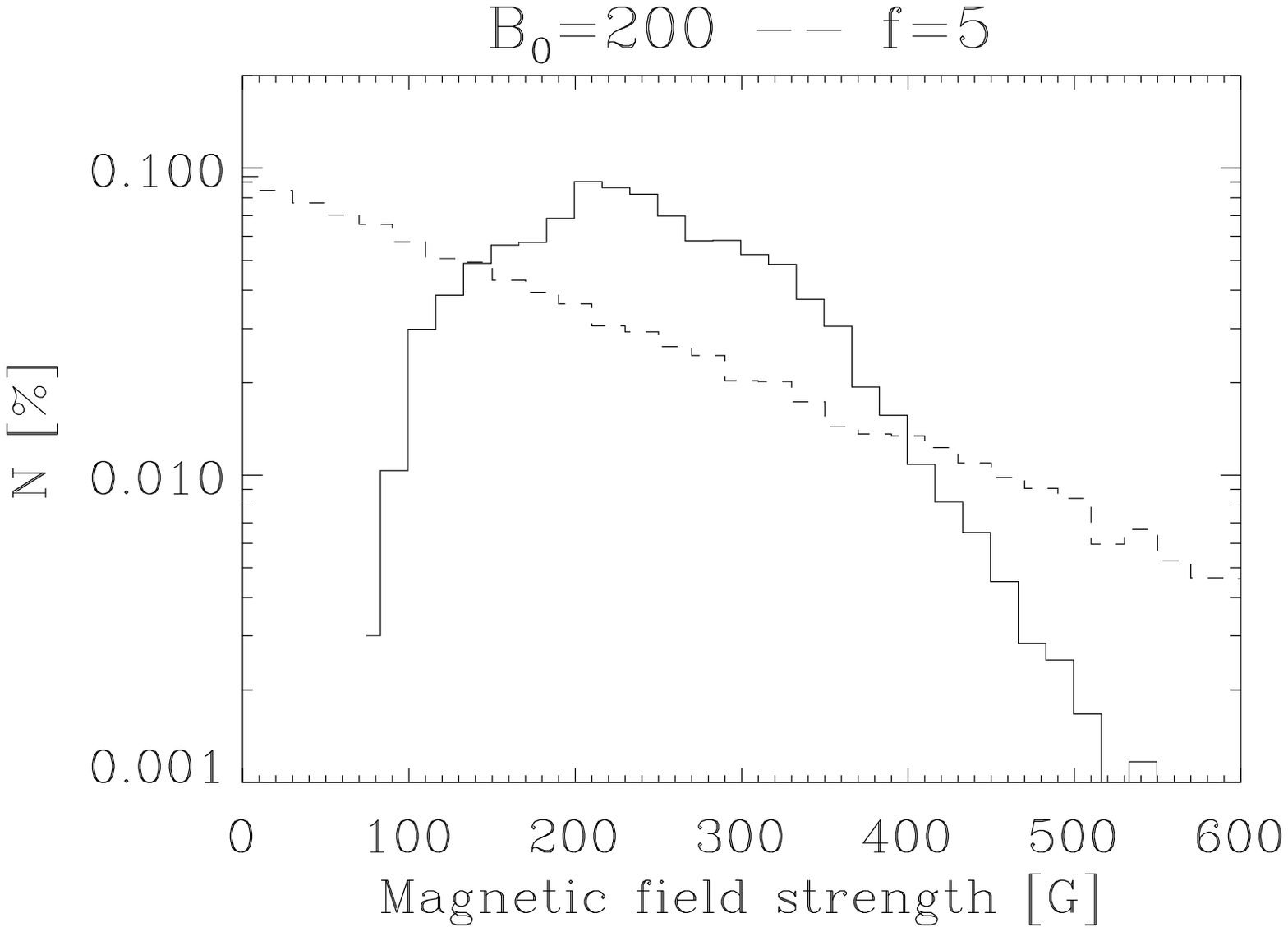}{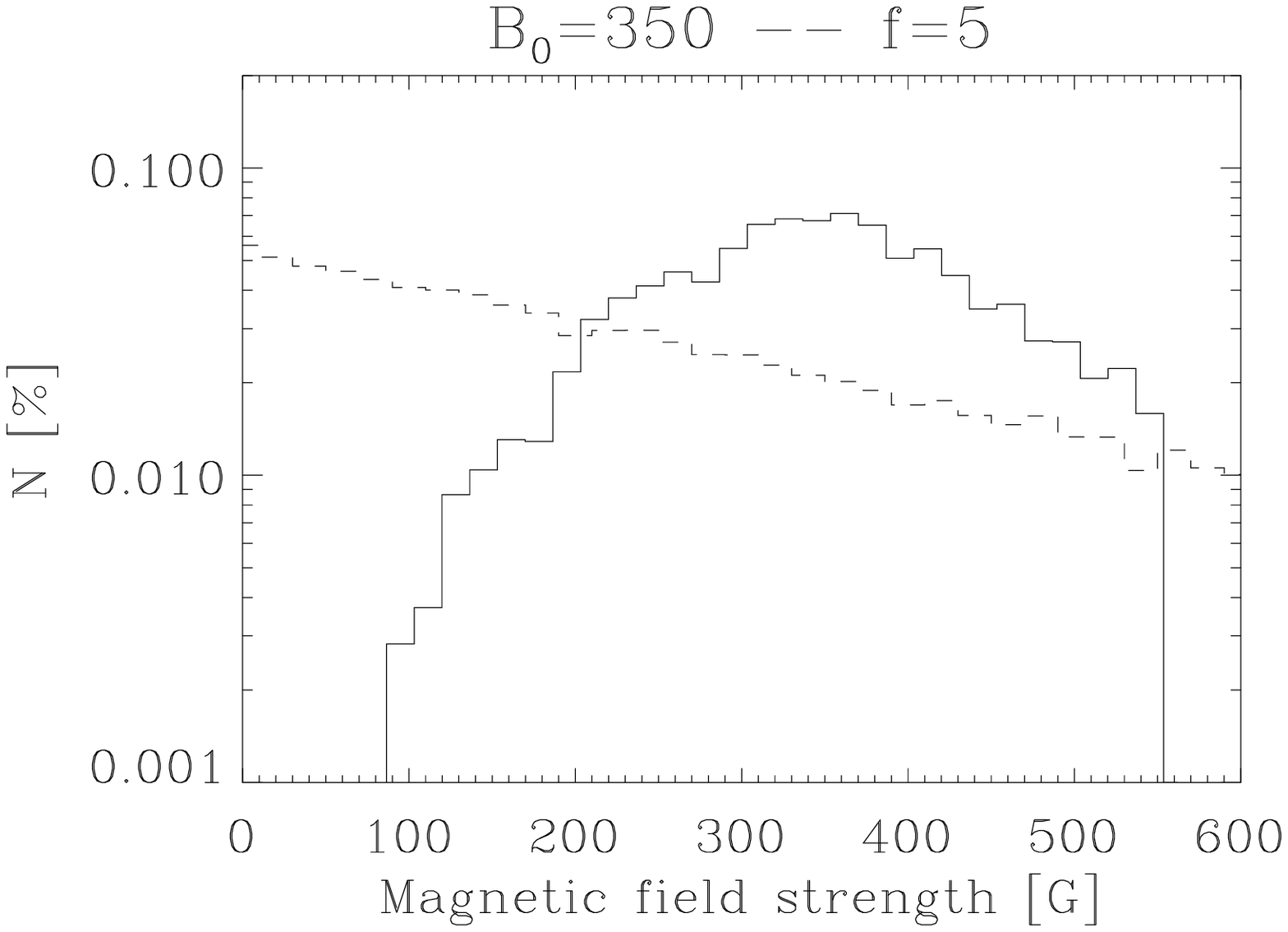}
\plottwo{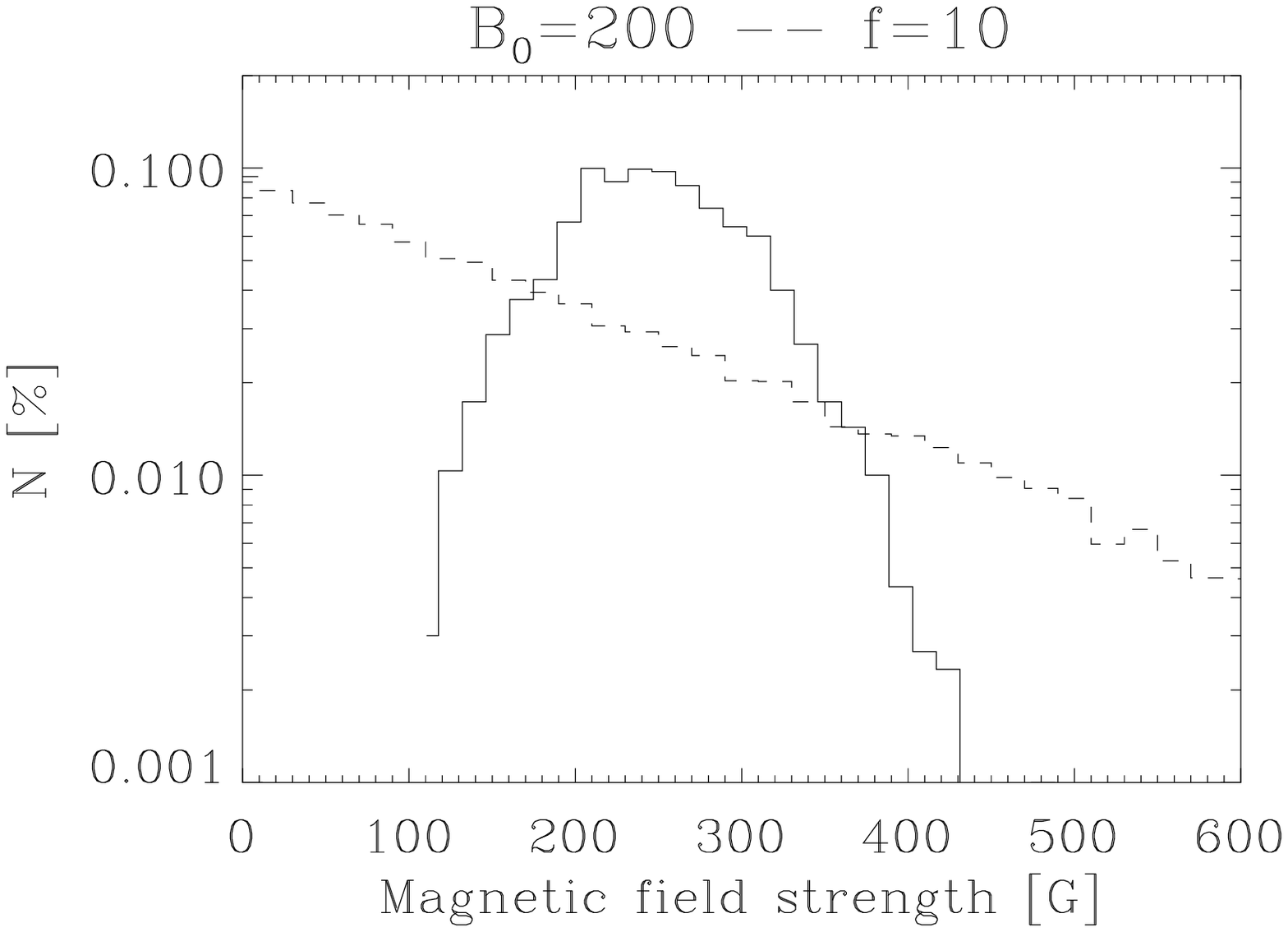}{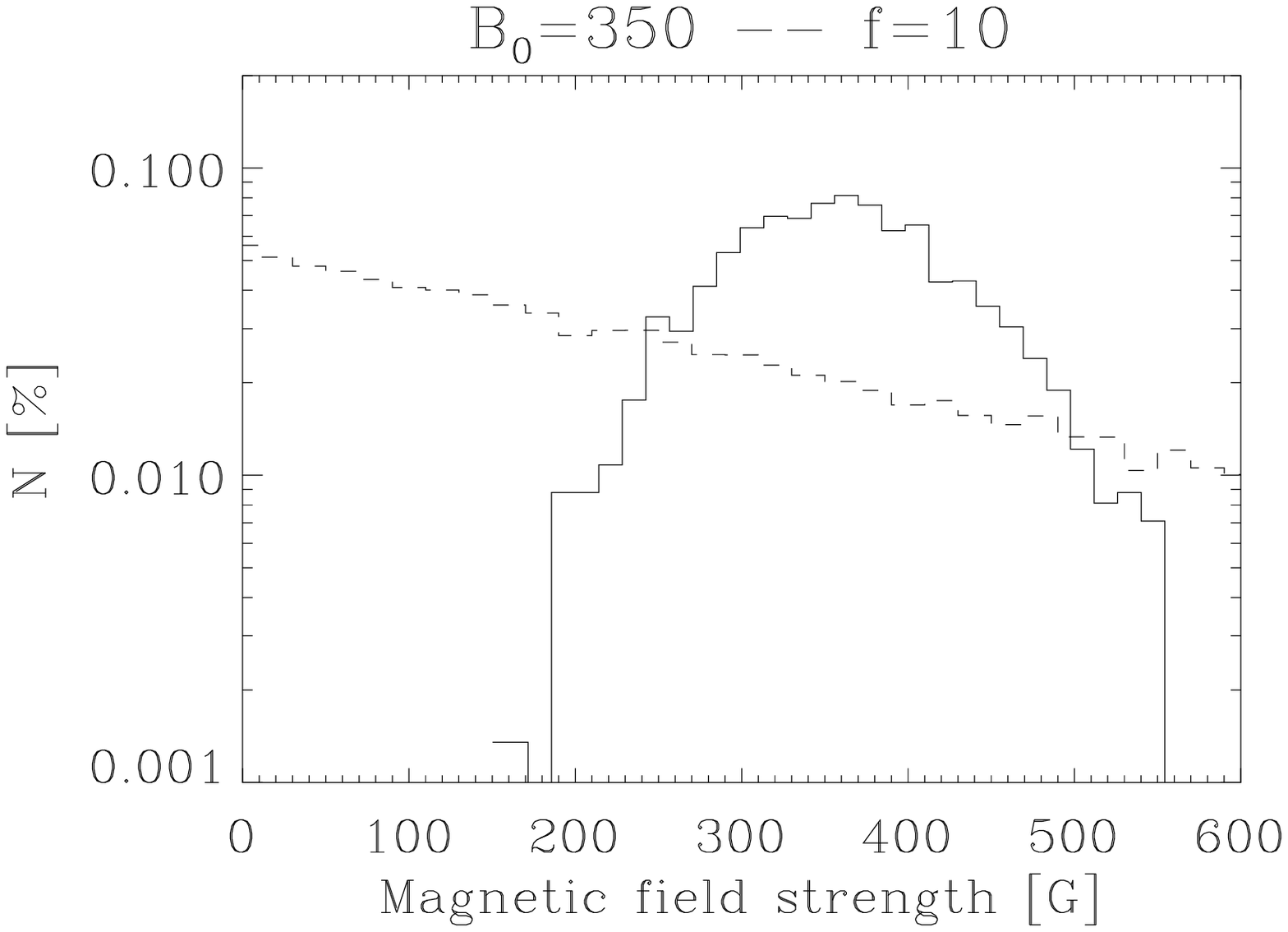}
\plottwo{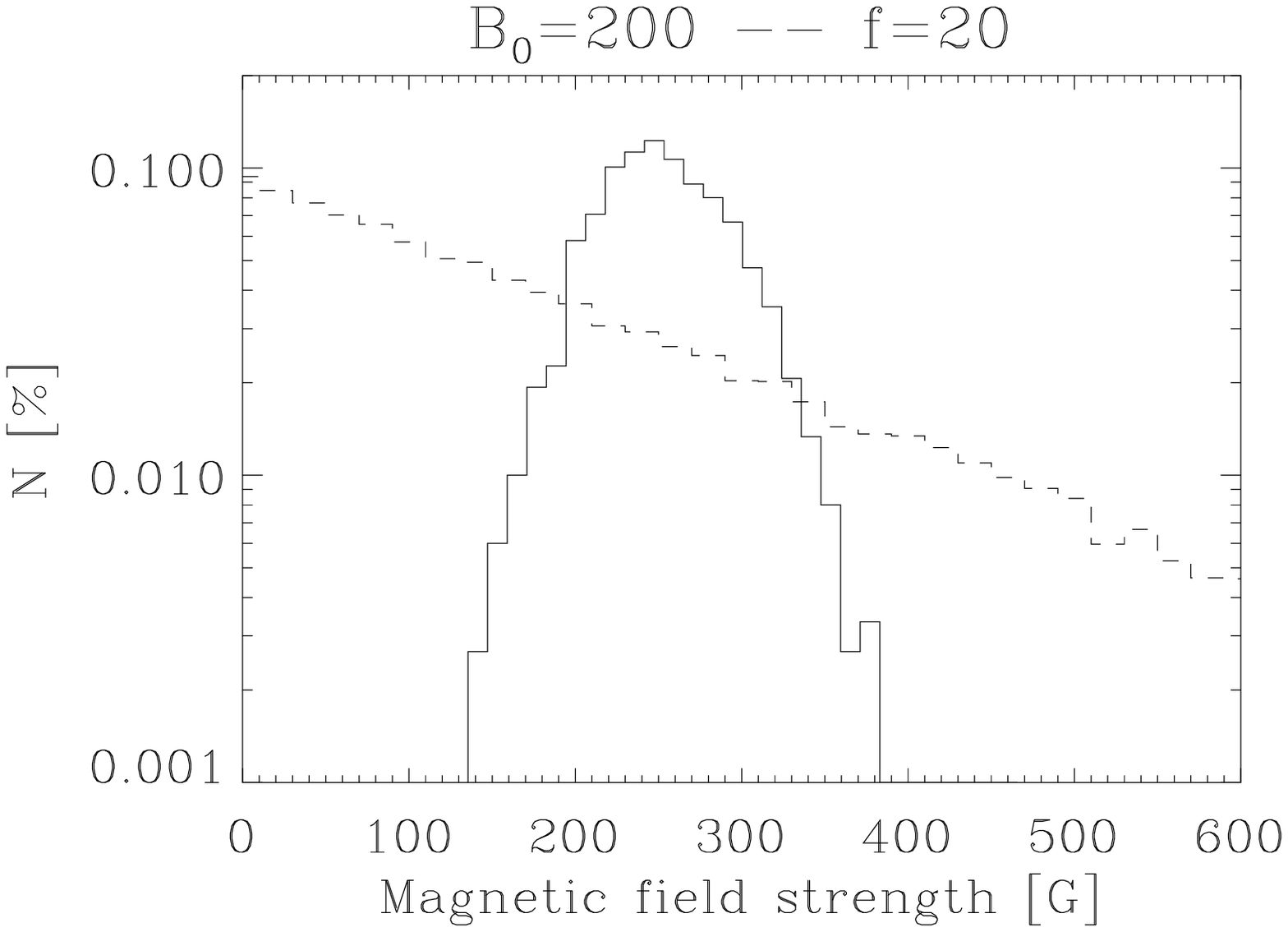}{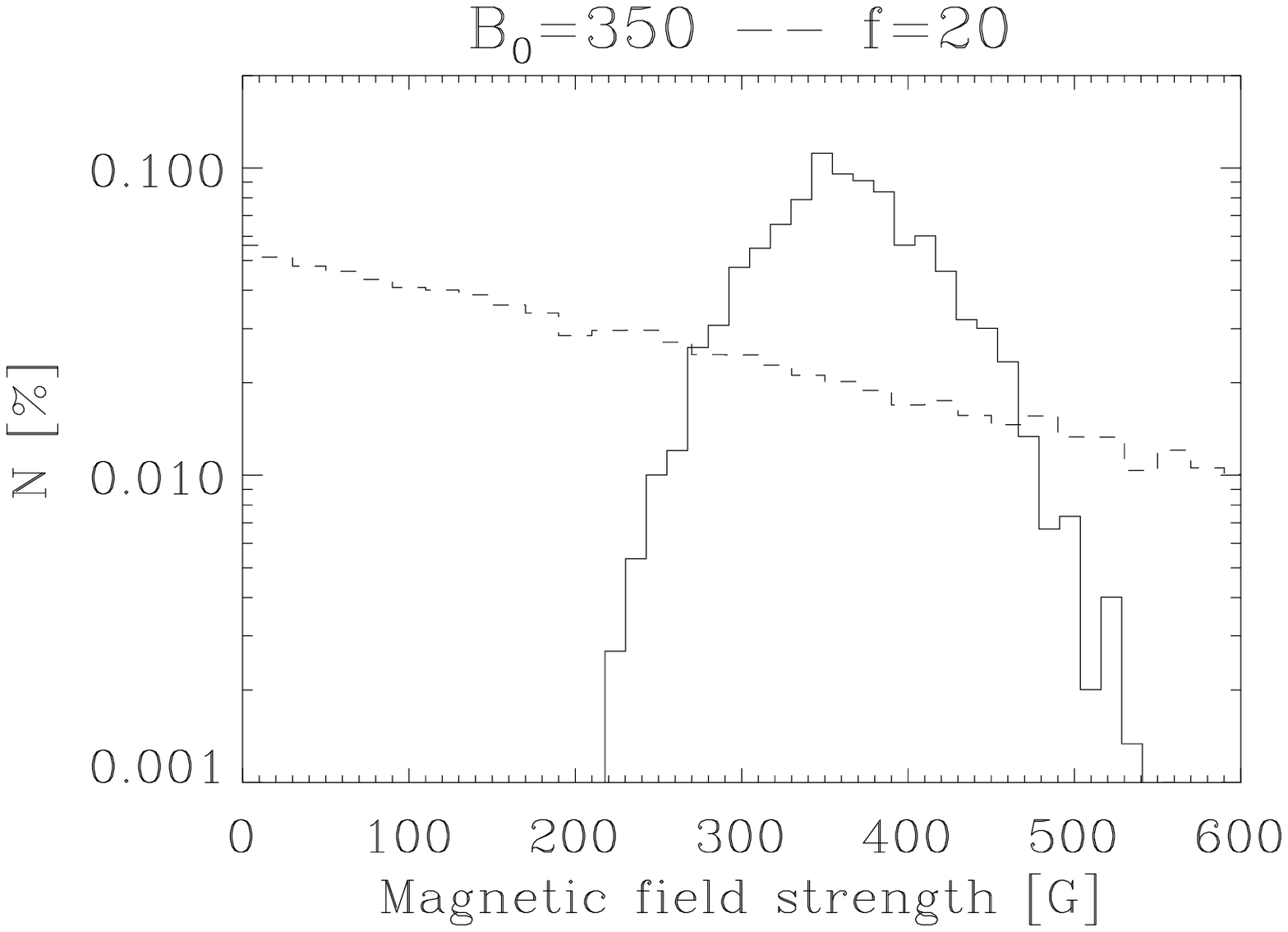}
\caption{Histogram of the magnetic field recovered by using the peak ratio of Stokes $I$ in the
numerical experiment. The original PDFs are exponentials with $B_0=200$ G and $B_0=350$ G
(shown in all the
panels with dashed lines) and is representative of the internetwork and network results, respectively. The solid line 
represents the recovered PDF using different values of $f$
that reproduces the loss of spatial resolution. Note that the slope and the behavior for weak fields
is correctly recovered when all the points are taken into account. When the resolution gets
worse, the retrieved PDF tends to a Gaussian centered at $B_0=200$ G for the first case and $B_0=350$ G
for the second case.\label{fig:experiment}}
\end{figure*}
\clearpage
\setlength{\voffset}{0mm}
\begin{figure*}[!ht]
\plottwo{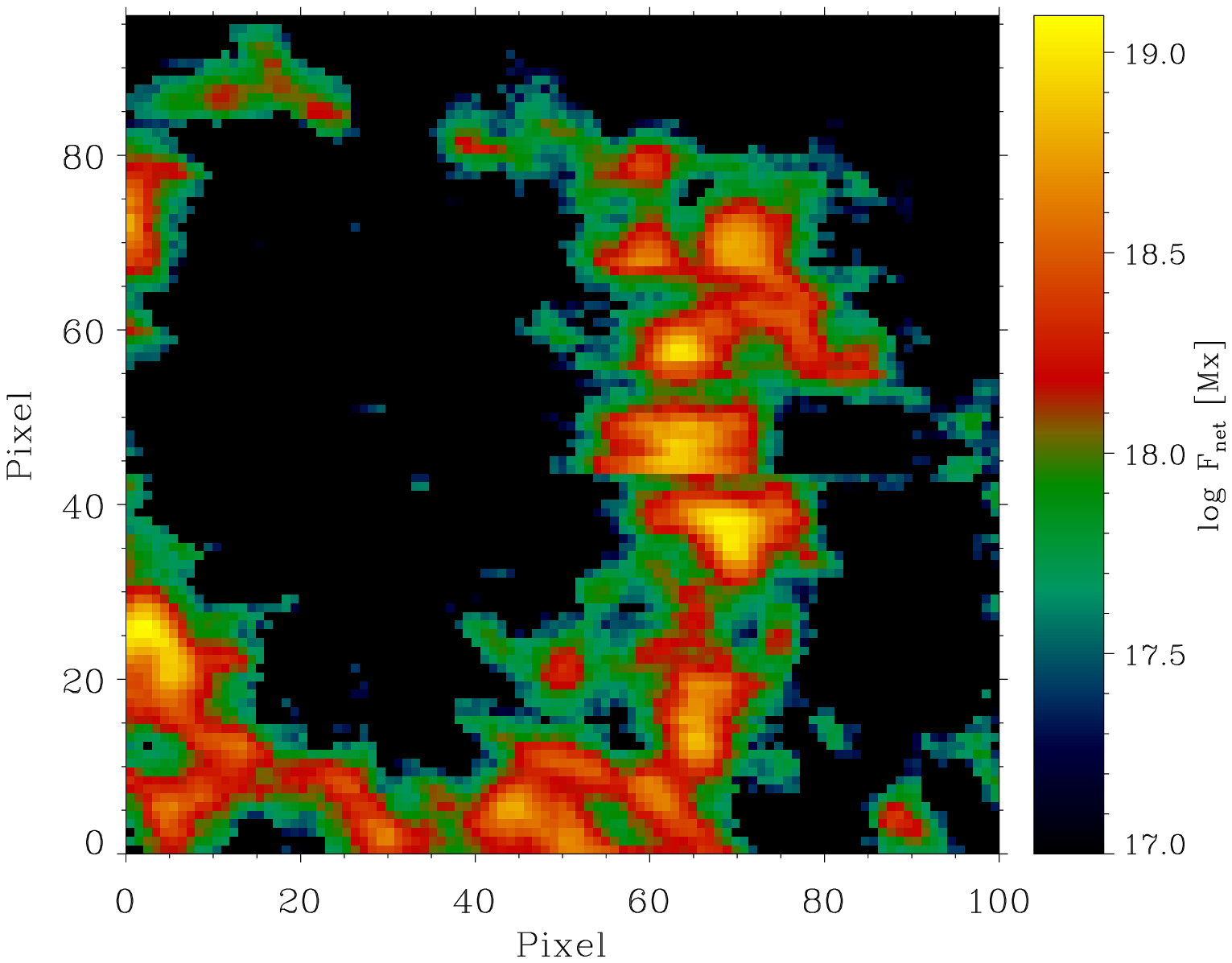}{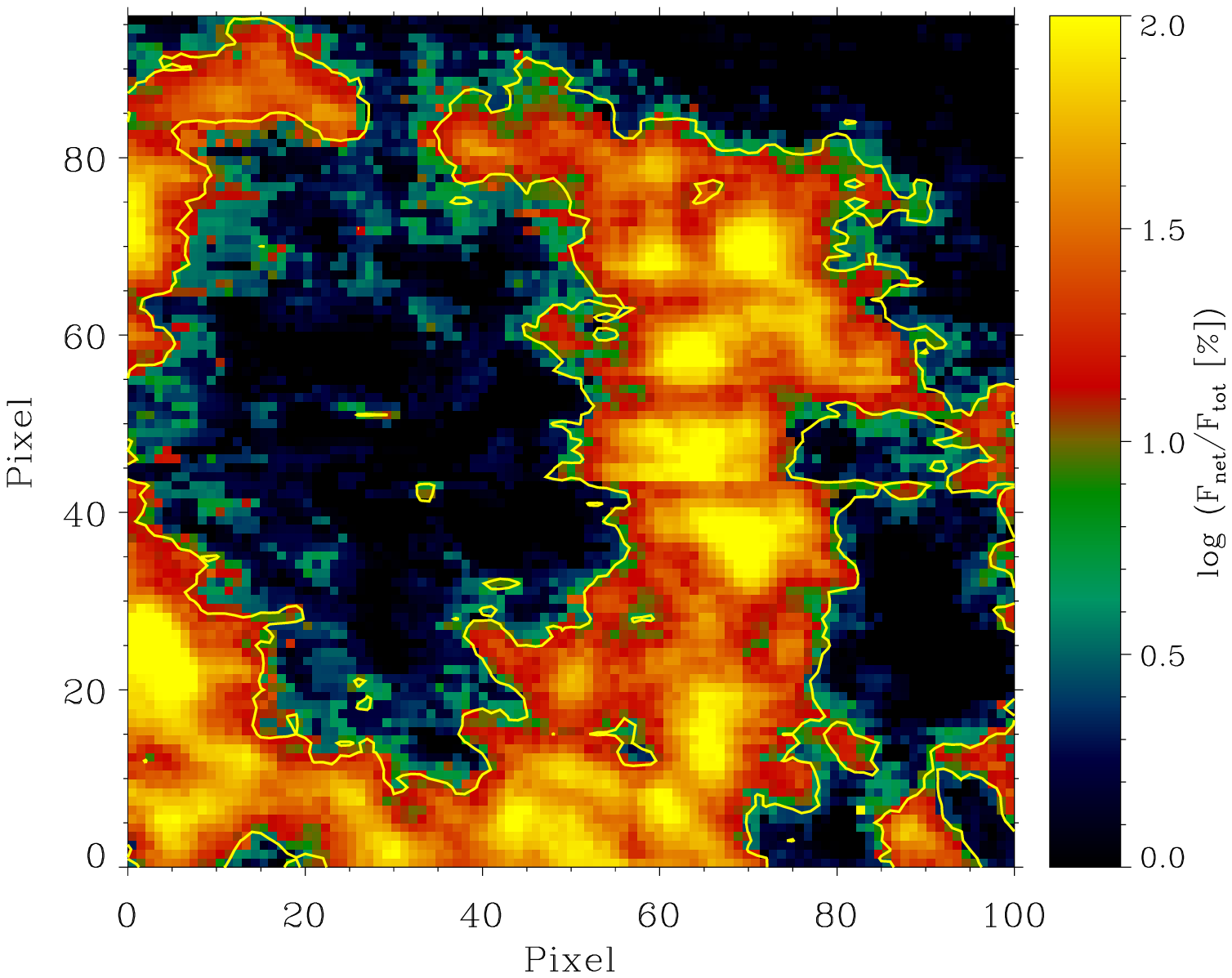}
\caption{The left panel shows the flux in the network in a logarithmic scale. Note that the values of the flux
are in accordance with the typical values. The right panel shows the ratio (in percentage) between the net flux and the total
flux in the network and in the internetwork. A value of 2 (i.e., 100\%) means that the no flux is cancelled, while
a small value means that some flux is cancelled. The total flux is obtained from the field strength calculated 
from the Stokes $I$ peak ratio. The net flux has been obtained in two different ways. The points inside
the contour, having Stokes $V$ profiles that clearly show clear indications of Zeeman saturation, we use 
Eq. (\ref{eq:ff_strong}). For
the rest of points, we use a net flux associated with each class of the SOM classification using the
weak-field approximation.\label{fig:net_total_flux}}
\end{figure*}
\begin{figure*}[!ht]
\plotone{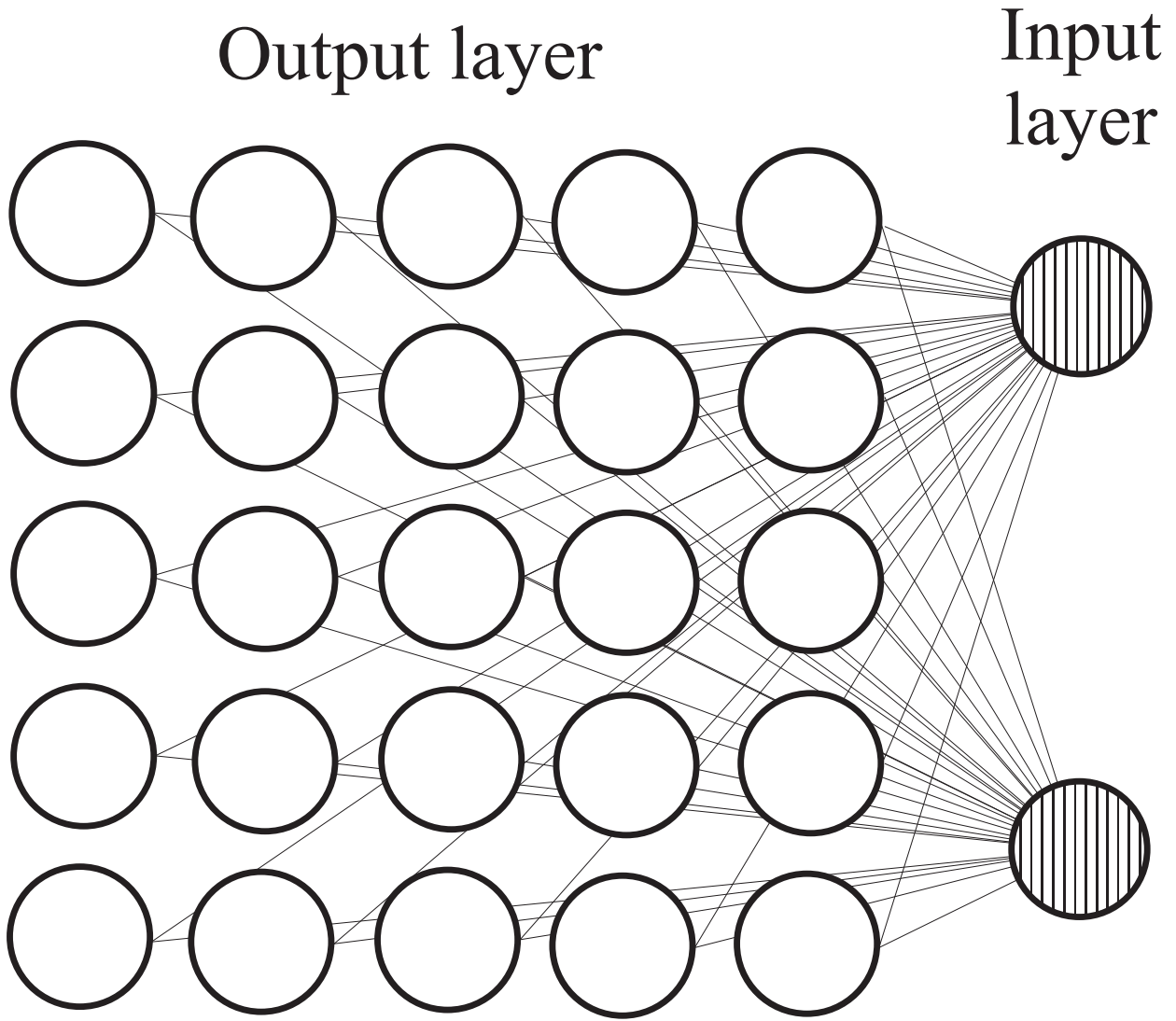}
\caption{Topology of the self-organizing map. The input layer has dimension $N_\mathrm{in}$ and the 
output layer consists on a $N_\mathrm{out} \times N_\mathrm{out}$ map of neurons. A weight
vector of dimension $N_\mathrm{in}$ is associated with each neuron.\label{fig:som}}
\end{figure*}

\end{document}